\documentclass[%
 reprint,
 amsmath,amssymb,
 aps,
]{revtex4-1}
\pdfoutput=1

\usepackage{url,comment}
\usepackage{times}
\usepackage{latexsym}
\usepackage{graphicx, graphics, hyperref, amsmath, amssymb, slashed, xcolor, bbm,amsthm, array,color}
 \usepackage{subfigure}

\usepackage{rotating}
\usepackage{afterpage}

\def\ifb{{\ \rm fb}^{-1}}

\newcommand{\nc}{\newcommand}
\nc{\beq}{\begin{equation}}
\nc{\eeq}{\end{equation}}
\nc{\barray}{\begin{eqnarray}}
\nc{\earray}{\end{eqnarray}}
\nc{\barrayn}{\begin{eqnarray*}}
\nc{\earrayn}{\end{eqnarray*}}
\nc{\bcenter}{\begin{center}}
\nc{\ecenter}{\end{center}}
\nc{\mc}{\mathcal}
\nc{\er}[1]{(\ref{eq:#1})}
\nc{\onehalf}{\frac{1}{2}} 
\nc{\partialbar}{\bar{\partial}}
\nc{\psit}{\widetilde{\psi}}
\nc{\Tr}{\mbox{Tr}}
\nc{\hc}{\mbox{H.c.}}
\nc{\ev}{\;\mathrm{eV}}
\nc{\mev}{\;\mathrm{MeV}}
\nc{\gev}{\;\mathrm{GeV}}
\nc{\tev}{\;\mathrm{TeV}}

\def\chii0{\chi_i^0}
\def\chij0{\chi_j^0}

\newcommand{\gsim}{\lower.7ex\hbox{$\;\stackrel{\textstyle>}{\sim}\;$}}
\newcommand{\lsim}{\lower.7ex\hbox{$\;\stackrel{\textstyle<}{\sim}\;$}}
\nc{\ttbar}{t\bar t}
\def\ifb{{\ \rm fb}^{-1}}

\newcommand{\fref}[1]{Fig.~\ref{f.#1}}
\newcommand{\eref}[1]{Eq.~(\ref{e.#1})}

\newcommand{\sref}[1]{Section~\ref{s.#1}}
\newcommand{\ssref}[1]{Section~\ref{ss.#1}}
\newcommand{\sssref}[1]{Section~\ref{sss.#1}}
\newcommand{\cref}[1]{Chapter~\ref{c.#1}}
\newcommand{\tref}[1]{Table~\ref{t.#1}}

\definecolor{darkgreen}{rgb}{0,0.5,0}

\graphicspath{{figs/}}

\mathchardef\mhyphen="2D 
\def \nonisotoiso{\mathrm{non\mhyphen iso\to iso }}
\newcommand{\deltaphi}{\Delta \phi(\mathrm{MET}, \mathrm{DV})}
\def\eg{{\it e.g.}}

\begin{document}

\title{
Data-driven Model-independent Searches for Long-lived Particles at the LHC
}

\author{Andrea Coccaro}
\thanks{andrea.coccaro@unige.ch}
\affiliation{Section de Physique, Universit\'e de Gen\`eve}

\author{David Curtin} 
\thanks{dcurtin1@umd.edu} 
\affiliation{Maryland
  Center for Fundamental Physics, University of Maryland, College
  Park, MD 20742}

\author{H. J. Lubatti}
\thanks{lubatti@u.washington.edu}
\affiliation{Department of Physics, University of Washington, Seattle, WA 98195}

\author{Heather Russell}
\thanks{heathrus@uw.edu}
\affiliation{Department of Physics, University of Washington, Seattle, WA 98195}

\author{Jessie Shelton} \thanks{sheltonj@illinois.edu}
\affiliation{Department of Physics, University of Illinois at
  Urbana-Champaign, Urbana, IL, 61801}

\begin{abstract}

Neutral long-lived particles (LLPs) are highly motivated by 
  many BSM scenarios, such as theories of
  supersymmetry, baryogenesis, and neutral naturalness, and present
   both tremendous discovery opportunities and experimental challenges for the LHC.
A major bottleneck for current LLP searches is
  the prediction of SM backgrounds, which are often impossible to
  simulate accurately.   In this paper, we propose a general strategy for obtaining
  differential, data-driven background estimates in LLP searches,
  thereby notably extending the range of LLP masses and lifetimes that
  can be discovered at the LHC. We focus on LLPs decaying in the ATLAS
  Muon System, where triggers providing both signal and control
  samples are available at the LHC Run-2.
  While many existing searches require two displaced decays, a detailed
  knowledge of backgrounds will allow for very inclusive searches that require just one
  detected LLP decay. As we demonstrate for the $h \to X X$ signal model of
  LLP pair production in exotic Higgs decays, this results in dramatic
  sensitivity improvements for proper lifetimes $\gtrsim 10$ m. In theories of Neutral Naturalness, this extends reach to glueball masses far below the $\bar b b$ threshold.
  Our strategy readily generalizes to other signal models, and other
  detector subsystems.
  This framework therefore lends itself to the development of a
  systematic, model-independent LLP search program, in analogy to the
  highly successful simplified-model framework of prompt searches.

\end{abstract}
\maketitle

\setcounter{equation}{0} \setcounter{footnote}{0}

\section{Introduction}
\label{s.intro}

The discovery of the Higgs boson has filled in the last missing piece
of the Standard Model (SM). It also has focused attention on the many
open issues the SM does not address, including dark matter, the
matter-antimatter asymmetry of our universe, and the naturalness of
the electroweak scale in the absence of obvious TeV-scale signals of
physics beyond the SM (BSM) to date.
Many extensions of the SM that address these issues allow or require for
long-lived particles (LLPs) that decay at some macroscopic distance
from the pp interaction point.  
Examples include mini-split supersymmetry (SUSY)
\cite{Arvanitaki:2012ps, ArkaniHamed:2012gw}, gauge mediation
\cite{Giudice:1998bp}, RPV SUSY \cite{Barbier:2004ez, Csaki:2013jza},
Stealth SUSY \cite{Fan:2011yu}, models of baryogenesis
\cite{Bouquet:1986mq, Campbell:1990fa, Cui:2012jh, Barry:2013nva,
  Ipek:2016bpf}, Hidden Valleys
\cite{Strassler:2006im,Strassler:2006ri,Strassler:2006qa,Han:2007ae,Strassler:2008bv,Strassler:2008fv},
dark photons \cite{Curtin:2014cca, Clarke:2015ala, Arguelles:2016ney},
and theories of Neutral Naturalness \cite{Burdman:2006tz, Cai:2008au,
  Chacko:2005pe}.
The proper decay lengths of such particles can range from the
mesoscopic (about $ 100 \mu$m) to on the order of kilometers, far in
excess of the detector scale.

As SM events overwhelmingly yield prompt signatures, displaced decays
can be extremely powerful probes of new physics at the LHC
\cite{
  Cui:2014twa, Craig:2015pha, Liu:2015bma,  Csaki:2015uza, Curtin:2015fna, Evans:2016zau}.  Using LHC data collected in Run-1, the
ATLAS, CMS, and LHCb collaborations have performed many searches for
LLPs.  A variety of signatures have been explored targeting neutral
particles decaying to visible particles within the detector volume,
including such detector signatures as displaced vertices (DVs) in the
inner tracker or in the muon spectrometer and jets with an unbalanced
energy deposit \cite{
ATLAS:2012av,
Aad:2012kw,
Aad:2012zx,
Aad:2015uaa, 
Aad:2015asa, 
Aad:2014yea, 
Aad:2015rba, 
Chatrchyan:2012sp,
Chatrchyan:2012jna,
CMS:2014wda, 
CMS:2014hka, 
Aaij:2014nma}.  Many of these searches have been constructed to be
background-free by applying stringent selection cuts or by requiring
two reconstructed objects that target two displaced decays. This
strategy guarantees tiny contributions from rare or mis-reconstructed
SM events
at the expense of limiting the kinematic region being explored.

While the increase in center-of-mass energy offered by Run-2 will certainly extend
the sensitivity of these searches to a broader range of masses and
proper lifetimes, substantially increasing the Run-2 mass-lifetime
reach for LLPs will require search strategies where the SM
background is no longer negligible in comparison to the expected
signal sample.  This is challenging, as the SM background to most
displaced signatures is notoriously difficult to model reliably.  SM
backgrounds to LLP searches generally can be understood as arising
from a combination of unusual physics in an event (such as a jet
giving rise to multiple tracks in the muon system) and unusual
detector response (such as the hadron calorimeter (HCAL) failing to
register a substantial fraction of the energy of a jet).  While such
events are extremely rare, they can nevertheless occur in appreciable
numbers due to the extremely high rates of SM processes such as jet
production. Reliably simulating these backgrounds in Monte Carlo is
not possible.  A data-driven approach to determine these backgrounds
is thus required. While a data-driven approach to measuring backgrounds for LLP searches
has been pioneered in some Run-1 searches \cite{Aad:2012kw,
  Aad:2014yea, Khachatryan:2014mea}, pursuing this approach in the LLP program
in general is highly nontrivial, since for most displaced signatures,
it is a challenge simply to record the data sets that would allow the
relevant background(s) to be measured.

In this paper, we describe a background estimation strategy which
depends on defining a primary trigger that selects decays of
long-lived particles in an LHC detector subsystem, and an accompanying
trigger that selects a set of events with analogous kinematic
properties but  consisting of mainly background events.

For example, a general trigger selecting displaced decays of neutral
objects to hadronic jets in the calorimeter would include an isolation
criterion (``iso'') to ensure there is little or no activity in a
$\Delta R$ cone upstream of the jet, in order to reject backgrounds
from QCD jets. A trigger that selects an orthogonal non-isolated set
of events would not include an isolation criterion (``non-iso''). The
details of the primary and orthogonal triggers will depend on the
detector and available trigger-level information. In principle,
implementing such primary and accompanying orthogonal triggers is
possible in the inner tracker, calorimeters and muon systems of both
the CMS and ATLAS detectors. The orthogonal, non-iso trigger selects a
background-rich sample of events that can be compared to the
potentially signal-rich iso sample. In this paper we show that this
approach, when combined with a signal-like-region (SR$_Y$) vs
control-like-region (CR$_Y$) split using an independent variable $Y$
based on expected signal properties, can significantly extend the
range of proper lifetime sensitivity.

We concentrate on DV searches in the ATLAS Muon Spectrometer (MS) in
order to illustrate these ideas. This has the advantage of building on
established trigger options of the ATLAS experiment.  We present a
simple analysis demonstrating how the iso and non-iso trigger samples
in the MS can be used to obtain a differential estimate of the SM rate
for single isolated DVs in a signal region of interest.  Using this
estimate, we can greatly improve the sensitivity of LHC searches for
LLPs with proper lifetimes greater than a meter, as we show using the
challenging and well-motivated example of Higgs decays into LLPs.

Our approach lends itself to the formulation of a model-independent
LLP search program in the Muon System, where different signal
topologies can be grouped together by the LLP production mode. We also
expect the principles of our approach to transfer to LLP searches in
different detector subsystems and hence LLP lifetimes.

This paper is structured as follows.  Sec.~\ref{s.overview} lays out
our proposed general strategy for obtaining data-driven background
estimate for LLP searches in the Muon System.  In Sec.~\ref{s.htoXX}
we apply this strategy to the example of SM Higgs decays to pairs of
LLPs, demonstrating substantially improved reach at long lifetimes,
and interpret the gain in sensitivity for theories of Neutral
Naturalness.  Sec.~\ref{s.futuredirections} outlines extensions of our
general strategy to model-independent searches for other signal
topologies, and application to other detector systems. We conclude in
Sec.~\ref{s.conclusions}.

\section{Estimating Background in the Muon System}
\label{s.overview}

In this section we establish a general strategy to obtain 
background estimates for LLP searches.
Our strategy is a generalization of the `ABCD' method that relies on having both a trigger that targets displaced
signal objects, and a trigger that can record a suitable
background-dominated control sample.  While implementing a suitable
pair of triggers is a challenge, in the case of LLPs that decay in the
MS, such trigger streams are available at ATLAS.  The Muon Region of
Interest (RoI) Cluster (``iso'') trigger, used in Run-1 searches for
LLPs that decay near the outer region of the HCal or in the MS
\cite{Aad:2015uaa, ATLAS:2012av}, selects an isolated cluster of muon tracks (muon
RoIs) in a $\Delta R =0.4$ cone with little or no activity in the
inner tracker or calorimeter~\cite{MuonRoITrigger}.  The isolation
requirement reduces backgrounds from punch-through jets and muon
bremsstrahlung. An example of an LLP event signature that could pass
the Muon RoI Cluster trigger is shown in \fref{basicprinciple} (a),
and \fref{basicprinciple} (b) shows a SM background punch-through
topology that has no inner tracker (IT) or calorimeter signal and thus
survives the isolation requirement.
New in Run-2 is an `orthogonal' trigger that is identical to
the Muon RoI Cluster trigger except that isolation requirements are
not imposed. This trigger can provide the necessary orthogonal,
non-isolated control sample. A typical event topology selected by this
orthogonal trigger is shown in \fref{basicprinciple} (c). Note that we refer to any standard detector objects (jets, leptons, etc.,) not directly connected to the displaced vertex as \emph{Associated Objects} (AOs). 

The ATLAS Run-1 search performed using the Muon RoI Cluster trigger
used an off-line, custom-built, standalone vertex reconstruction
algorithm to reconstruct DVs in the MS~\cite{MSVXRECO}, and required
two reconstructed vertices in the MS or one in the MS and one in the
IT. Requiring two reconstructed displaced vertices effectively
eliminated SM backgrounds at the price of reducing signal
efficiencies.  Since LLPs are often pair-produced (for example, in a
model yielding exotic Higgs decays $h \to X X$, where $X$ is
long-lived), the search has excellent sensitivity for proper lifetimes
($c\tau$) of tens of meters, but the requirement that both particles
decay in the MS degrades limits when the proper lifetime $c\tau$ is
longer, with exclusions on cross-sections scaling as $(c
\tau)^{-2}$. In addition, a search for two DVs is completely
insensitive to singly produced LLPs.

A search requiring only one reconstructed vertex in the MS would
significantly extend the sensitivity for longer-lived or singly
produced LLPs, with limits scaling as $(c \tau)^{-1}$.  However,
relaxing the requirement of two reconstructed vertices requires that
the no longer negligible backgrounds from jet punch-through and other
sources can be properly estimated. This means that we need to estimate
$\sigma^{SM}_\mathrm{iso}$, the effective cross-section for objects
produced in SM processes that (i) fake a displaced decay by passing
the isolation criteria of the Muon RoI Cluster trigger and (ii)
reconstruct a displaced vertex.  In fact, what is needed is not simply
the total cross-section, but the differential cross-section
\begin{equation}
\frac{d \sigma^{SM}_\mathrm{iso}}{d x_1 d x_2 \ldots} \,
\end{equation}
where the $x_i$ are kinematic variables computed using AOs, such as
$H_T=\sum_i|p_{T,j_i}|+$MET, jet $p_T$, etc., in order to allow for
the use of kinematic cuts on such variables to enhance sensitivity to
BSM physics.

The major contribution to $\sigma^{SM}_\mathrm{iso}$ comes from QCD
processes such as \fref{basicprinciple} (b), where each jet has a
small probability, $\epsilon^\mathrm{fake}_\mathrm{iso}$, to pass the
isolation criteria of the MS RoI cluster trigger and be reconstructed
as a displaced vertex in the MS. Parametrically, ignoring jet
multiplicity factors etc.,
\begin{equation}
\label{e.epsiloniso}
\frac{d \sigma^{SM}_\mathrm{iso}}{d x_1 d x_2 \ldots} \  \ \sim  \ \ \frac{d \sigma_\mathrm{QCD}}{d x_1 d x_2 \ldots} \ \cdot \ \epsilon^\mathrm{fake}_\mathrm{iso}(x_1, x_2, \ldots) ,
\end{equation}
where $\sigma_{QCD}$ is the inclusive multi-jet production cross
section, which can be calculated or measured directly from
data. Parameterizing a rare background as a known process rescaled by
some empirically determined fake rate such as
$\epsilon_\mathrm{iso}^\mathrm{fake}$ is most reliable when that known
process is a very close match to the background process. Otherwise,
the fake rate may have a strong dependence on kinematic variables or
other event properties that would be difficult to capture
reliably. Simply rescaling standard QCD cross-sections is likely to
miss important effects. The use of the orthogonal (non-iso)
trigger avoids these problems by providing a very closely related
sample of background-dominated events.

\begin{figure*}
\begin{center}
\includegraphics[width=17cm]{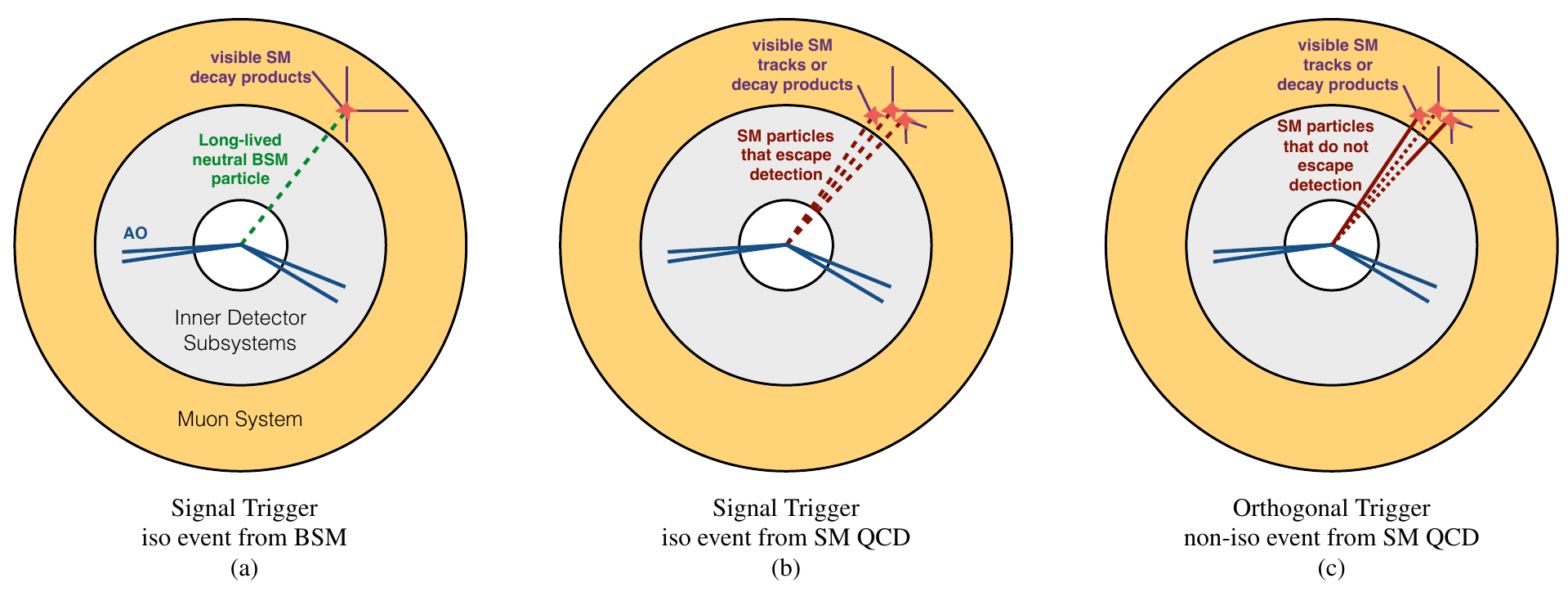}
\end{center}
\caption{{\small Schematic representation of signal and background
    events that pass the signal and orthogonal triggers. Dashed lines
    indicate invisible or undetected particles. All regular detector
    objects (prompt leptons, jets, etc.) that are produced in
    association with the LLP(s) are referred to as \emph{Associated
    Objects} (AOs).  } }
\label{f.basicprinciple}
\end{figure*}

Henceforth we refer to events that pass the iso trigger and 
have a reconstructed displaced vertex as events in the \emph{iso-region}
(or iso-events), and events that pass the non-iso trigger and do not 
pass isolation criteria with a reconstructed displaced vertex as events in the
 \emph{non-iso-region} (or non-iso-events). Because events that
pass the non-iso trigger are SM-dominated, the non-iso trigger rate
will be significantly larger than the rate of the isolated trigger,
ensuring a suitably large control sample for estimating the number of
expected \emph{iso-region} events due to SM backgrounds~\footnote{In
  case of bandwidth limitations, the non-iso trigger rate can be
  adjusted to populate the control regions with the necessary
  statistics.}. Specifically,
\begin{equation}
\frac{d \sigma^{SM}_\mathrm{iso}}{d x_1 d x_2 \ldots} \  \ \approx  \ \ \frac{d \sigma_\mathrm{non\mhyphen iso}}{d x_1 d x_2 \ldots} \ \cdot \ r_\nonisotoiso(x_1, x_2, \ldots)
\end{equation}
and the rescaling function $r_\nonisotoiso$ is related to the ratio of
probabilities
$\epsilon^\mathrm{fake}_\mathrm{iso}/\epsilon^\mathrm{fake}_\mathrm{non\mhyphen
  iso}$, where $\epsilon^\mathrm{fake}_\mathrm{non\mhyphen iso}$ is
the probability for a QCD jet (or other SM event) to fire the non-iso
orthogonal trigger.

The differential rescaling function $r_\nonisotoiso$ allows us to
obtain a prediction of the SM background events in the \emph{iso-region} 
event sample by using the \emph{non-iso-region} events.
The differential determination of $r_\nonisotoiso$ is important for
enabling the imposition of additional cuts. For example, when requiring a
high-$p_T$ jet and/or an isolated lepton in the \emph{iso-region}
event sample, we can obtain a background prediction by applying the
same criteria to the \emph{non-iso-region} events and rescaling.

\subsection{Determining the rescaling function}

The function $r_\nonisotoiso$ needs to be measured from data. This can
be achieved by identifying some variable $Y$ (e.g.,~the number of
identified leptons, or the angle between the MET vector and the
displaced vertex) that fulfills two requirements:
\begin{enumerate}
\item for fixed $x_i$, the rescaling function $r_\nonisotoiso$ is
  independent of $Y$, and
\item $Y$ can be used to split the iso and non-iso-events into a
  signal-like region SR$_Y$, and a control-like region CR$_Y$. SR$_Y$
  contains the BSM signal of interest while CR$_Y$ is by comparison
  SM-enriched.
\end{enumerate}
The separation of the \emph{iso-region} and \emph{non-iso-region}
events into SR$_Y$ and CR$_Y$ by using the variable $Y$ results in
one signal region A and three control regions B,C,D as shown
schematically in \fref{SRCR}.  The BSM signal events dominantly
populate region A.  As noted above, by design $r_\nonisotoiso$ is the
same in SR$_Y$ as in CR$_Y$ and consequently can be determined from
data in regions C and D,
\begin{equation}
\label{e.rmeasurement}
r_\nonisotoiso(x_1, x_2, \ldots) \ \equiv \ 
\frac{d \sigma_\mathrm{C}^\mathrm{data}}{d x_1 d x_2 \ldots} \ \cdot \ \left[\frac{d \sigma_\mathrm{D}^\mathrm{data}}{d x_1 d x_2 \ldots}\right]^{-1} \ ,  
\end{equation}
thus making it possible to obtain a background prediction for region
A:
\begin{equation}
\label{e.sigmaAprediction}
\frac{d \sigma_\mathrm{A}^\mathrm{prediction}}{d x_1 d x_2 \ldots} \  \ =  \ \ \frac{d \sigma_\mathrm{B}^\mathrm{data}}{d x_1 d x_2 \ldots} \ \cdot \ r_\nonisotoiso(x_1, x_2, \ldots) .
\end{equation}
Having the background prediction enables a search for BSM signals with
just one DV in the Muon Spectrometer.

\begin{figure}
\begin{center}
\includegraphics[width=7.5cm]{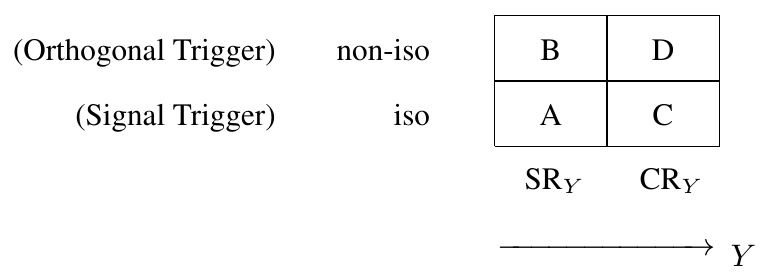}
\end{center}
\vspace*{-3mm}
\caption{ {\small Schematic representation of the four regions A, B, C, D into
  which the \emph{iso-region} and the \emph{non-iso-region} events are
  divided. By construction, region A is significantly enriched with
  BSM signal events compared to region C.}}
\label{f.SRCR}
\end{figure}

There is the practical question of how to parametrize the function
$r_\nonisotoiso$. The SM contribution to both the iso-region and non-iso-region 
is dominated by events where a single jet reconstructs a vertex in the
MS.  Both the probability that a jet will reconstruct a vertex in the
MS and the probability that the vertex will pass isolation criteria
depend on the local properties of the jet itself. Thus
we expect $r_\nonisotoiso$ to be a function of the jet's
$p_T$ and $\eta$ (assuming azimuthal symmetry). In general we would expect the $\eta$
dependence to be non-negligible, resulting in, \eg, different values
of $r_\nonisotoiso$ in the barrel and in the endcaps.  For our present
purposes, we approximate $r_\nonisotoiso$ as independent of $\eta$ for
simplicity, and focus on what we expect to be the most important
kinematic dependence, the jet $p_T$.  Given that an energy measurement
of the jet that fakes the DV is not available (by definition,
especially in the iso-region), we hypothesize that the kinematic
dependence of $r_\nonisotoiso$, to the extent that it exists, can be
captured mostly as a function of the variable $H_T^\prime$,
\begin{equation}
\label{e.HTprime}
H_T' \equiv \left( \sum_i |\vec p_{T \, i}^{\ (AO)}| \right) + \mathrm{MET}^\prime
\end{equation}
where $\vec p_{T \, i}^{\ (AO)}$ are the transverse momenta of the
associated objects (regular prompt leptons, jets, etc.) and
\begin{equation}
\mathrm{MET}^\prime = |\vec E_T^{\  \prime \,\mathrm{miss}}|,
\phantom{move}
  \vec E_T^{\  \prime \,\mathrm{miss}}\equiv - \sum_i \vec p_{T \, i}^{\ (AO)}.
\end{equation}
Note that $\vec E_T^{\ \prime \,\mathrm{miss}}$ is just the regular
transverse missing energy 2-vector for iso-events, but for non-iso
events, jets between the interaction point and the Muon RoI are
treated as invisible to ensure the variable definitions are equivalent
for iso- and non-iso-region events.  We therefore assume for this paper that
$r_\nonisotoiso \approx r_\nonisotoiso(H_T^\prime)$.  In practice, an
experimental analysis adopting this approach will need to determine
the most useful parameterization~\footnote{For example, it may be that
  $\mathrm{MET}^\prime$ is the more useful variable, since it more
  directly isolates the properties of the individual jet producing the
  DV, but this detail does not affect our discussion.}.
	
In any analysis it will be important to assess both the applicability
of this background estimation technique and the systematic uncertainty
in the experimental determination of the function $r_\nonisotoiso$. This can  
be done by further subdividing CR$_Y$ and checking for consistency of
\eref{sigmaAprediction}. In addition, when splitting both iso- and
non-iso-region events into SR$_Y$ and CR$_Y$ using the kinematic variable
$Y$, care has to be taken that events in control region C populate the
same range of relevant kinematic variables (e.g., $H_T^\prime$) as BSM
events in the signal region A.

\subsection{Statistical Uncertainties and Cuts}

As usual, the uncertainty of the resulting data-driven prediction for
the background is limited by the statistical uncertainty in the
control regions.  We now discuss the statistical precision available
for a background estimate given a choice of signal and control
regions, first by ignoring non-$Y$ kinematic dependencies for
simplicity, and then extending the discussion to the case of interest
where signal and control regions are considered differentially in
$H_T'$.

The first task is to choose the kinematic variable $Y$, defining the
signal-like and control-like regions of \fref{SRCR}. The number of iso
BSM and SM events in SR$_Y$ and CR$_Y$ must satisfy
\begin{equation}
\frac{N_\mathrm{C}^\mathrm{BSM}}{N_\mathrm{A}^\mathrm{BSM}}  
<
\frac{N_\mathrm{C}^\mathrm{SM}}{N_\mathrm{A}^\mathrm{SM}}  \ .
\end{equation}
Ideally, the inequality is actually $\ll$, but in either case one
might have to deal with BSM contamination of CR$_Y$, which we will
discuss below. Because $r_\nonisotoiso$ is by assumption the same in
SR$_Y$ and CR$_Y$, and the non-iso-events are highly SM-dominated, we
can write this condition as
\begin{equation}
\label{e.SRCRcondition}
\frac{N_\mathrm{C}^\mathrm{BSM}}{N_\mathrm{A}^\mathrm{BSM}}  
<
\frac{N_\mathrm{D}}{N_\mathrm{B}}  \ ,
\end{equation}
where the LHS can be computed from the Monte Carlo signal prediction,
and the RHS is determined purely from data.

Note that satisfying \eref{SRCRcondition} does \emph{not} imply that
$N_\mathrm{D} > N_\mathrm{B}$, or equivalently
$N_\mathrm{C}^\mathrm{SM} > N_\mathrm{A}^\mathrm{SM}$. It merely
requires that CR$_Y$ contain a larger fraction of iso SM events than
SR$_Y$. Ignoring kinematic dependence,
\begin{equation}
\label{e.rdef}
r_\nonisotoiso = \frac{N_C}{N_D} , 
\end{equation}
with relative uncertainty
\begin{equation}
\label{e.deltar}
\frac{\delta r_\nonisotoiso}{r_\nonisotoiso} = \frac{1}{\sqrt{N_C}} \ .
\end{equation}
(For simplicity, we here ignore contamination from BSM events in the
various control regions, ignore systematic uncertainties in the determination
of $r_\nonisotoiso$, and assume that all event numbers are
sufficiently large that the Poisson fluctuation for $N$ events is
simply $\sqrt{N}$. We also ignore the subdominant contribution to
$\delta r_\nonisotoiso$ from the statistical uncertainty in region D,
since the non-iso-region is much more populated than the iso-region.)
This gives the expected number of background events in region A,
\begin{equation}
\langle N_A^\mathrm{SM}\rangle \ \ =  \ \ r_\nonisotoiso  \, N_B  \ \ =  \ \ \frac{N_C}{N_D} \, N_B.
\end{equation}
Therefore, if the ideal scenario of $N_\mathrm{D} \gg N_\mathrm{B}$ is
realized (keeping in mind that $N_B \gg N_A$ and $N_D \gg N_C$), the
95\% CL limit on the number of BSM events in region $A$ is
approximately
\begin{equation}
N_A^\mathrm{BSM}  \ \ <  \  \ 2 \sqrt{N_A^\mathrm{SM}} \ ,
\end{equation}
which is determined by the Poisson fluctuations of the SM background
in region A, with no significant added uncertainty from
$r_\nonisotoiso$.  If, conversely, CR$_Y$ is not as populated as
SR$_Y$, i.e., $N_\mathrm{D} \ll N_\mathrm{B}$ and hence
$N_\mathrm{C}^\mathrm{SM} \ll N_\mathrm{A}^\mathrm{SM}$, then the
rescaling uncertainty in \eref{deltar} is much larger than the Poisson
fluctuations of the SM background in signal region A. Therefore, the
95\% CL limit on the number of BSM events in region $A$ is
approximately
\begin{equation}
N_A^\mathrm{BSM}  \  \ <  \ \  2 \sqrt{N_A^\mathrm{SM}} \ \sqrt{
\frac{N_B}{N_D}
}
\ .
\end{equation}
The sensitivity is degraded by the square root of the factor by which
CR$_Y$ has worse statistics than SR$_Y$.

We now restore the kinematic dependence to make explicit how cuts on
kinematic variables are performed. Since we parameterize
$r_\nonisotoiso$ as a function of $H_T^\prime$, we will treat all
events, whether simulated, predicted, or from data, as binned in
$H_T^\prime$, with bins $H_{T,i}^\prime$ and bin occupations $N_i$.

All of the above expressions apply in each $H_T'$ bin, i.e., taking $N \to
N_i$, etc. So, for example, the rescaling function is defined bin-by-bin as
\begin{equation}
\label{e.rdefi}
r_\nonisotoiso(H_{T,i}^\prime) = r_\nonisotoiso^i = \frac{N_{C,i}}{N_{D,i}} , 
\end{equation}
with relative uncertainty
\begin{equation}
\label{e.deltari}
\frac{\delta r_\nonisotoiso^i}{r_\nonisotoiso^i} = \frac{1}{\sqrt{N_{C,i}}} \ .
\end{equation}
The background prediction in region A is given by
\begin{equation}
\label{e.NAiSM}
\langle N_{A,i}^\mathrm{SM}\rangle \ \ =  \ \ \left(\frac{N_{B,i}}{N_{D,i}}\right) \, N_{C,i},
\end{equation}
where the statistical uncertainty of $N_{C,i}$ dominates the
uncertainty of $\langle N_{A,i}^\mathrm{SM}\rangle$. In particular, if
no events are observed in a control region C bin, $N_{C,i} = 0$, then we only
have an upper bound on $\langle N_{A,i}^\mathrm{SM}\rangle$. To
perform cuts on the events in region $A$, the corresponding SM
prediction after cuts can be obtained by performing those cuts on the
non-iso-events:
\begin{equation}
\label{e.NAiSMaftercuts}
\langle N_{A,i}^\mathrm{SM, \ after\ cuts}\rangle \ \ =  \ \ \left(\frac{N_{B,i}^\mathrm{after\ cuts}}{N_{D,i}}\right) \, N_{C,i}
\end{equation}
The corresponding predictions $N_{A,i}^\mathrm{BSM}$ or
$N_{A,i}^\mathrm{BSM, \ after \ cuts}$ for the signal can be obtained
from Monte Carlo.

\subsection{Important Considerations}
\label{ss.importantconsiderations}

The best choice of the observable $Y$ used to define CR$_Y$ and SR$_Y$
will depend on the signal model.  Choosing a CR$_Y$ is very easy if,
for example, the LLPs are always or frequently produced in
association with certain specific AOs, such as a lepton. In that case,
a good CR$_Y$ would simply invert the lepton requirement, ensuring
very large CR$_Y$ statistics, and thereby allowing the CR$_Y$ sample
to be subdivided to further reduce systematic uncertainties.  By
contrast, when the LLPs are dominantly produced with few or no AOs, as
occurs for exotic Higgs decays to LLPs, choosing a CR$_Y$ becomes more
challenging.  We discuss this in greater detail in the next Section.

As discussed above, one of the requirements that $Y$ must satisfy is
that $r_\nonisotoiso(H_T^\prime)$ is independent of $Y$ for fixed
values of $H_T'$.  When dealing with a binned rescaling function
$r_\nonisotoiso^i$, \eref{rdefi}, this becomes the requirement that,
in a given $H_T' $ bin, $r_\nonisotoiso(H_T^\prime)$ is a sufficiently
slowly varying function that any correlations of $Y$ with $H_T'$, and
therefore any differences of $r_\nonisotoiso^i$ between SR$_Y$ and
CR$_Y$, are negligible.  A violation of this requirement would
introduce a systematic uncertainty in an \emph{individual bin's} background
prediction in region A that, unlike the statistical uncertainties
discussed above, does not scale with luminosity. Fortunately, since
increased statistics allow for smaller bin sizes, the \emph{overall}
effect of this systematic error on the search sensitivity will
actually decrease with luminosity. As outlined below, we therefore
expect weak correlations to be manageable in a real analysis.

To quantify how slowly varying $r_\nonisotoiso$ needs to be in order
for this systematic error to be negligible, consider a single
$H_T^\prime$ bin $(H_{T,1}^{\prime},H_{T,2}^{\prime})$ with bin
occupation $N_{C}$ in region C, and similarly for region A. Define
\begin{equation}
\label {e.deltahtpdef}
\Delta H_T^\prime = \frac{| \langle H_T^\prime \rangle_C - \langle H_T^\prime \rangle_A |}{H_{T,2}^{\prime} - H_{T,1}^{\prime}}  < 1 \ ,
\end{equation}
which is the difference, between region A and region C, of the mean
$H_T^\prime$ in this bin, normalized to the bin width.  The limit of no correlations between $H_T^\prime$ and $Y$ corresponds to $\Delta H_T^\prime \to 0$.
Assuming the bin is narrow enough
that $r_\nonisotoiso(H_T^\prime)$ is approximately linear
across the bin, the condition that the systematic error in
$r_\nonisotoiso$ is negligible compared to its statistical
uncertainty can then be written as 
\begin{eqnarray}
\nonumber
\frac{|r_\nonisotoiso(H_{T,2}^{\prime})-r_\nonisotoiso(H_{T,1}^{\prime})|}{r_\nonisotoiso(H_T^\prime)} \ll \frac{1}{\sqrt{N_C} \ \Delta H_T^\prime} , 
\\
\label{e.systerrorcondition}
\end{eqnarray}
where the denominator on the LHS is the average value of
$r_\nonisotoiso$ in this bin. (Note that this condition is trivially
satisfied if there are no correlations between $H_T^\prime$ and $Y$.)
In the limit of large statistics, both $\Delta H_T^\prime$ and
$r_\nonisotoiso(H_T^\prime)$ are approximately invariant with
bin-size. On the other hand, the numerator on the LHS and $N_C$
decrease if we shrink the bin.  In optimizing a given analysis, we can
therefore hope to satisfy this condition by choosing the smallest
possible bin size that still ensures each bin in region C is populated
with at least a few events~\footnote{If a bin is so small that its   expected occupation is much less than 1, then the experimentally derived upper limit on its expected occupation will significantly  overestimate $r_\nonisotoiso$.
The minimum useful bin width thus   depends on the background rate in a given search.}.
In that case one
expects $\Delta H_T^\prime \gtrsim \mathcal{O}(0.1)$ purely due to
random scatter. In order for systematic error to be significant and
\eref{systerrorcondition} to be violated, $r_\nonisotoiso(H_T^\prime)$
would have to vary by at least an $\mathcal{O}(1)$ factor across a
single $H_T^\prime$ bin.
Whether this is the case depends on the statistics in $CR_Y$, but
based on our toy analysis of a single DV in the MS in \sref{htoXX}, we
expect this effect to be negligible or controllable in a real
analysis.  
Furthermore, the need to contend with weak correlations when using the ABCD method is a familiar issue. For example, the ABCD analysis of Ref.~\cite{Aad:2015sms} obtains good results in the presence of correlations of typically 6\%--10\% between their control variables, which are handled by marginalizing over nuisance parameters in a likelihood function.

One may also have to contend with BSM contamination of the control
region. We will account for this in our estimate in Sec.~\ref{s.htoXX}
 by including BSM contributions in $N_C$, but we will
underestimate the strength of the obtainable exclusions by not using
that knowledge when deriving a limit on the number of BSM events in
region $A$. In a fully self-consistent analysis, a given hypothesis
for the BSM cross-section could be tested by subtracting the (known)
BSM contribution from the measured $N_A$ and $N_B$, then deriving
$r_\mathrm{\nonisotoiso}(H_T^\prime)$ in both SR$_Y$ and CR$_Y$ and
checking for inconsistencies, which may arise if significant amounts
of BSM signal are present that, by construction, populate SR$_Y$ and
CR$_Y$ in different proportions. (We discuss more model-independent
approaches in \sref{futuredirections}.)

We now demonstrate this data-driven technique by computing a toy
sensitivity estimate for a simple and well-motivated benchmark signal
model in \sref{htoXX}. This will clarify many of the practical details
of how such an analysis would be performed. In
\sref{futuredirections}, we discuss generalizations for other signal
models, and LLP searches in other detector subsystems.

\section{Example: $h \to X X$ analysis}
\label{s.htoXX}

In this section we demonstrate how the background estimation strategy
of \sref{overview} is applied in practice. The signal we consider is
the production of a scalar $\Phi$ via gluon fusion, followed by its
decay to two identical unstable particles $X$. Such decays are among
the leading signatures of, for example, theories of Neutral Naturalness or
Hidden Valleys
where the decaying scalar is the 125 GeV SM-like Higgs boson.
As we discuss below, this signal model is also one of the most
\emph{challenging} for which to implement our analysis strategy, since
the inclusive production mode of the LLPs prohibits the most obvious
choices of $Y$ to define a SR$_Y$/CR$_Y$ split.
Even so, we show in a toy model estimate that significant sensitivity
gains at long proper lifetimes are possible compared to a search for two
DVs in the MS \cite{Aad:2015uaa}.

We assume the $X$ decays to pairs of SM particles via a small mixing
with the SM-like Higgs. The parameters of the signal model are
therefore:\vspace*{-2mm}
\begin{itemize}
\itemsep=-1mm
\item $\sigma_\Phi \cdot \mathrm{Br}(\Phi \to X X) \equiv \sigma_{\Phi \to XX}$
\item $m_\Phi$
\item $m_X < m_\phi/2$
\item $c \tau_X$
\end{itemize}
\vspace{-2mm} 
For simplicity we assume $\Phi$ has a narrow decay
width.  The sensitivity of a search is quantified as the value of
$\sigma_{\Phi \to XX}$ that can be excluded for a given $(m_\Phi, m_X,
c \tau_X)$.  Since exotic decays of the SM-like 125 GeV Higgs boson
are particularly well-motivated \cite{Curtin:2013fra}, we set $\Phi = h$
and hence $m_\Phi = m_h$ in our estimate, but the analysis generalizes
easily to other cases.

\subsection{Control regions for $h\to XX$}
\label{ss.SRCRdiscussion}
The particular challenge posed by the $h\to XX$ signal model is the
lack of any distinctive AOs produced in association with the LLPs.  Thus, the
non-iso sample that is most closely related to the iso sample of
interest is the entire inclusive sample.  Defining a separate control
region that includes an AO, for instance an identified lepton, would
indeed define three control and one signal region as shown \fref{SRCR}. 
However, the small statistics of CR$_Y$ and resulting large uncertainty on
the region A background prediction, $N_B/N_D$,
would typically be so large that no useful limit can be extracted. The
best one could do by requiring an AO, from the point of view of
relative event rates, is to define a CR$_Y$ that requires a single
$b$-tag. In this case the resulting signal and control
regions do not satisfy \eref{SRCRcondition}, because, taking the
$p_T$-dependence of realistic $b$-tagging algorithms into account,
both BSM and SM processes include contributions of similar relative
size with associated tagged $b$-jets. 
Fortunately, this signal model pair-produces LLPs, and the LLP that is not reconstructed as a DV can be used to inform choices for the variable $Y$, which depend on the LLP lifetime. 

We will focus here on long lifetimes of $X$, since this is the regime
where a single-DV search in the MS has unique sensitivity compared to
other displaced searches.
In this case, 
events with one DV will typically feature one $X$ decaying in the MS,
while the other $X$ escapes the detector. The MET in signal events is
then sensitive to the Higgs $p_T$, which is typically only of order
tens of GeV.  Given that the typical MET resolution is $\mathcal{O}(10
\gev)$~\cite{ATL-PHYS-PUB-2015-023}, the MET vector in signal events
will be highly sensitive to soft jet and pileup activity, and not
preferentially aligned with the DV.
The dominant SM background is dijet production, with one jet faking a
DV. In these events, there is no source of `truth-level' MET aside
from the energy of the mis-measured jet. Since harder jets are
expected to be more likely to reach the MS and fake a DV, the MET is
expected to be peaked at higher values than for BSM events, and will
preferentially point along the DV.
Therefore, the angle in the transverse plane $\deltaphi$ between the
DV in the Muon Spectrometer and the missing transverse energy 2-vector
is a useful choice for the variable $Y$. This variable is also
suitable because it is not strongly correlated with the fake jet
energy, and therefore with $H_T^\prime$.  A CR$_Y$ can be defined by
the requirement $\deltaphi <\Delta\phi_{min}$.

For shorter $X$ proper lifetimes, $c\tau \lesssim \mathcal{O}(1)$ m,
such an analysis is obviously not optimal, because the $X$ that does
not decay in the MS is now
most likely to decay in one of the other detector systems closer to the IP, causing
 the MET to again be aligned with the DV.
In this regime, the best choice of variable to split SR$_Y$ from
CR$_Y$ is likely to be some unusual 
property of objects in the tracker or calorimeters. In order 
for one $X$ to reach the muon system, even with
high luminosity to allow access to the tail of the boost distribution,
its lifetime must be more than a few centimeters. With such proper lifetimes, the
$X$ 
decaying in the other detector systems will
have signatures such as trackless jets and/or displaced vertices
in the inner tracker. This offers a control region given
by a {\em veto} on unusual objects in the inner detector, such as only
allowing events where each AO passes a stringent quality cut to ensure
it originates at the primary vertex, and thus lead to greater
signal acceptance than would be obtained by requiring the
identification and reconstruction of a second displaced vertex.  We
expect that this strategy would significantly enhance the
sensitivity of a search for a single DV in the MS for short $X$ proper
lifetimes. However, explicitly modeling the impact of such vetos with
publicly available tools is difficult to do with any quantitative
reliability, and since the unique advantage of the MS search is the
long-lifetime regime, we will not discuss this short-lifetime case in
detail.

\subsection{Toy Sensitivity Estimate}
\label{ss.toyestimate}

We now perform a Monte Carlo study to compute the potential
sensitivity of our analysis strategy. This has to be regarded as a toy
estimate, since the difficulty of accurately simulating SM
contributions that fake DVs was the very motivation for developing our
data-driven background estimation. QCD background will be estimated in
two ways: one that is more optimistic, and one that is extremely
pessimistic. As explained below, we expect that the optimistic
estimate is the more realistic of the two, but we show results for
both possibilities, since they are likely to bracket the achievable
sensitivity. The obtained limit projections differ only by an
$\mathcal{O}(1)$ factor, which gives us confidence that these rough
estimates are robust within their understood precision. In each case,
we expect significant improvements compared to the background-free
search for 2 DVs in the MS.

\subsubsection{Computation of BSM contributions}
\label{sss.bsmcontribution}

Concentrating on the case where $\Phi = h$ is the SM-like Higgs, we
normalize the inclusive Higgs production cross-section to the value
computed by the LHC Higgs Cross-Section Working Group
\cite{Heinemeyer:2013tqa} and parametrize limits as reach projections
of $\mathrm{Br}(h \to X X)$, assuming SM production.  The Hidden Abelian Higgs Model (HAHM)
\cite{Curtin:2014cca} is used to generate gluon-fusion $h \to X X$
events in Madgraph \cite{Alwall:2014hca}, with matched production of
up to one extra jet, that are showered and hadronized in Pythia 6
\cite{Sjostrand:2007gs}~\footnote{In \cite{Curtin:2014cca}, it was
  shown that the matched production of $h$ with one extra jet within
  the EFT framework of the Madgraph model gives a surprisingly
  accurate representation of the Higgs $p_T$ spectrum.}.

The probability that each signal event passes the trigger and yields a
reconstructed DV in the MS can be estimated by calculating the
probability of decaying within the sensitive regions of the MS, see
\tref{detectoruseful}, for a given lifetime, and convolving with
trigger and DV reconstruction efficiencies \cite{Aad:2015uaa}, where
we ignore an $\mathcal{O}(1)$ $m_X$-dependence of that efficiency for the
purpose of this simple estimate. 

\begin{table*}
\begin{center}
\begin{tabular}{| l ||  c | c | c || c | c|}
\hline 
& $r$ (m) & $|z|$ (m) & $|\eta|$ & $\epsilon_\mathrm{trigger}$ & $\epsilon_\mathrm{DV}$ \\
\hline \hline
Muon Spectrometer (barrel)
   & (4, 6.5) & --- & $< 1.1$ &  0.40 & 0.25
\\
\hline
Muon Spectrometer (endcaps)
  & --- & (7, 12) & (1.1, 2.4) &  0.25 & 0.50
\\
\hline
\end{tabular}
\end{center}
\caption{ {\small
    Regions of the ATLAS Muon Spectrometer that have sensitivity to LLP decays. We assume uncorrelated efficiencies  $\epsilon_\mathrm{trigger}$ and $\epsilon_\mathrm{DV}$ for an LLP decaying in the given detector region to pass the Muon RoI Signal trigger and be reconstructed as a DV offline, respectively. For simplicity we ignore a modest dependence on $m_X$. The geometrical definition of sensitive detector regions and approximate trigger/reconstruction efficiencies for displaced $h\to XX \to 4f$ decays are taken from efficiency curves in 
    \cite{Aad:2015uaa}.
}
}
\label{t.detectoruseful}
\end{table*}

Due to the unusual nature of our signal we do not use any detector
simulation, but manually include relevant detector effects by
analyzing Pythia-clustered, truth-level events. For each event:
\begin{itemize}
\item Any $X$ that decays before reaching the Muon Spectrometer is
  treated as a regular jet.
\item Jets $j_i$, ordered by $p_T$, are counted if they have
  $p_T^{j_i} > 20 \gev$ and $|\eta| < 2.4$.
\item The above set of jets determines the two-vector
  $\vec{E}_T^{miss} = - \sum_i \vec p_T^{j_i}$. The most important
  detector effect to include is the resolution on both the size and
  direction of this vector.  

  To accurately model the MET resolution, we need to take into account
  the effects of pile-up.  For each event, we choose a number $N_{PV}$
  of primary vertices. This is done as follows:
\begin{itemize}
\item For $\sqrt{s} = 8 \tev$, we use the LHC Run-1 distribution
  \cite{8tevpileup} of the mean number of interactions per crossing
  $\langle \mu \rangle $, which is 20.7 on average. The resulting
  number of primary vertices can be obtained from the parameterization
\begin{equation}
\langle N_{PV} \rangle = 0.73 \langle \mu \rangle (1-0.008 \langle \mu \rangle)
\end{equation}
in Ref.~\cite{ATL-PHYS-PUB-2015-008}, which results in a distribution of
expected $\langle N_{PV} \rangle$ that is peaked around 17. That
$\langle N_{PV} \rangle $ distribution is sampled to obtain an
\emph{expected} $\langle N_{PV} \rangle$ for each event, which in turn
defines a Poisson distribution that is sampled to obtain the
\emph{observed} $N_{PV}$ for that event.
 
\item For $\sqrt{s} = 13 \tev$ with $30 \ifb$ or $300\ifb$, we use the
  13 TeV $\langle \mu \rangle$ distribution given
  in Ref.~\cite{13tevpileup}, which has an average of 13.5. Since that
  distribution was obtained from a low-luminosity run, we shift it
  upwards by doubling $\langle \mu \rangle$ (without increasing the
  width of the curve) to more realistically model the higher pile-up
  conditions of the full LHC run 2. Using the $\langle \mu
  \rangle$-distribution thus defined, we follow the same steps as for $8
  \tev$.

\item For $\sqrt{s} = 13 \tev$ with $3000 \ifb$,
  \cite{PLOT-UPGRADE-2014-003} shows explicit distributions of
  $N_{PV}$ for different assumptions of $\langle \mu \rangle$. We
  choose the curve with an average $\langle \mu \rangle$ of $140$
  scenario as our benchmark point, and sample that curve directly to
  obtain $N_{PV}$ for each event.

\end{itemize}
Ref.~\cite{ATL-PHYS-PUB-2015-023} contains Track-based soft term (TST) 
MET resolution curves as a function of $N_{PV}$. For each event, the
chosen $N_{PV}$ defines an RMS uncertainty (typically about $\ 10 - 20 \gev$)
for each $(E_T^{miss})_{x,y}$ component. This in turn defines the
variance of a Gaussian distribution that is again sampled to generate
the spurious $(E_T^{miss})_{x,y}$ components, which are added to the
truth-level $\vec{E}_T^{miss}$.  

\item $\vec{E}_T^{miss}$ is used to compute $\deltaphi$, as well as
  $\mathrm{MET} = |\vec{E}_T^{miss}|$.

  Note that, for small values of MET, the finite MET resolution means
  that the angle $\deltaphi$ is largely unrelated to its value at
  truth level, while for large MET values $\deltaphi$ will be peaked
  at its truth-level value. As
  explained above, we exploit this in our analysis.

\item $H_T^\prime$ is computed as in \eref{HTprime} using the AOs,
  i.e., the above-defined jets $j_i$ and $\vec{E}_T^{miss}$.

\end{itemize} 
Since $\deltaphi$ will be used to define SR$_Y$ and CR$_Y$, the above
variables will allow BSM predictions to be computed in regions A and C
of \fref{SRCR}, i.e., in the iso-regions.
We have checked that our results are robust under different modeling
of the MET resolution.

Note that we were very careful to model experimental resolution for
MET-related quantities, because $\deltaphi$ is vital for the
definition of our signal and control regions, but we did not account
for detector effects in the computation of $p_T^{j_i}$ and therefore
in $H_T^\prime$. This is acceptable for our toy estimate, since
whenever we make use of these variables we only exploit the coarse
structure of their distributions. Fine details in these distributions,
arising from finite jet energy resolution, do not affect our results. 
We also assume that trigger efficiencies remain constant at high luminosities.

\subsubsection{Computation of SM Contributions}
\label{sss.smcontribution}

QCD contributions to the iso- and non-iso-regions are very difficult
to model reliably---this is exactly the reason why a data-driven
approach is necessary. Even so, we can perform some estimates of the
QCD contributions that are sufficient to demonstrate that our analysis
strategy will improve sensitivity to very long-lived BSM particles.

For estimating sensitivity, the two most important questions are: 
\begin{enumerate}
\item What is the size of the SM contribution in the signal region A? 
\item What is the precision with which the SM contribution in region A can be determined from data?
\end{enumerate}
Answering both of these questions only requires simulating events in
the iso-region. This can be easily seen by rewriting
\eref{NAiSMaftercuts} for the data-driven prediction of the SM
contribution in region A:
\begin{eqnarray}
\label{e.NAiSMaftercutsMC2}
\langle N_{A,i}^\mathrm{SM, \ after\ cuts}\rangle  & = &  \left(\frac{N_{B,i}^\mathrm{after\ cuts}}{N_{D,i}}\right) \, N_{C,i}
\\ \nonumber \\ \nonumber 
& = &  \ \ \left(\frac{N_{A,i}^\mathrm{SM_\mathrm{MC}, after\ cuts}}{N_{C,i}^\mathrm{SM_\mathrm{MC}}}\right) \, N_{C,i}^\mathrm{SM_\mathrm{MC}+BSM_\mathrm{MC}} 
\ ,
\end{eqnarray}
where the MC subscript indicates the quantity is computed from Monte
Carlo. 
The second equality occurs because the kinematic variable $Y$ in
\fref{SRCR} is assumed to be uncorrelated with the isolation condition,
and $N_{C,i} = N_{C,i}^\mathrm{SM_\mathrm{MC}+BSM_\mathrm{MC}}$ simply
reflects the fact that we are simulating events in the iso-region~\footnote{Systematic uncertainties from weak correlations will be subdominant, as discussed in 
\ssref{importantconsiderations}.}.  This conservative estimate of $\langle N_{A,i}^{SM}\rangle$ does not account for signal contamination in control-like regions.
The quantity in brackets can be assumed to be known to extremely high
precision, because of the much higher statistics in the non-iso-region
than in the iso-region (even though we use a ratio of quantities
computed from Monte Carlo in the iso-region to describe it for our
sensitivity estimate).
For finite statistics in iso-regions A and C, the dominant contribution to the
uncertainty of $\langle N_{A,i}^\mathrm{SM, \ after\ cuts}\rangle$
is the statistical uncertainty of
$N_{C,i}^\mathrm{SM_\mathrm{MC}+BSM_\mathrm{MC}}$. For the purpose of
our sensitivity estimate, we can therefore define
\begin{equation}
\label{e.NAiSMaftercutsMC3}
\langle N_{A,i}^\mathrm{SM, \ after\ cuts}\rangle \ \ =   \ \ \left(\frac{N_{A,i}^\mathrm{SM_\mathrm{MC}, after\ cuts}}{N_{C,i}^\mathrm{SM_\mathrm{MC}}}\right) \, (\tilde N_i)^{+\delta_{+i}}_{-\delta_{-i}}\ ,
\end{equation}
where $\tilde N_i$ is the number of events we observe in $H_T^\prime$
bin $i$ of region C, and $\delta_{\pm i}$ are the Poisson uncertainties for
the observation of $\tilde N_i$ events. We will take $\tilde N_i$ to
be given $N_{C,i}^\mathrm{SM_\mathrm{MC}+BSM_\mathrm{MC}}$ rounded up
to the nearest integer, which is a conservative choice.

We now discuss how to simulate QCD jets generating DVs in the Muon
Spectrometer. The ATLAS Run-1 analysis \cite{Aad:2015uaa} observed
about $1.0 \times 10^5$ events that fired the Muon RoI Signal
trigger. The chance that one of those events, which are assumed to
stem dominantly from QCD, also results in a reconstructed DV in the
corresponding Muon RoI is about $1.3 \times 10^{-2}$ in the MS Barrel
and $8.0 \times 10^{-2}$ in the endcaps. The numbers of events that
both fired the trigger and reconstructed a DV in the barrel versus in
the endcaps are not separately given. Consequently the \emph{total}
number of QCD events with a single reconstructed DV in the ATLAS MS 
in Run-1 could range between $\sim 1300$ for all vertices in the
barrel, and $~8000$ for all vertices in the endcaps.  We take as our
estimate the total number to be $\sim 3000$. This will at most be off
by a factor of $\sim 3$ from the real value in either direction, which
will not significantly affect our conclusions. We therefore assume
that, at truth level,
\begin{equation}
\label{e.NA8tev}
\sum_i N_{A,i}^\mathrm{SM} = 3000
\end{equation}
for $\sqrt{s} = 8 \tev$ with $25 \ifb$,
where the sum is over all $H_T^\prime$ bins. For simplicity, we assume
the cross-section to produce fake DVs from QCD does not change
significantly between $\sqrt{s} = 7 \tev$ and 8 TeV.  The ATLAS
analysis also estimated the number of QCD background events in the
signal region of their search, which required \emph{two} displaced
vertices in the MS. It was found to be
\begin{equation}
\label{e.NA2DV8tev}
N_\mathrm{2DV}^\mathrm{SM} \sim \mathcal{O}(0.1), \ \ \ \  \ \ \ \ \mbox{for $\sqrt{s} = 8 \tev$ with $25 \ifb$}.
\end{equation}

These two data points allow us to normalize our Monte Carlo prediction
for generating fake DVs with QCD. To obtain concrete simulated events,
we take an approach related to \eref{epsiloniso} and assume each jet
has a $p_T$-dependent chance
$\epsilon^\mathrm{fake}_\mathrm{iso}(p_T)$ of faking a DV in the
MS. For simplicity we assume
$\epsilon^\mathrm{fake}_\mathrm{iso}(p_T)$ to have a linear dependence,
	\begin{equation}
	\epsilon^\mathrm{fake}_\mathrm{iso}(p_T) = \left\{
		\begin{array}{ll}
			\epsilon_\mathrm{iso}^0 \times  \displaystyle \frac{(p_T - p_T^\mathrm{min})}{\gev}
					& \mbox{ for $p_T \geq  p_T^\mathrm{min}$} \\
			0 & \mbox{ for $p_T <  p_T^\mathrm{min}$} 
		\end{array}\right. \ ,
	\end{equation}
and consider two possibilities:
\begin{enumerate}

\item \emph{Optimistic choice:} because we expect harder jets to be
  dominantly responsible for the DVs, a reasonable modeling of the QCD
  fake rate is to require a relatively large
  $p_T^\mathrm{min}$ that, along with $\epsilon_\mathrm{iso}^0$, is
  chosen to satisfy both Eqns. (\ref{e.NA8tev}) and
  (\ref{e.NA2DV8tev}). With this assumption, the fake DV background
  is dominated by relatively energetic jets, allowing BSM and
  SM events to be effectively distinguished.

\item \emph{Pessimistic choice:} assume that \emph{every} jet is able
  to fake a DV in the MS by setting $p_T^\mathrm{min} = 0 \gev$. The
  constant $\epsilon_\mathrm{iso}^0$ is then chosen to satisfy
  \eref{NA8tev}. This is a pessimistic choice for the shape of the SM
  background, because the fake DVs are dominated by very soft
  QCD jets with few kinematic features to distinguish them from Higgs
  production events with exotic decay to LLPs. \eref{NA2DV8tev} is also
	not satisfied, since the high production rate of soft jets
  means that $\epsilon_\mathrm{iso}^0$ is so small that
  $N_\mathrm{2DV}^\mathrm{SM}$ in \eref{NA2DV8tev} is predicted to be
  many orders of magnitude less than unity. We include this
  possibility in our analysis to demonstrate that even with these
  extremely pessimistic assumptions, our data-driven approach has more
  sensitivity at long lifetimes compared to a standard search requiring
  2 DVs.

\end{enumerate}
We will derive limit projections for both the pessimistic and the more
realistic optimistic choice. The sensitivity of a real analysis will
likely lie somewhere in between these possibilities.

Both QCD samples were simulated in MadGraph and showered and hadronized in Pythia~6.
An unmatched dijet sample was used for the optimistic choice to
adequately sample hard jets with $p_T > 100 \gev$.  Matched generation
of 2 + 3 jets was used for the pessimistic choice to give sensible
distributions of the soft jets.  We use the tree-level QCD cross
sections supplied by MadGraph, since any NLO effects are included in
the normalization of the $\epsilon^\mathrm{fake}_\mathrm{iso}$ fake
rate to Eqns. (\ref{e.NA8tev}) and (\ref{e.NA2DV8tev}). This results in
\begin{eqnarray*}
\epsilon_\mathrm{iso}^0 &=& 7.6 \ \times \ 10^{-12} \ ,\ \ \ \ \ p_T^\mathrm{min} = 0 \gev
\end{eqnarray*}
for the pessimistic\ QCD\ scenario, and 
\begin{eqnarray*}
\epsilon_\mathrm{iso}^0 &=& 1.1 \times 10^{-8} \ , \ \ \ \ \ \ \ \ p_T^\mathrm{min} = 120 \gev
\end{eqnarray*} for the optimistic \ QCD \ scenario, which also gives  $N_\mathrm{2DV}^\mathrm{SM} \sim 0.1$.

In each QCD event, any jet with rapidity $|\eta| < 2.4$ (so it can
reach the MS) is considered as a possible fake DV, with the
event weighted according to $\epsilon^\mathrm{fake}_\mathrm{iso}$. The
remaining jets are used to reconstruct the event in an identical
fashion as the signal events above.

We now discuss possible systematic errors in the data-driven
determination of $r_\nonisotoiso$.  As might be expected from the
presence of additional energy scales in the event (e.g., from pile-up),
some slight correlation between $H_T^\prime$ and $Y = \deltaphi$ is
indeed present. Empirically, we determine $\Delta H_T^\prime$ in
\eref{deltahtpdef} to be $\lesssim 0.1$ for both our optimistic and
pessimistic QCD background estimates. At the 13 TeV LHC with $30
\ifb$, SM background rates are high enough that $H_T^\prime$ bins as
narrow as 1 or 2 GeV are sufficiently populated in region C.
\eref{systerrorcondition} therefore implies that the systematic error
is negligible unless $r_\nonisotoiso(H_T^\prime)$ varies by a factor
of about  $5 - 10$ over an $H_T^\prime$ range of only a few GeV.
Therefore, the systematic error should in general be negligible. However, as a consistency check, it should be verified that
\eref{systerrorcondition} is satisfied for the chosen binsize of a
realistic analysis.

\subsubsection{Analysis and Projected Limits: Search with Two Displaced Vertices}

\newcommand{\widthbla}{8cm}
\begin{figure*}
\begin{center}
\begin{tabular}{c}
\includegraphics[width=15cm]{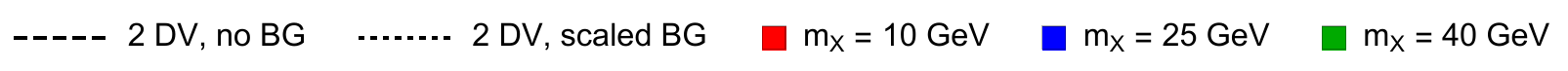}
\\
\begin{tabular}{cc}
\includegraphics[width=\widthbla]{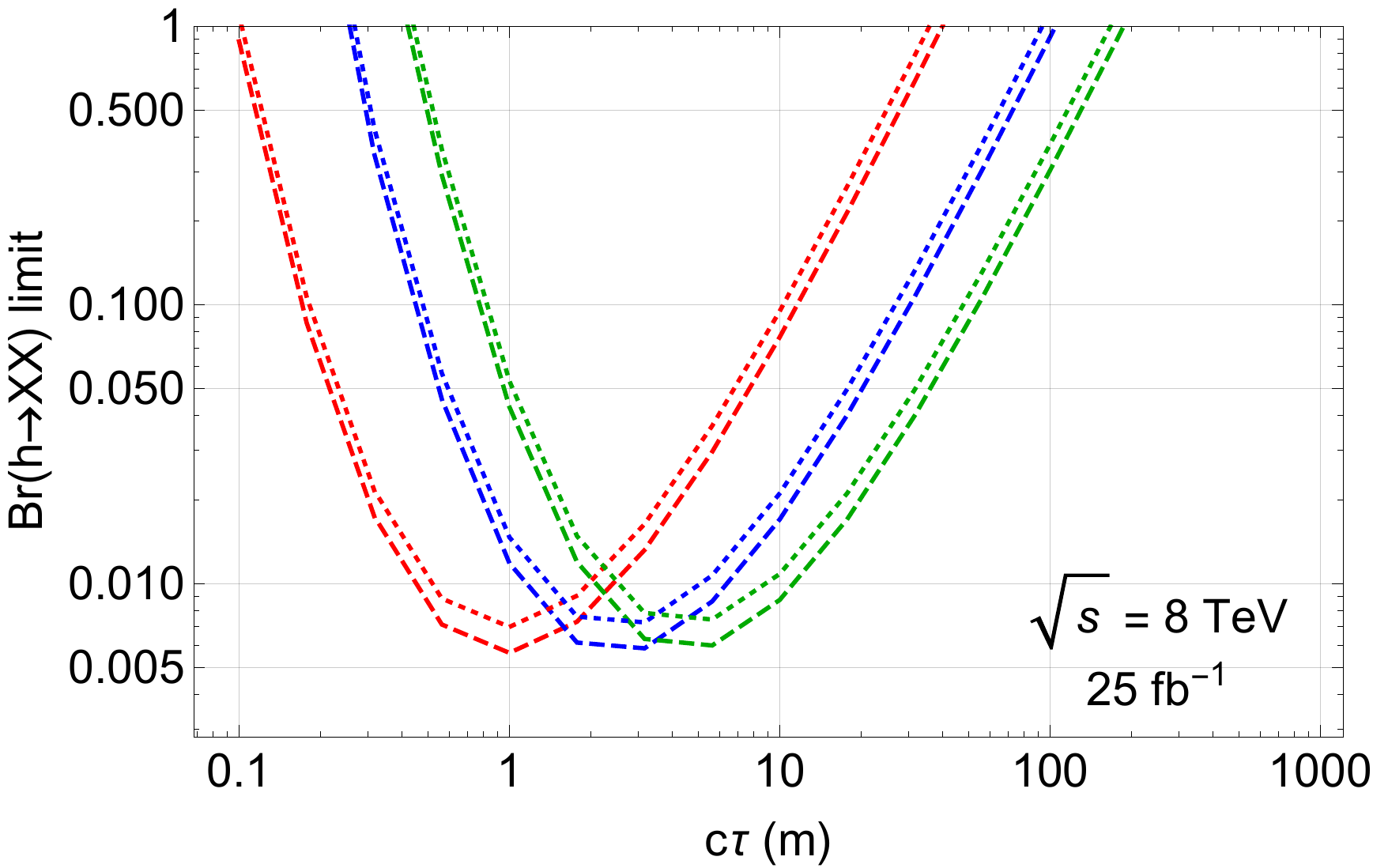}
&
\includegraphics[width=\widthbla]{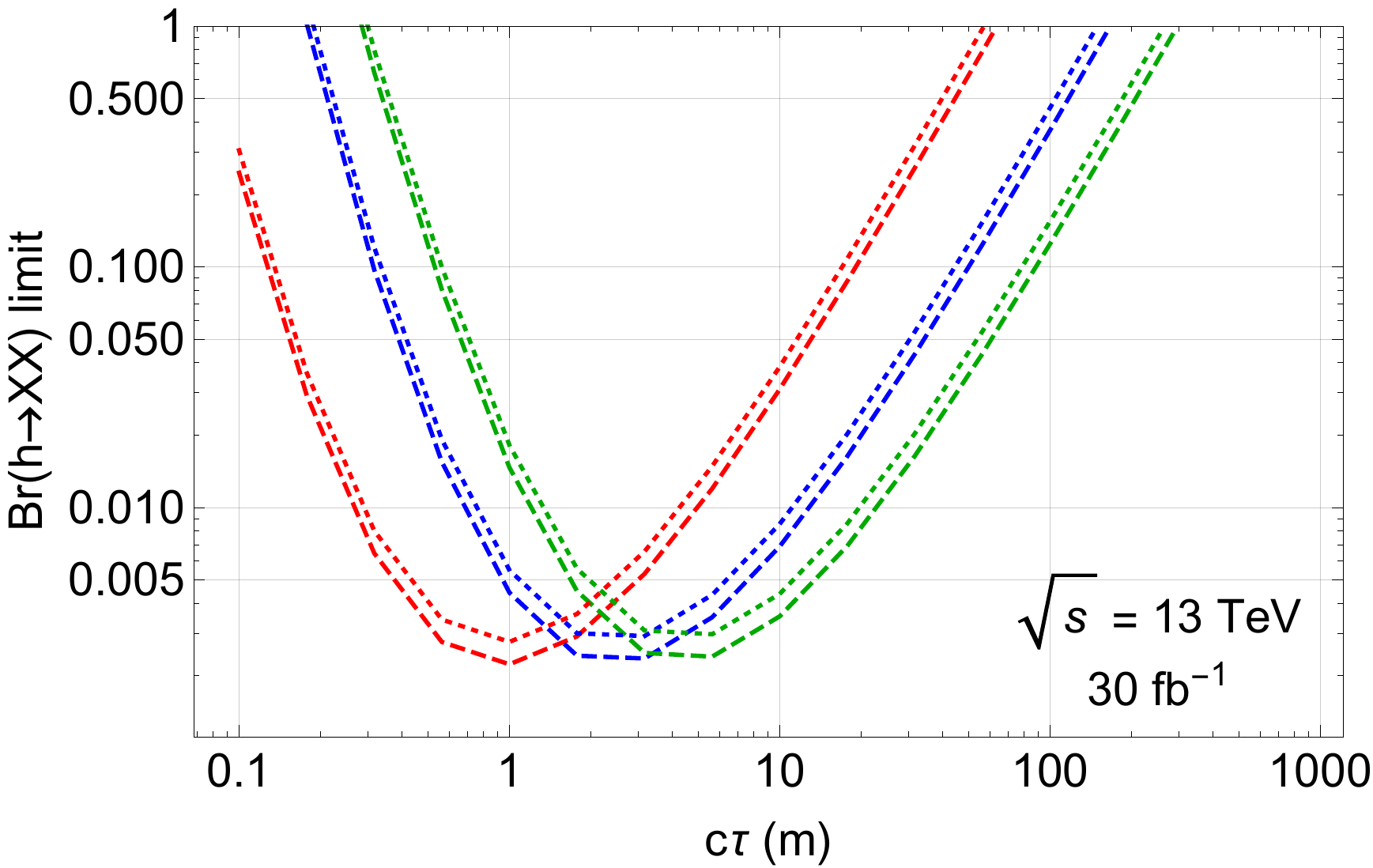}
\\
\includegraphics[width=\widthbla]{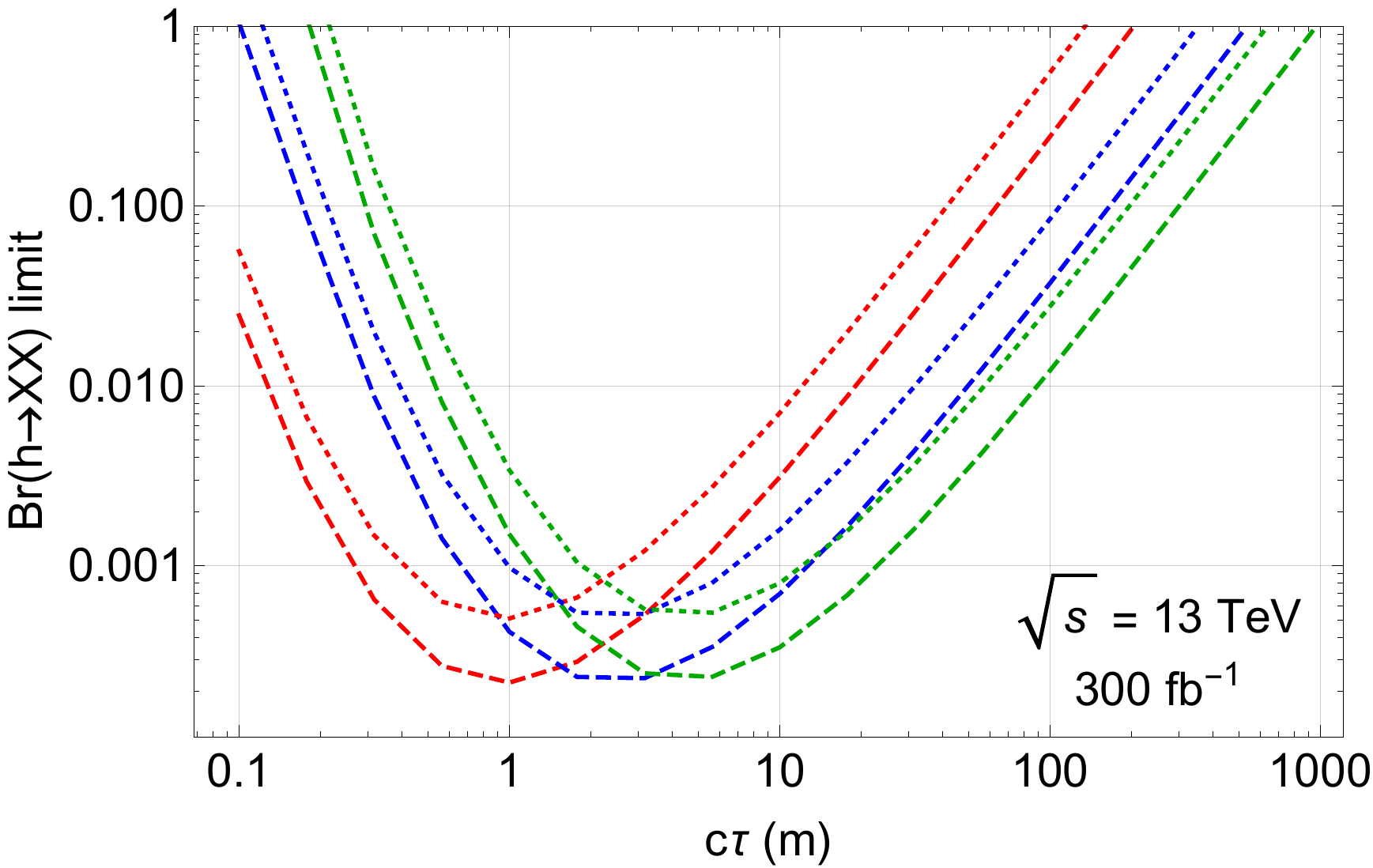}
&
\includegraphics[width=\widthbla]{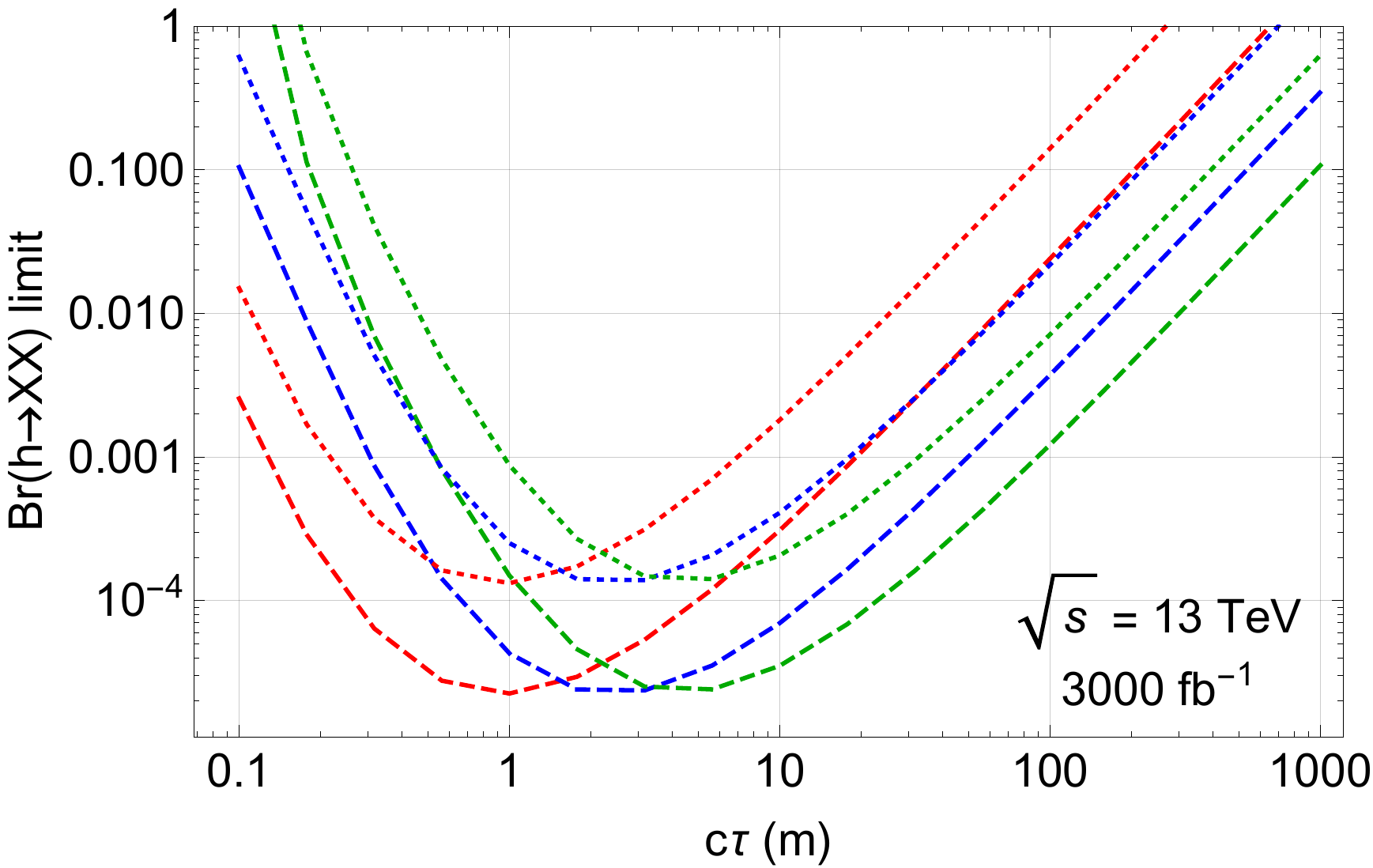}
\end{tabular}
\end{tabular}
\end{center}
\caption{ {\small Simulated limits on $\mathrm{Br}(h \to X X)$ as a function
  of $X$ lifetime for $m_X = 10, 25, 40 \gev$ at the 8 and 13 TeV LHC
  from a search that requires two DVs in the ATLAS Muon Spectrometer, in
  analogy to \cite{Aad:2015uaa}. This result serves as a baseline against
  which to compare our projections for a data-driven search requiring
  just a single DV.} }
\label{f.limits2DV}
\end{figure*}

We first derive estimated limits on $\mathrm{Br}(h \to X X)$ for an
ATLAS search analogous to \cite{Aad:2015uaa} that requires two DVs in
the MS at the 13 TeV LHC. We also produce limit projections for $25
\ifb$ of 8 TeV data to compare with \cite{Aad:2015uaa} (even though we
do not include the optional reconstruction of a second DV in the inner
tracker instead of the MS). These limits for a search with two DVs
will serve as a baseline against which we compare the sensitivity of
our proposed data-driven search for a single DV.

These limit projections for the two-DV-search are derived under two
assumptions. First, we show limits for zero background, which simply
corresponds to about 4 signal events. Second, we show limits for
non-zero background, derived by a naive rescaling of the LHC Run-1
background prediction in \cite{Aad:2015uaa}. At the 8 TeV LHC with $25
\ifb$ this corresponds to 1 background event (rounded up from 0.4). At
13 TeV, this scales to about 1, 10 and 100 events respectively for $30
\ifb$, $300 \ifb$ and $3000\ifb$.

Once again, in a realistic search, the actual limits will lie
somewhere between these two cases.  However, since it is likely that
improvements in DV reconstruction algorithms and other optimizations
for higher luminosity will enable future analyses to suppress
backgrounds to a greater extent than what is predicted by simply
rescaling the results from Ref.~\cite{Aad:2015uaa}, we will compare
projected sensitivity of single-DV searches to the background-free
two-DV limits. This also will give the most pessimistic assessment of
the relative gained sensitivity of the one-DV search, and demonstrate
the significance of these gains.

Our projected limits for the two-DV searches are shown in
\fref{limits2DV}. The 8 TeV projections reproduce the actual limits of
the ATLAS analysis \cite{Aad:2015uaa} up to a $\mathcal{O}(1)$
factor. This modest difference is not surprising since we used a very
simple parametrization of the DV reconstruction efficiency, which
amongst other things neglected dependence on the LLP
mass. Nevertheless, since our limits for a single DV are derived under
the same assumptions, these sensitivities will serve as a valid base
of comparison for the proposed one-DV search.

\subsubsection{Analysis and Projected Limits: Search with One Displaced Vertex}

\fref{spectrabeforeSRCR} shows the distributions in $H_T^\prime$, 
$p_T^{j_1}$, MET and $\deltaphi$ of BSM and SM events in the
iso-region at $\sqrt{s} = 13 \tev$ before SR$_Y$ and CR$_Y$ are
defined. This illustrates that while BSM events for long lifetimes are
relatively uniformly distributed in $\deltaphi$, QCD events are peaked
at $\deltaphi = 0$, especially for the more  optimistic QCD
assumption of $p_T^\mathrm{min} = 120 \gev$. We therefore define
CR$_Y$ to be $\deltaphi < 1.5$ for the pessimistic analysis with the
$p_T^\mathrm{min} = 0 \gev$ QCD background sample, and $\deltaphi <
1.0$ for the optimistic analysis with the $p_T^\mathrm{min} = 120
\gev$ QCD background sample. In both cases, it is clear that the
sensitivity will decrease with shorter lifetimes, since those signal
events are also peaked at small $\deltaphi$, as discussed above.

\begin{figure*}
\begin{center}
\includegraphics[width=17cm]{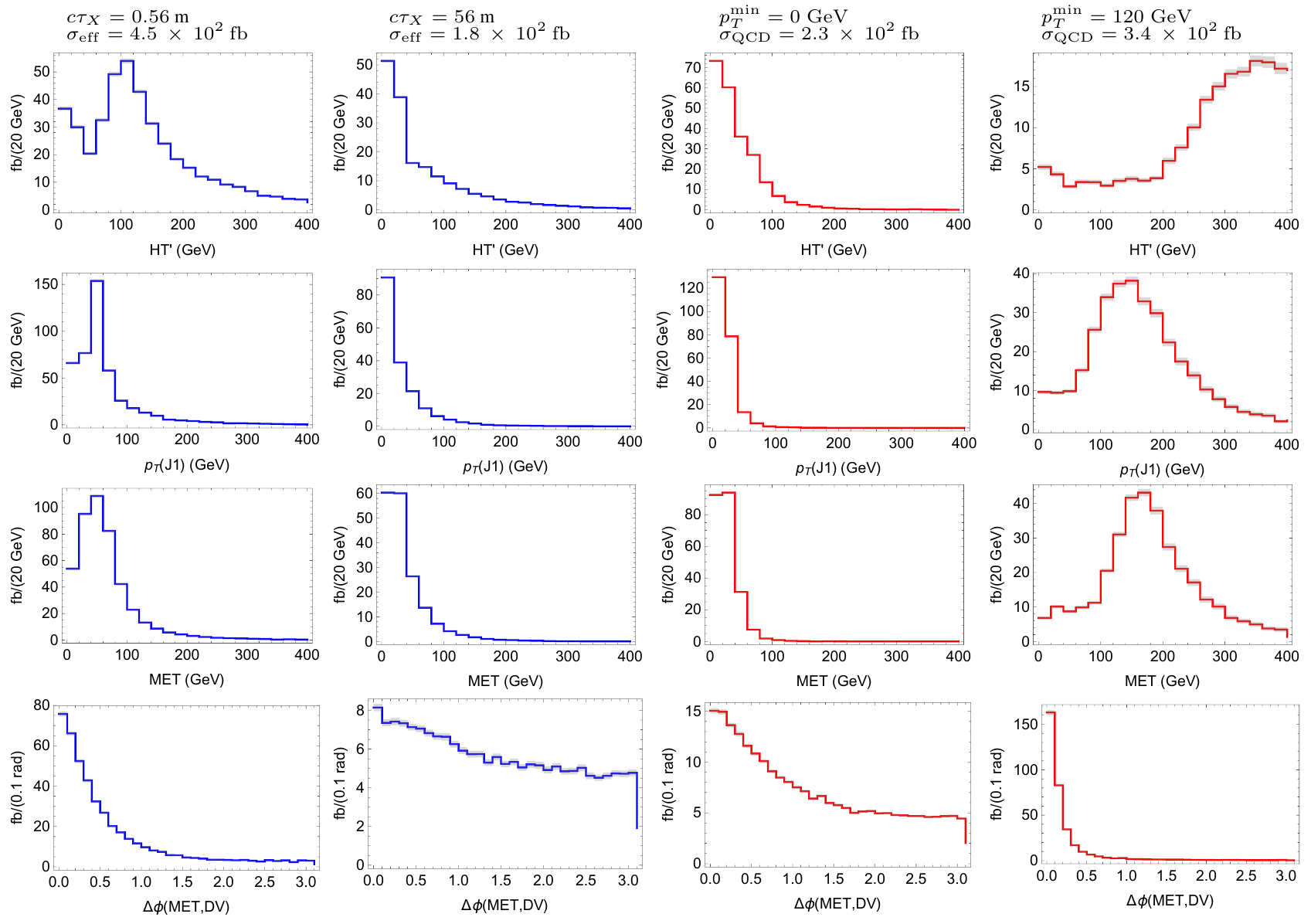}
\end{center}
\caption{{\small Distributions of events with one DV in iso-region before
    defining a SR$_Y$/CR$_Y$ split, with $\sqrt{s} = 13 \tev$ and
    $m_X = 25 \gev$. $p_T(J_1)$ refers to the highest jet $p_T$ in the
    event.  First column: BSM events with short lifetime (56~cm). Second
    column: BSM events with long lifetime (56 m). Third column: QCD background
    events under pessimistic assumption ($p_T^\mathrm{min} = 0
    \gev$). Fourth column: QCD background events under optimistic assumption
    ($p_T^\mathrm{min} = 120 \gev$).   $\sigma_\mathrm{eff}$ is the effective
    signal cross-section of events with at least one detected DV in
    the Muon System, after geometric, trigger and detection
    efficiencies are taken into account (setting $\mathrm{Br}(h\to X X) = 1$), see  \sssref{bsmcontribution}. $\sigma_\mathrm{QCD}$ is the
    cross-section of SM QCD background in the shown kinematic region as estimated using the methods
    outlined in \sssref{smcontribution}.  } }
\label{f.spectrabeforeSRCR}
\end{figure*}

\newcommand{\tempheight}{4cm}
\begin{figure*}
\begin{center}
\begin{tabular}{m{14cm}m{1cm}}
\begin{tabular}{cm{0.5cm}c}
\includegraphics[height=\tempheight]{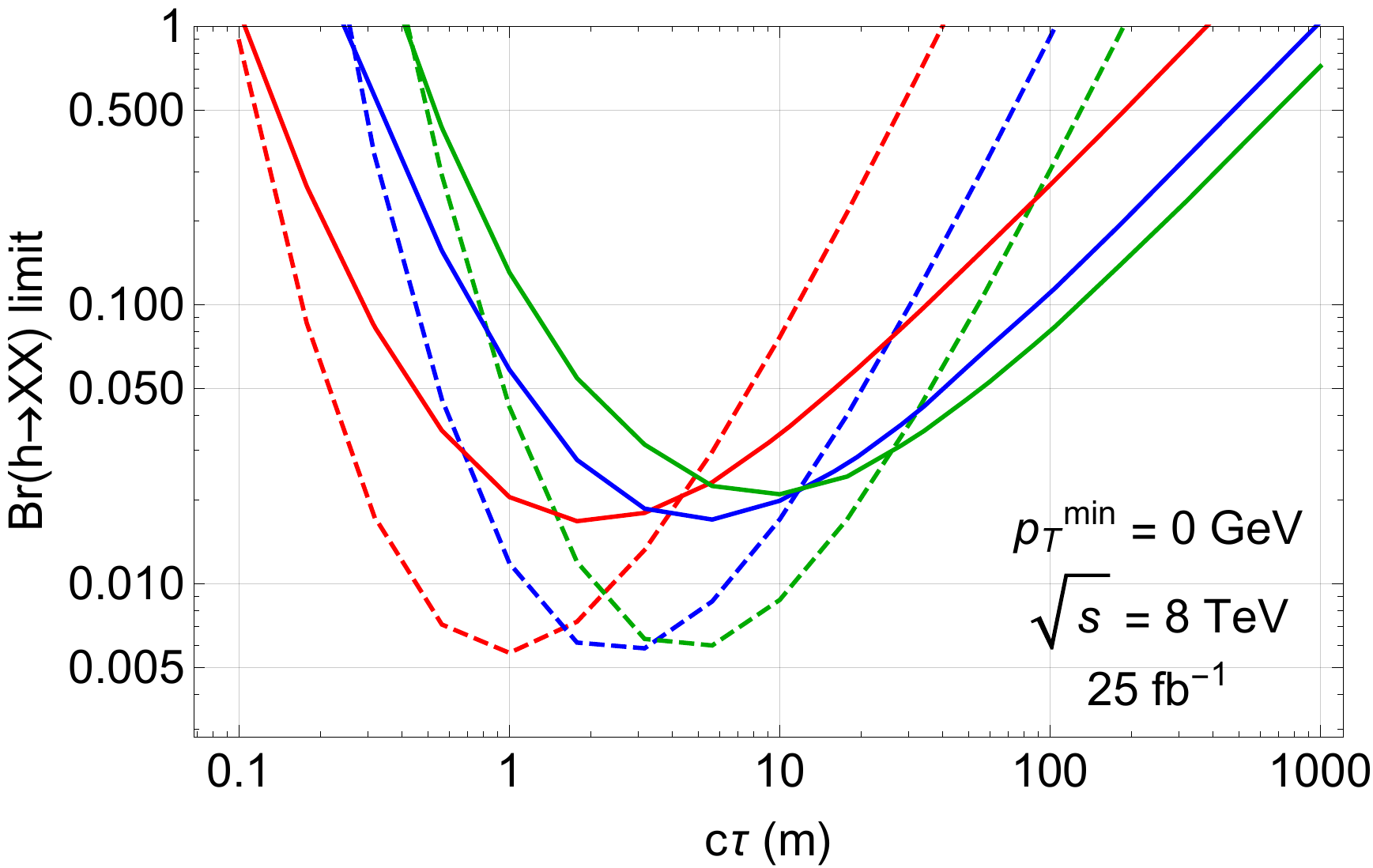}
& &
\includegraphics[height=\tempheight]{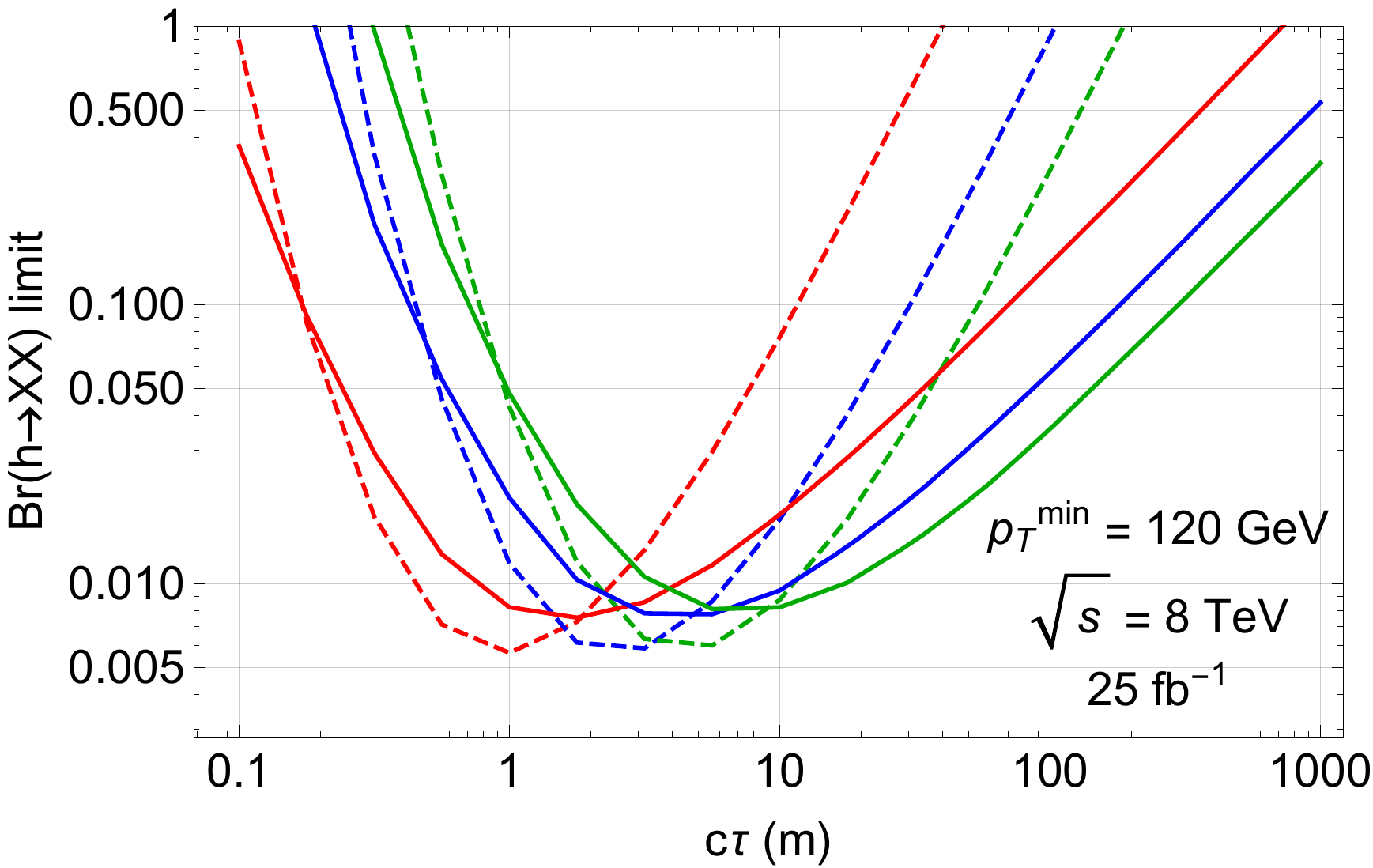}
\\
\includegraphics[height=\tempheight]{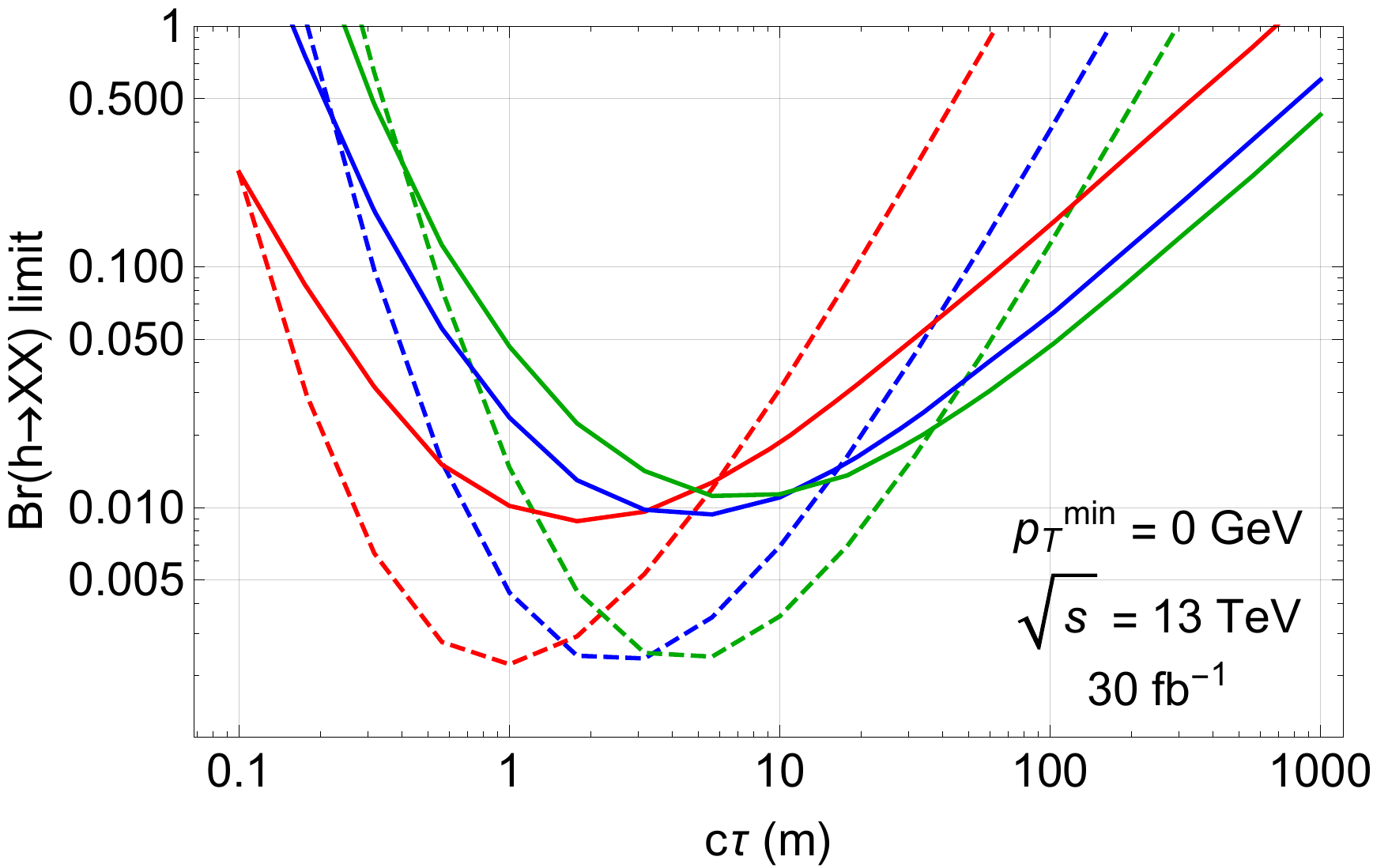}
& &
\includegraphics[height=\tempheight]{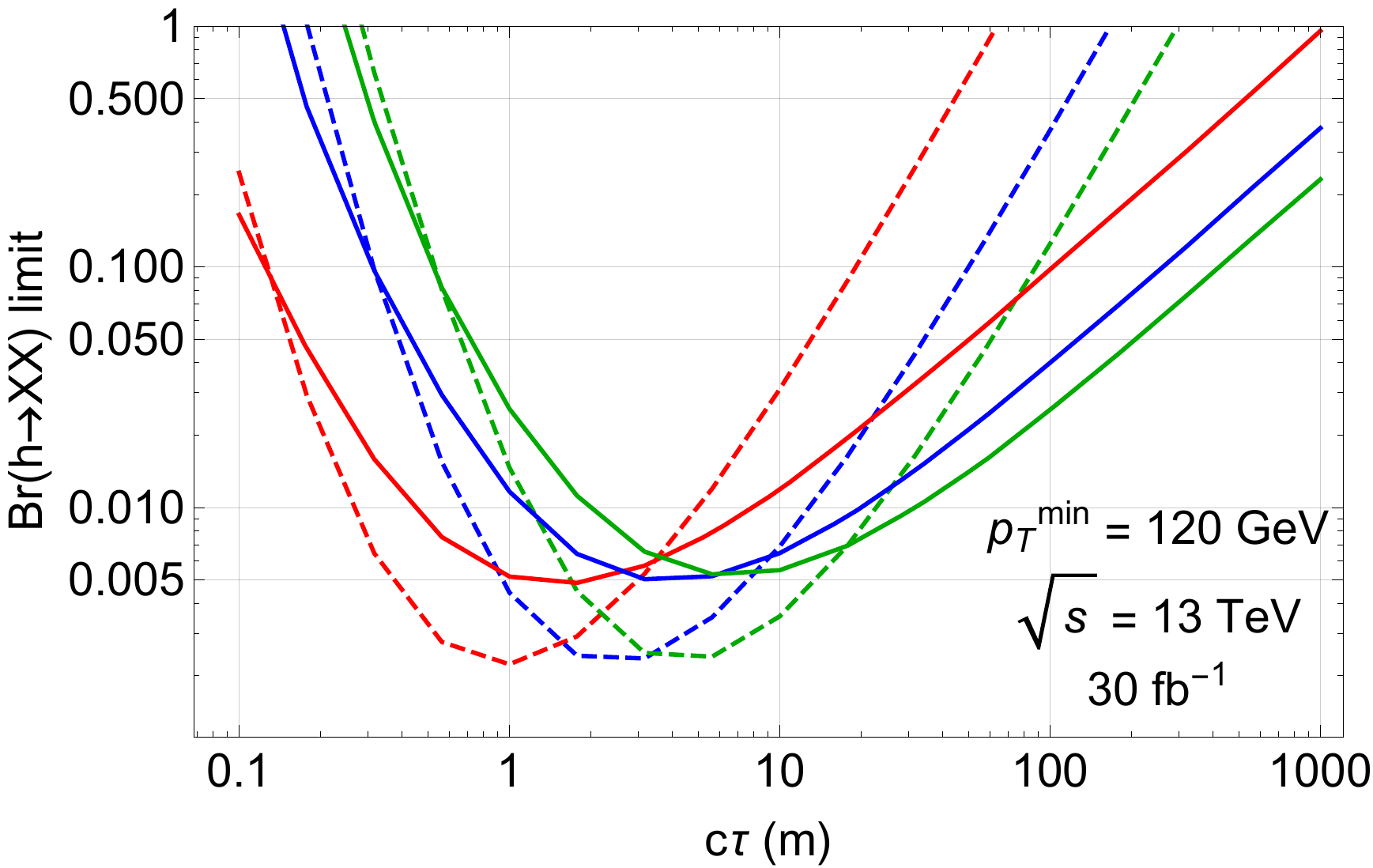}
\\
\includegraphics[height=\tempheight]{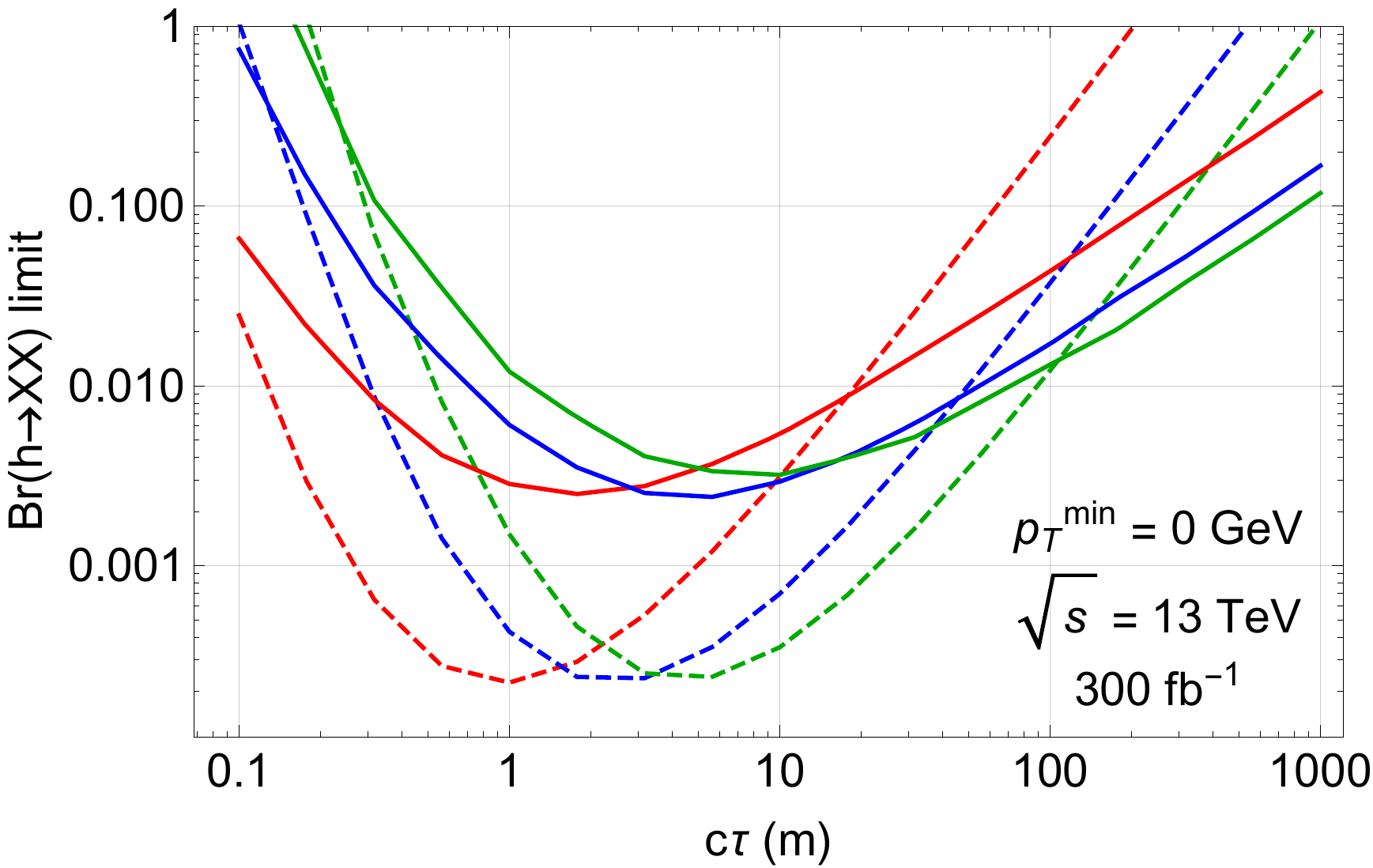}
& &
\includegraphics[height=\tempheight]{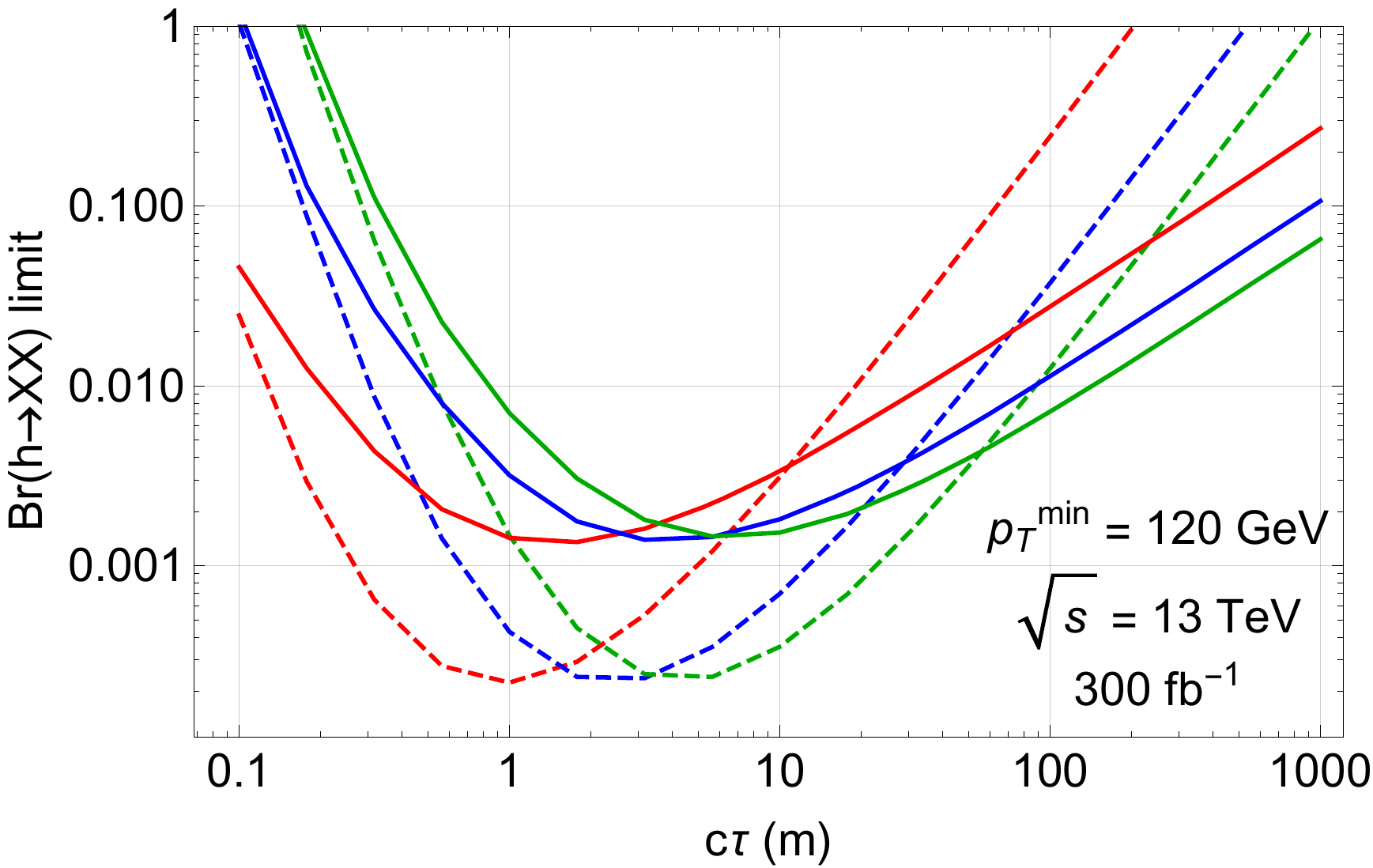}
\\
\includegraphics[height=\tempheight]{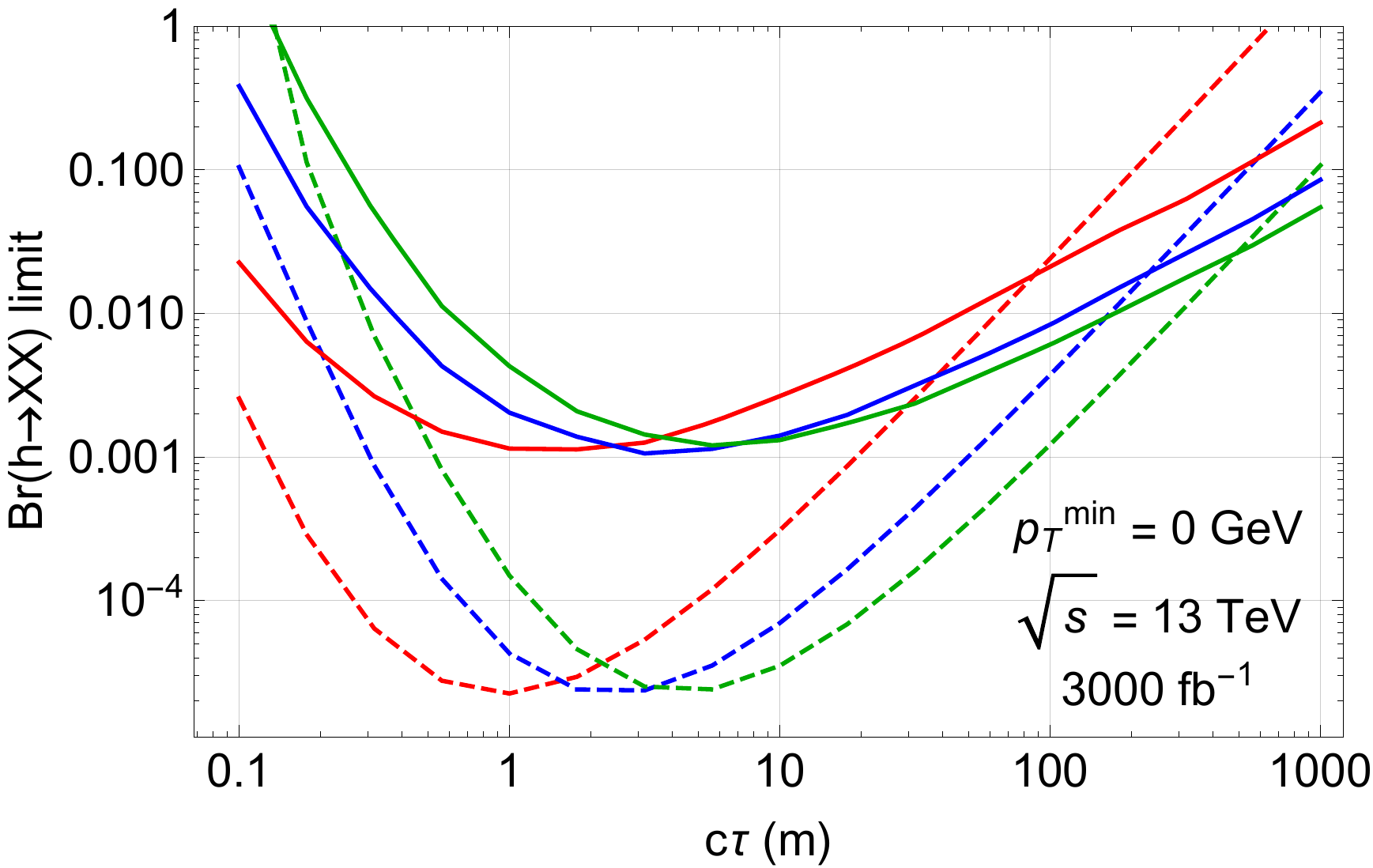}
& &
\includegraphics[height=\tempheight]{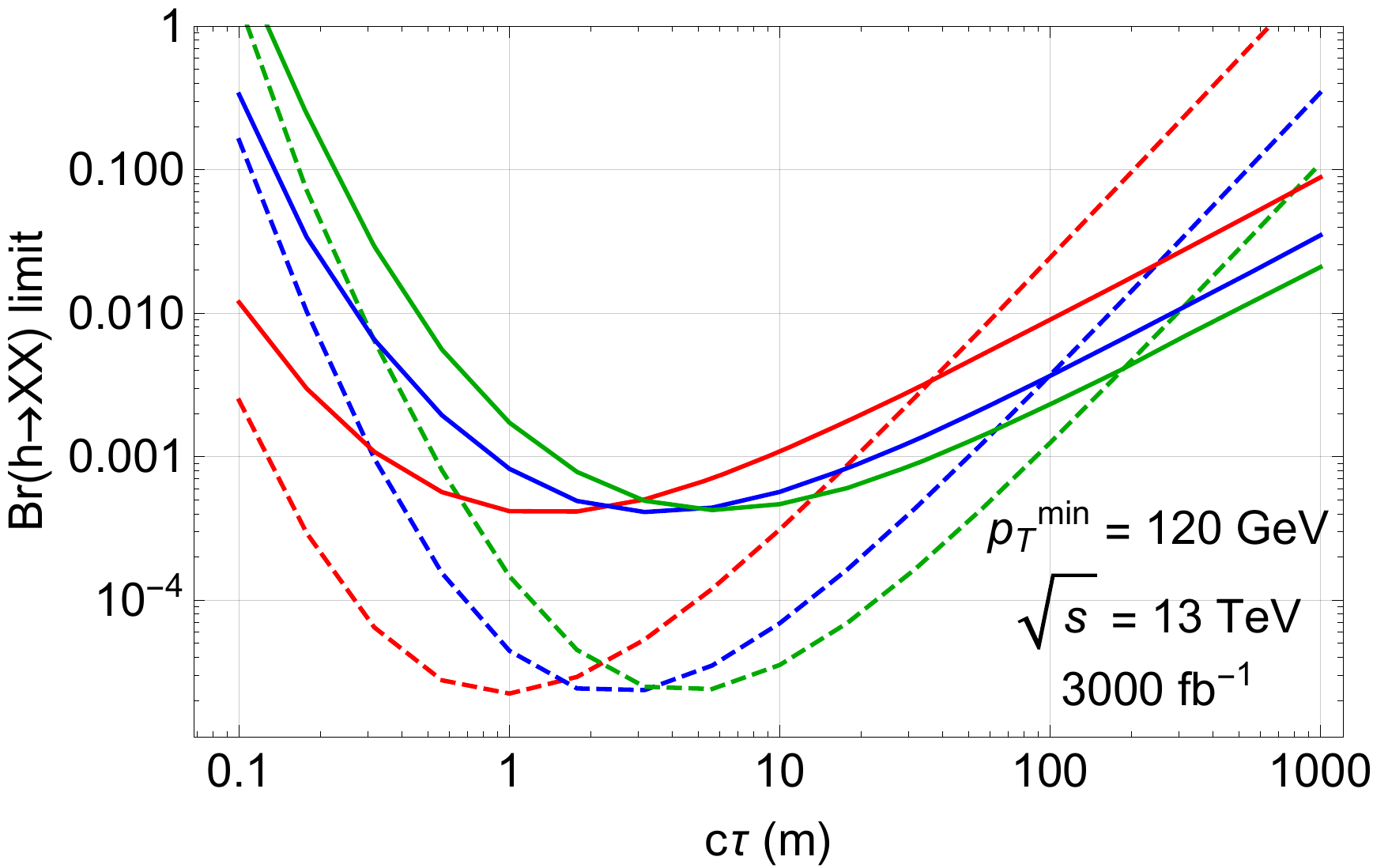}
\\
\end{tabular}
&
\includegraphics[height=15cm]{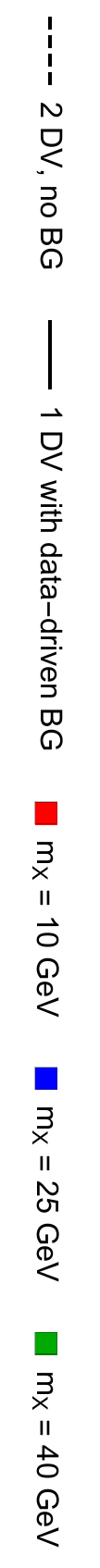}
\end{tabular}
\end{center}
\caption{{\small Limit projections for the data-driven search for a single DV
  in the ATLAS Muon Spectrometer (solid lines) compared to an assumed
  background-free search for two DVs (dashed lines). For comparison
  with existing limits \cite{Aad:2015uaa}, the top row shows limits
  that may have been achieved by performing this search at the LHC run
  1. The left (right) column corresponds to the pessimistic
  (optimistic) choice of QCD background, both normalized to give 3000
  background events at the LHC run 1.}}
\label{f.limits}
\end{figure*}

For the pessimistic analysis, additional kinematic cuts on the events
in SR$_Y$ with $\deltaphi > 1.5$ (corresponding to region A in
\fref{SRCR}) can lead to very slight increases in sensitivity compared
to a simple counting experiment in SR$_Y$ with the background
prediction derived from the observed events in CR$_Y$ (region C in
\fref{SRCR}). We find the two most useful strategies to be (a) no
further cuts, and (b) $H_T^\prime > 80 \gev$.  This may be indicative
of the kinds of cuts one might perform in a real experimental
analysis, but the details should be taken lightly, given
the crude nature of our fake-DV background simulation. The resulting
limits on $\mathrm{Br}(h \to X X)$ are shown in the left column of
\fref{limits}~\footnote{The actual experimental sensitivity in the short lifetime regime will also depend on any differences between the contributions to the MET from QCD jets versus $X$ particles that decay before the MS, as well as a more detailed treatment of the DV reconstruction efficiency for $X$ particles decaying in the outer regions of the HCAL.}.

For the optimistic analysis, SR$_Y$ with $\deltaphi > 1.0$ is so
signal-enriched that no cuts are necessary to enhance sensitivity even
for high luminosities. The resulting limits on $\mathrm{Br}(h \to X
X)$ are shown in the right column of \fref{limits}.

These limit projections confirm our expectation that the data-driven
search for one DV represents a great improvement at long $X$ proper
lifetimes, yielding limits orders of magnitude better than even a
background-free search for two DVs in the Muon Spectrometer. The
limits in the optimistic QCD case are noticably better than the
pessimistic QCD limits, especially for modest proper lifetimes less
than about one meter, due to the better intrinsic separation of signal
and background. However, the difference in the projected limits from
the two very different modelings of the QCD background is only about a
factor of $2$, indicating that our strategy is quite robust.
The background-free sensitivity projections of the two-DV search
scales with luminosity $\mathcal{L}$, while for the one-DV search with
data-driven background estimates the limit scales like
$\sqrt{\mathcal{L}}$. At high luminosities, the sensitivity gain of
the one-DV search relative to the two-DV search is therefore reduced,
but even in this case the former has superior reach at long
lifetimes. (Our estimates for the two-DV search are also likely to be
optimistic in the case of the HL-LHC due to the different running
conditions.)

\newcommand{\tempheighttwo}{4cm}
\begin{figure*}
\begin{center}
\hspace*{-1cm}
\begin{tabular}{c}
\includegraphics[width=12cm]{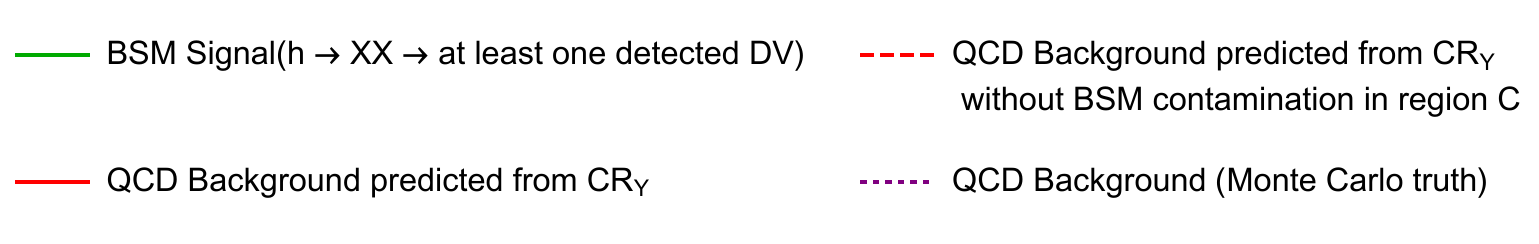}
\\
\begin{tabular}{ccc}
\includegraphics[height=\tempheighttwo]{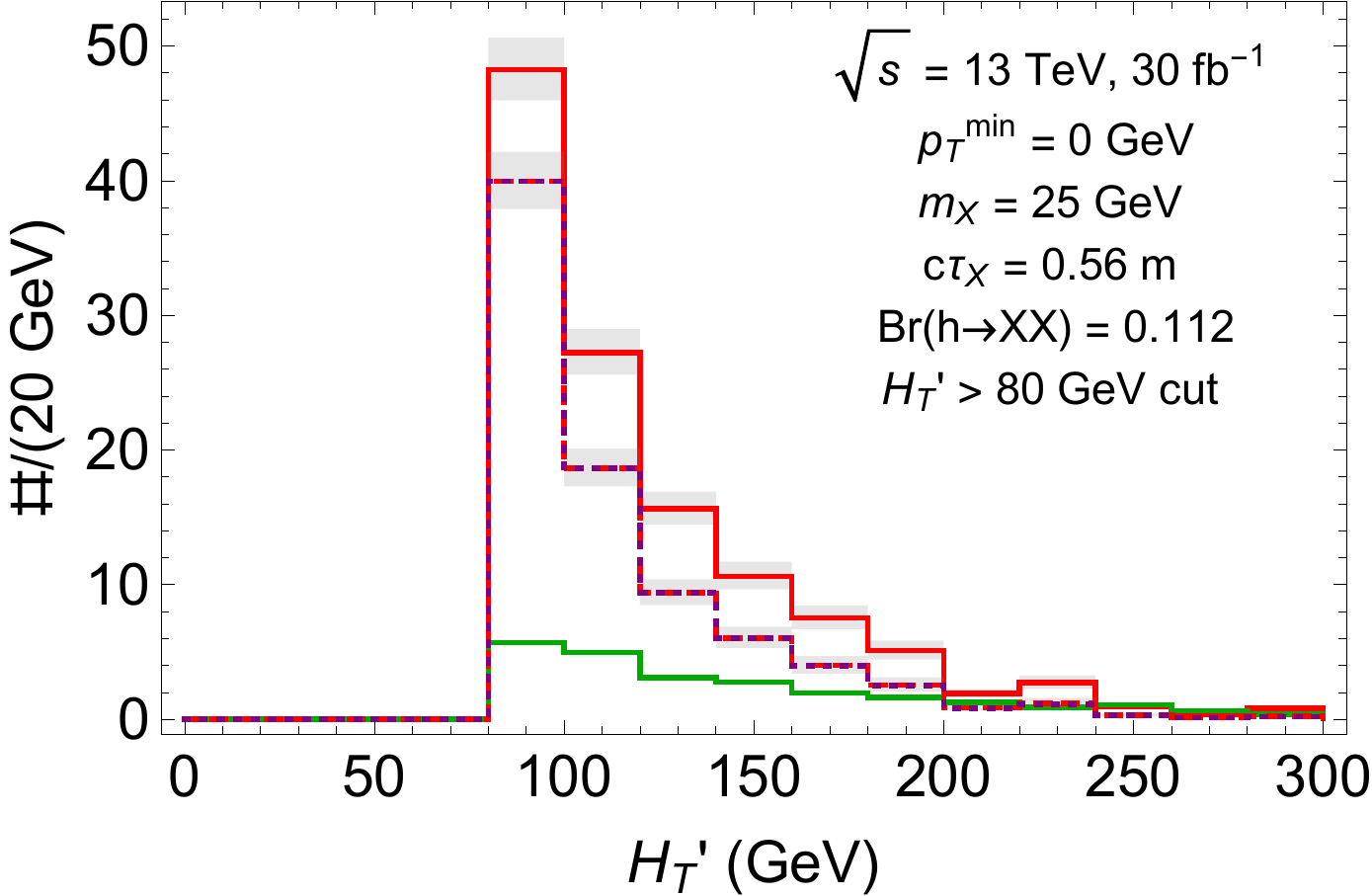}
& &
\includegraphics[height=\tempheighttwo]{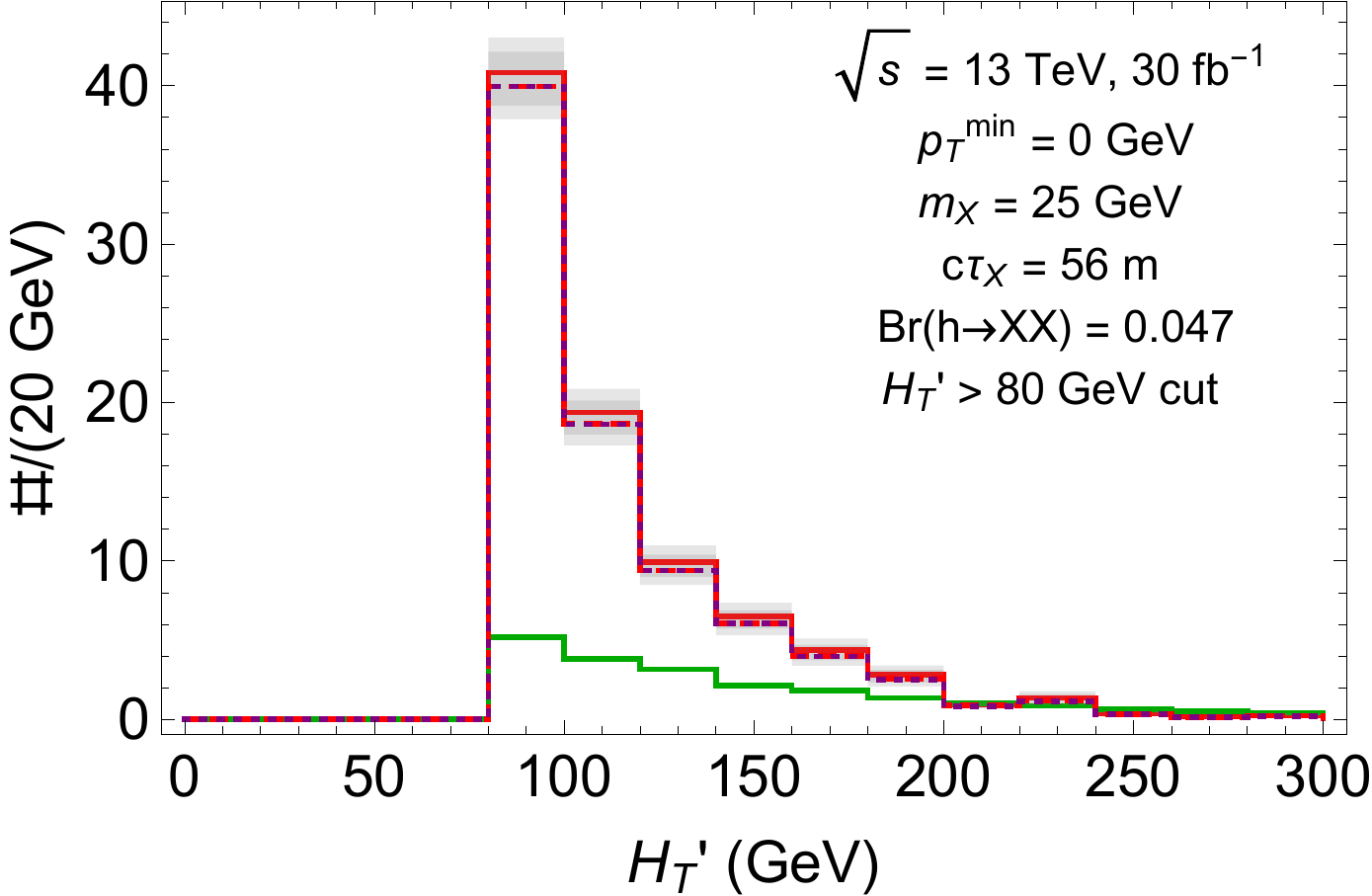}
\\
\includegraphics[height=\tempheighttwo]{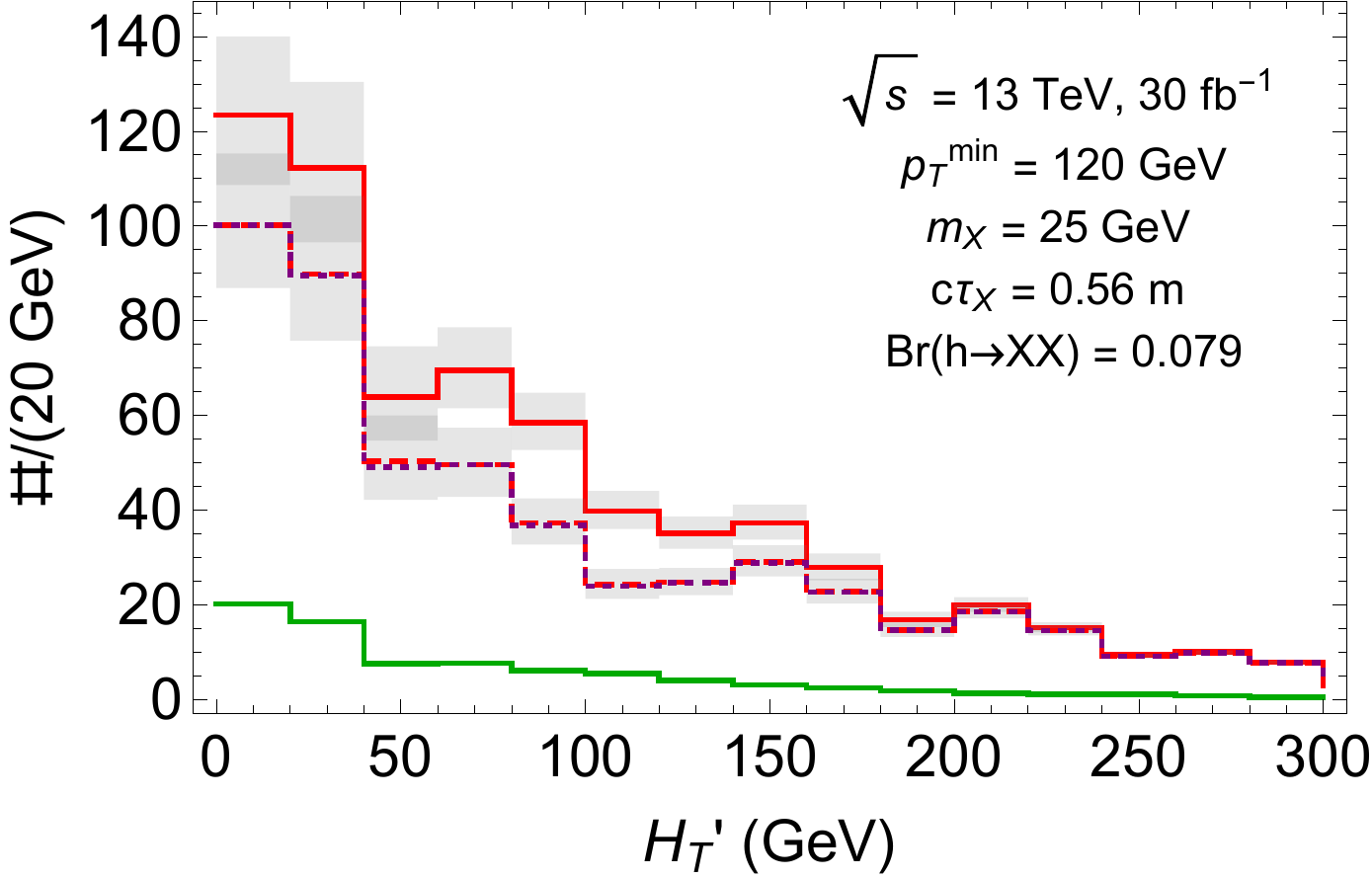}
& & 
\includegraphics[height=\tempheighttwo]{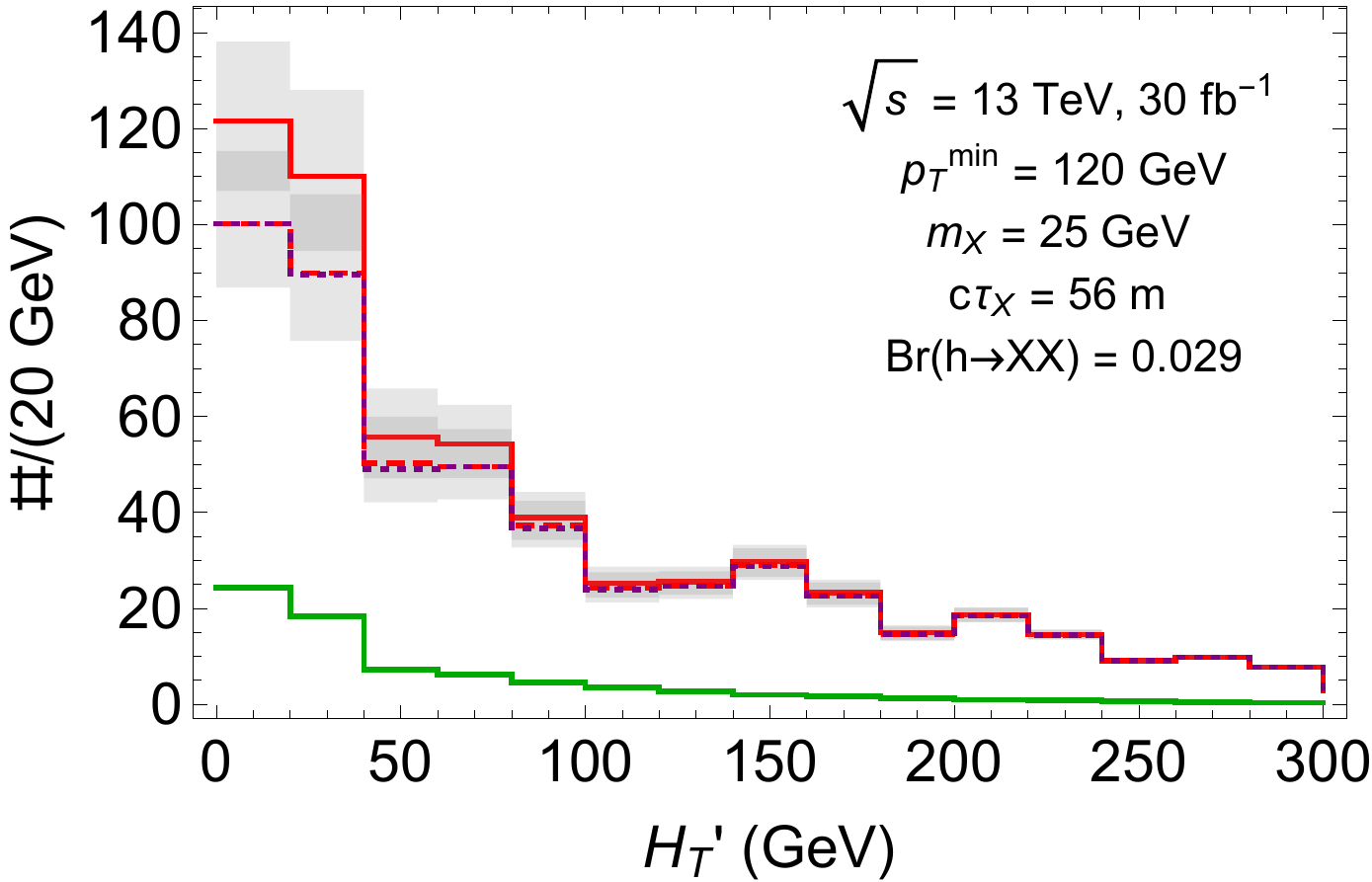}
\end{tabular}
\end{tabular}
\end{center}
\caption{{\small $H_T^\prime$ distributions, at the 13 TeV LHC with $30\ifb$
    of luminosity, in signal region A. For the pessimistic QCD case
    with $p_T^\mathrm{min} = 0 \gev$ and an additional $H_T^\prime >
    80 \gev$ cut (top), and for the optimistic QCD case with
    $p_T^\mathrm{min} = 120 \gev$ (bottom).  The mass of the LLP is
    $m_X = 25 \gev$, with a short lifetime of $c \tau = 56
    \mathrm{cm}$ on the left and a long lifetime of $c\tau = 56 m$ on
    the right. Green: $h \to X X$ signal, Red: QCD background
    prediction from CR$_Y$, including BSM contamination of region
    C. Dashed red: QCD background prediction if there were no BSM
    contamination in region C. Purple dotted: truth-level QCD in region A. 
    Gray shading indicates the $2\sigma$
    uncertainty in the SM prediction from limited region C statistics,
    which is significant in the lowest $H_T^\prime$ bin in the
    optimistic QCD case.}
}
\label{f.spectra}
\end{figure*}

We close this discussion by commenting on BSM contamination in the
control regions. \fref{spectra} shows the distribution of BSM $h\to X
X$ and SM events in region A, for long and short lifetimes with $m_X =
25 \gev$. In each case, $\mathrm{Br}(h \to X X)$ is chosen to be at
the 95\% CL limit projection. The solid red histogram shows the QCD
prediction derived from the observation in region C, while the dashed
red histogram shows what the prediction would be if there were no BSM
events in the CR$_Y$. In the short lifetime case, a significant
fraction of the background prediction results from  BSM events
 falling into the control region. This underscores why
sensitivity decreases sharply for proper lifetimes less than a
meter. As discussed in \ssref{SRCRdiscussion}, we expect alternative
definitions for the CR$_Y$ to be more useful in this case.

In deriving our limit projections for the single-DV search, we simply
took the expected observation in the CR$_Y$ at face-value to predict
the SM background. In a full analysis, sensitivity would be further
improved by taking into account the CR$_Y$
contamination for each assumption of $\mathrm{Br}(h \to X X)$, as
discussed in Sections \ref{ss.importantconsiderations} and
\ref{s.futuredirections}.

\subsection{Reach Improvement for Theories of Neutral Naturalness}

Theories of Neutral Naturalness, so-called because they solve the
little hierarchy problem through top partners that are neutral under
the SM strong force, are among the best-motivated theories that give
rise to Higgs decays to LLPs.  These theories predict (sub-)weak-scale
degrees of freedom that may carry either electroweak charges, as in
Folded SUSY \cite{Burdman:2006tz} and the Quirky Little Higgs
\cite{Cai:2008au}, or no SM charges at all, as realized in the Twin
Higgs model \cite{Chacko:2005pe}.  
 As LHC Run-1 results have reduced significantly the viable natural parameter space for colored top partners, models of Neutral Naturalness, which generalize the usual assumptions about top partner phenomenology, have come into new prominence as viable solutions to the hierarchy problem.
 Most importantly for
our current purposes, Higgs decays to LLPs are among the leading
signatures of these models, in many cases offering the best window
into the physics of SM-neutral top partners, and thereby onto the
stability of electroweak scale.  This makes theories of Neutral
Naturalness one of the most exciting motivations for LLP searches at
the LHC in general, and for the signal $h\to XX$ in particular.  In
this subsection we demonstrate how the sensitivity gains from the
search for a single DV in the MS proposed above translate to expanded
reach in the parameter space of the Fraternal Twin Higgs model
(FTH)~\cite{Craig:2015pha}.

The most important low-lying fundamental degrees of freedom in the FTH
are SM singlet top and bottom partners $T$ and $B$, which are charged
under a mirror QCD gauge group and couple to the Higgs via a
mixing-suppressed Yukawa interaction.  The Higgs boson acts as a
portal between the SM and the mirror QCD sector, through both its
direct Yukawa couplings to mirror quarks and the resulting effective
coupling to mirror gluons. 
These couplings enable low-lying mirror hadron states to be produced
in exotic Higgs decays.  These mirror hadrons decay back to the SM via
an off-shell Higgs, and are generically long-lived
\cite{Craig:2015pha, Morningstar:1999rf, Juknevich:2009ji,
  Juknevich:2009gg}.
The phenomenology of exotic Higgs decays in the FTH model depends in
detail on the relative values of the mass of the mirror bottom, $m_B$,
and the strong coupling scale of mirror QCD, $\Lambda_{QCD'}$, and can
be quite complicated.  The low-lying hadrons may be mirror
glueballs, mirror bottomonia, or a mixture of both; the total Higgs
branching fraction into mirror hadrons can be controlled by either the
mirror bottom Yukawa coupling or the (mirror-top-induced) effective
coupling to mirror gluons; the lifetime of the mirror glueballs depends
on both $\Lambda_{QCD'}$ and the mass of the mirror top $m_T$; and
non-perturbative physics describing hadronization in the mirror sector
can introduce large uncertainties.
Some of these issues were discussed in \cite{Craig:2015pha, Curtin:2015fna},
and will be explored in detail in an upcoming study
\cite{fthpheno}. Here, we merely give an abbreviated preview of those
results by focusing on a region of parameter space where the search
for a single DV in the MS offers obvious and unique advantages.

\begin{figure}
\begin{center}
\begin{tabular}{c}
\includegraphics[width=5.5cm]{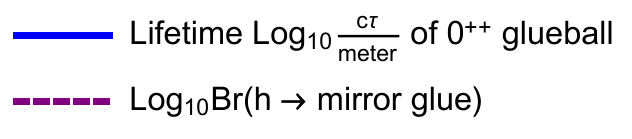}
\\
\includegraphics[width=8.5cm]{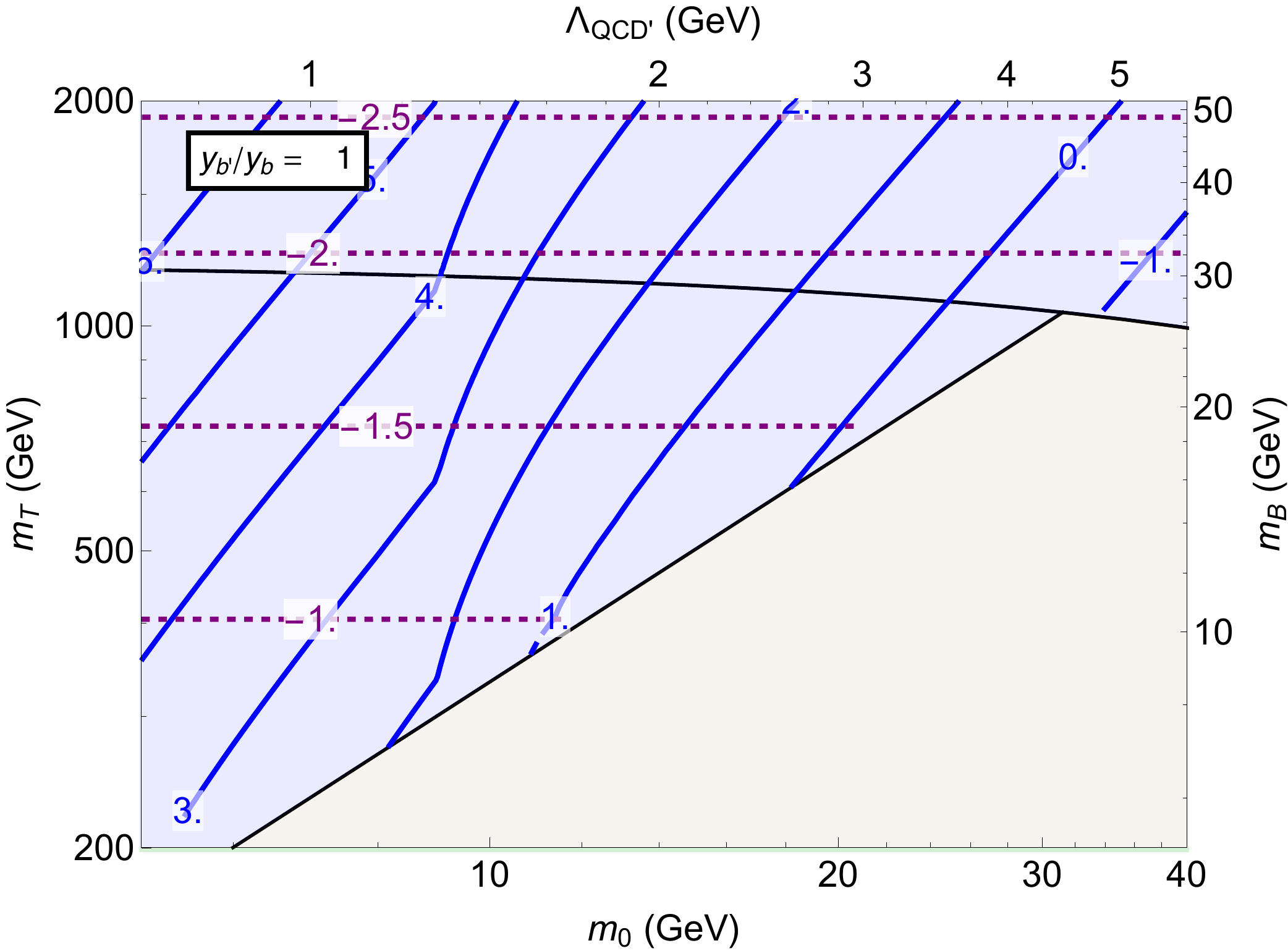}
\end{tabular}
\end{center}
\caption{ {\small Partial phase space of FTH model with $y_b^\prime =
  y_b^\mathrm{SM}$. 
  Blue background shading: YEGP phase of the FTH model.
  Brown shading, and other areas with $m_0 > 40 \gev$: more
    complicated exotic Higgs decay scenarios that will be explored in
    \cite{fthpheno}.
    Blue contours: proper lifetime $\log_{10}
    (c \tau/\mathrm{meter})$ of $0^{++}$ glueballs. 
    Purple contours:
    $\log_{10}$ of the perturbatively calculated exotic Higgs branching ratio to mirror gluons via intermediate mirror bottoms.  Below the horizontal black line the intermediate state can be multiple mirror bottomonia, while above the black line it can be an excited quirky bound state of two mirror bottoms which annihilates to glueballs. 
     }}
\label{f.NNlifetime}
\end{figure}

\begin{figure*}
\begin{center}
\hspace*{-9mm}
\begin{tabular}{c}
\includegraphics[width=18cm]{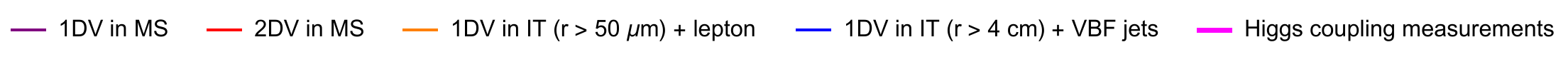}
\vspace{0mm} \\
\begin{tabular}{ccc}
\includegraphics[width=9.5cm]{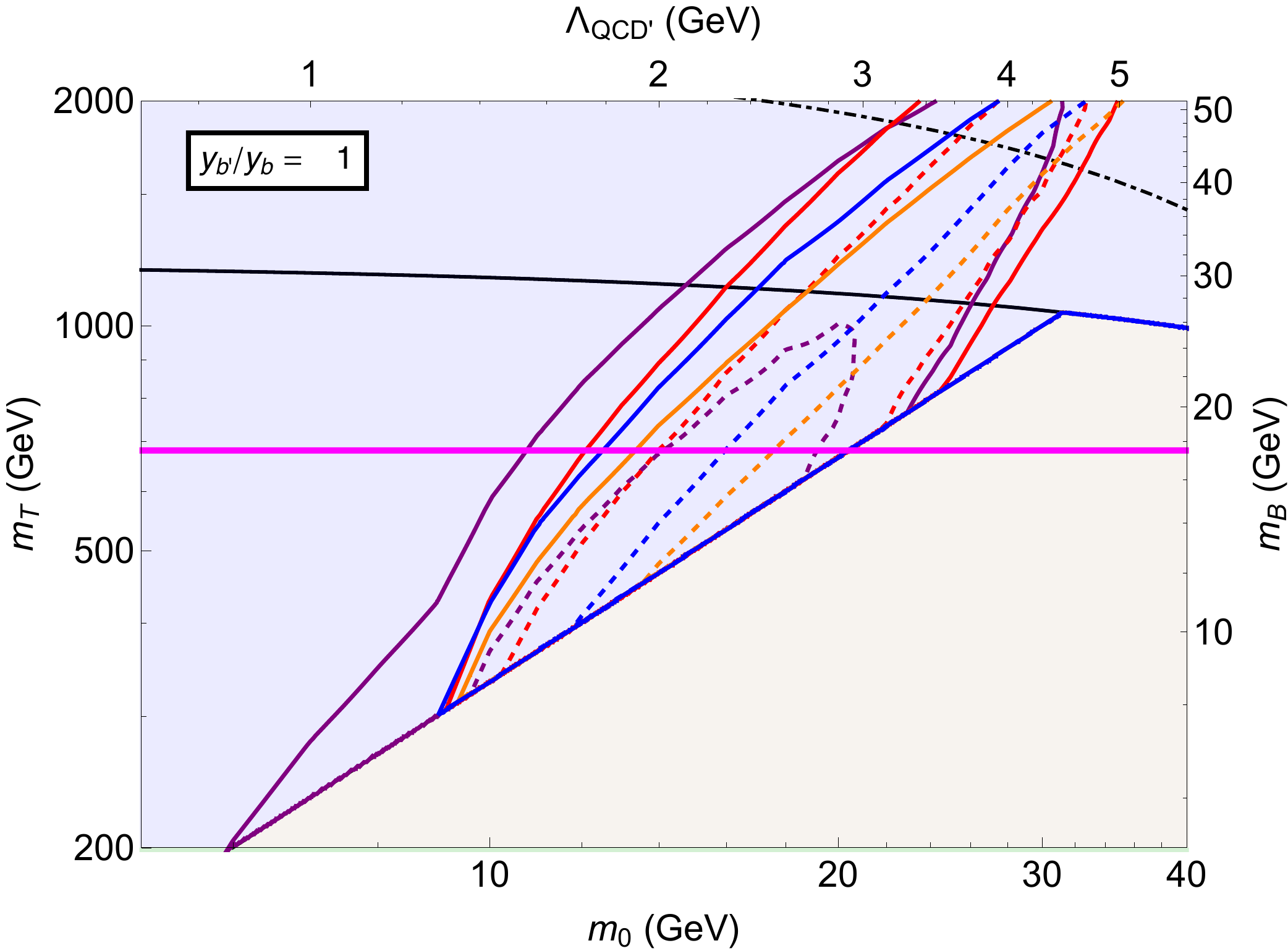}
& &
\includegraphics[width=9.5cm]{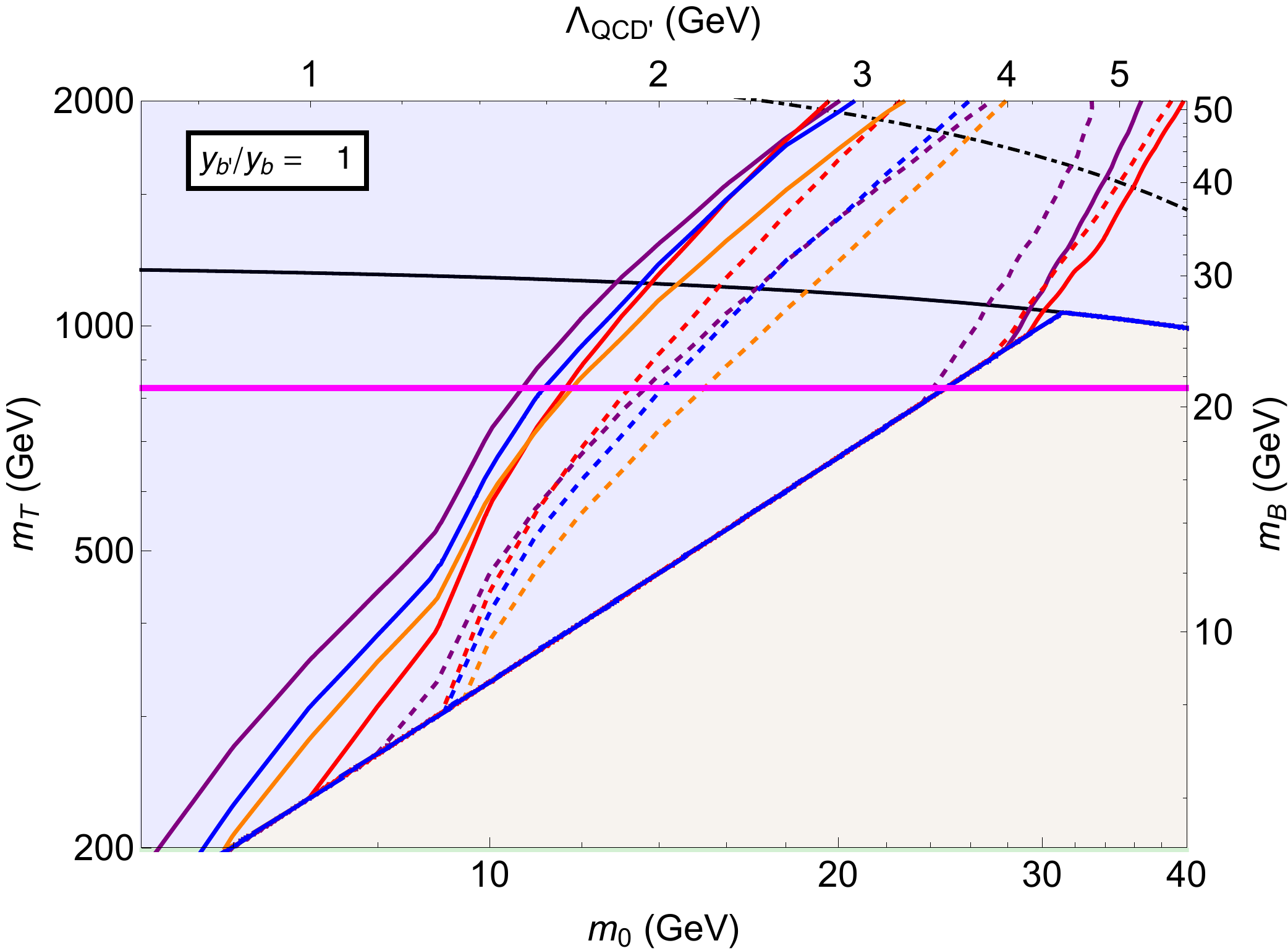}
\\ \\
LHC $300 \ifb$ & & HL-LHC $3000 \ifb$
\end{tabular}
\end{tabular}
\end{center}
\vspace*{-2mm}
\caption{ {\small Colored contours: reach of various proposed searches
    for LLPs in the YEGP phase of the FTH model (blue background
    shading, see \fref{NNlifetime}) at the LHC with $300\ifb$ (left) and $3000\ifb$ (right)
    of luminosity, using results from \cite{Curtin:2015fna} and the
    limits for searches in the MS from this paper.  Solid (dashed)
    contours indicate optimistic (pessimistic) assumptions for the
    number of $0^{++}$ glueballs produced, see text. Magenta line
    shows reach of Higgs coupling constraints for that luminosity.
    Projections above the black
    dot-dashed line should be treated with caution due to
    possible non-perturbative suppressions of the exotic Higgs decay
    branching ratio. Similar suppressions could occur for some
    glueball masses above 40 GeV.
    }}
\label{f.NN}
\end{figure*}

The plots in Figs.~\ref{f.NNlifetime} and \ref{f.NN} show part of the
phase space of exotic Higgs decays in the FTH model for the
$\mathbb{Z}_2$ symmetric choice of mirror bottom Yukawa $y_b^\prime =
y_b^\mathrm{SM}$. With this parameter fixed, the top partner mass
(left vertical axis) determines both the mirror Higgs vev $f$ and the
bottom partner mass (right vertical axis). The confinement scale
$\Lambda_{\mathrm{QCD}^\prime}$ (top horizontal axis) of mirror QCD is
an unknown parameter that depends on both the full mirror sector
spectrum and the UV completion, and determines the mass $m_0 \approx 7
\Lambda_{\mathrm{QCD}^\prime}$ of the lightest mirror glueball $G_0 =
0^{++}$ (bottom horizontal axis). At each point in this $(m_0,
m_T)$-plane, all mirror hadron masses, lifetimes, and exotic Higgs
branching fractions are determined (with the exception of additional
bottomonium decay modes if mirror leptons are
light). \fref{NNlifetime} shows the exotic Higgs decay branching
fraction to mirror glue, and the lifetime of $0^{++}$ glueballs. At
glueball masses below the $\bar b b$ threshold, the proper lifetimes
become extremely long, making this the most challenging regime for LLP
searches.

In the brown shaded regions, both glueballs and mirror bottomonia
$\eta$ are light enough to be produced in exotic Higgs decays, and can
potentially mix with each other. For $m_0 > 40 \gev$, there are
regions where glueballs are either not produced or decay to mirror
bottomonia, meaning that all exotic Higgs decays produce bottomonium
final states. These regions, as well as different choices of
$y_b^\prime$, will be explored in
\cite{fthpheno}.

The blue regions are the area of most interest for single DV searches
in the MS. Here, the decay $h \to B \bar B$ is perturbatively
allowed. The $\bar B B$ states either produce mirror bottomonia, which
can decay to glueballs, or a so-called quirk bound state
\cite{Kang:2008ea,Burdman:2008ek,Harnik:2008ax,Harnik:2011mv,Burdman:2014zta,
  Chacko:2015fbc}, which can be thought of as a single very excited
bottomonium, that promptly annihilates to glueballs. In either case,
the exotic Higgs decay branching fraction is dictated by the mirror
bottom Yukawa, and is therefore rather large ($\sim$ 1 - 10$\%$ for
$m_T \lesssim \tev$), but the final states are the long-lived
glueballs, which decay to the SM via the highly suppressed top partner
loop and an off-shell Higgs boson, with proper decay lengths ranging
from $\sim$ 1000 km to $\sim$ millimeters for glueball masses $\sim 5
- 60 \gev$.  The combination of relatively large LLP production rates
and long lifetimes makes this phase of the FTH, which we refer to as
``Yukawa-enhanced glueball production'' (YEGP), an ideal benchmark for
our single-DV searches.

Ref.~\cite{Curtin:2015fna} examined the reach of displaced searches at
the LHC for glueballs in theories of Neutral Naturalness, assuming the
exotic Higgs decays are mediated by the top partner loop. Three
searches were found to have great combined coverage of the parameter
space:
\begin{enumerate}
\item a search for 1 DV in the MS, with an additional DV in either the  MS or IT;
\item a search for 1 DV in the IT with minimum distance of 4~cm from
  the IP (modeled on current displaced vertex reconstruction
  capability at ATLAS), with VBF jets for triggering;
\item a search for 1 DV in the IT with a minimum distance of 50~$\mu
  m$ from the IP, with an additional lepton for triggering.
\end{enumerate} 
The first search has already been performed by ATLAS
at LHC Run-1 \cite{Aad:2015uaa}, while the other two are proposals for
future searches that could be performed by either general-purpose LHC
experiment. In particular, the third search demonstrates how much
sensitivity could be gained if very short displacements could be
reconstructed. The search projections are pessimistic in the sense
that ATLAS Run-1 reconstruction efficiencies are assumed for DVs in
the IT for the entire LHC program, and optimistic in the sense of
assuming no backgrounds.

We apply the same methodology as \cite{Curtin:2015fna} to the YEGP
phase of the FTH. This involves making the pessimistic assumption that
only two glueballs are produced per exotic Higgs decay (for small
$m_0$, showering will likely produce more). The projected exclusions
of the background-free (DV in IT + VBF) and (DV in IT + lepton)
searches in the YEGP phase are shown in \fref{NN} as blue and orange
contours. The corresponding exclusions from a background-free search
for 2 DVs in the MS and our proposed search for 1 DV in the MS (see
\fref{limits}) are shown as red and purple contours,
respectively. Solid contours indicate reach if all glueballs are the
lightest $0^{++}$ state, while dashed lines make the more pessimistic
assumption that $\sim 10\%$ of glueballs are $0^{++}$ when all states
are kinematically available~\footnote{Thermodynamic arguments
  \cite{JuknevichPhD} indicate that $\mathcal{O}(50\%)$ of produced
  glueballs end up in the $0^{++}$ state, but those predictions are
  highly uncertain due to our ignorance of pure-gauge
  hadronization. More sophisticated methods of parameterizing this
  ignorance have been proposed in \cite{Chacko:2015fbc} but they are
  not suitable for exotic Higgs decays where glueball masses cannot be
  neglected.}. Finally, limits above the dot-dashed black should be
treated with caution due to non-perturbative suppressions of the
exotic Higgs branching fraction compared to the perturbative rate we
assume, which will be discussed in more detail in \cite{fthpheno}.
Similar suppressions could occur for some glueball masses above 40
GeV, see \cite{Curtin:2015fna}. Finally, we also indicate the
exclusion reach of Higgs coupling measurements on the FTH model,
derived using the profile likelihood method \cite{Cowan:2010js} and
sensitivity projections for 300 and 3000 $\ifb$ from
\cite{Marono:2014jta, Flechl:2015foa}~\footnote{We use the most
  optimistic projections for the precision of Higgs coupling
  measurements (CMS) and assume they apply for both ATLAS and CMS, in
  order to demonstrate the large gain in sensitivity achieved by the
  direct searches for displaced vertices.}.

\fref{NN} makes clear that our proposed inclusive search for 1 DV in
the MS significantly extends the LHC reach in the FTH parameter space
to glueball masses as low as 6 GeV, and increases the top partner mass
reach by several hundred GeV compared to other searches. These
significant gains into the most challenging parts of FTH parameter
space, where glueballs have very long lifetime and mostly escape the
detector, strongly motivate implementation of this search. At the
HL-LHC, sensitivity improvements compared to the background-free (DV
in IT + VBF) search seem more modest. However, the background-free
assumption for the latter search is likely overly optimistic,
especially at high instantaneous luminosities, while our 
projections for the 1DV search already take different running
conditions into account in estimating backgrounds. Therefore, it is
very likely that the 1DV search will perform significantly better at
low glueball masses than the 1DV + lepton or jets searches.

\section{Directions for a Future Search Program}
\label{s.futuredirections}

In \sref{overview}, we explained how a data-driven ABCD method can be
used to obtain differential background estimates for LLP searches in
the MS. We demonstrated the technical details of such an analysis and
the resulting potential gains in sensitivity, using a particularly
well-motivated (and challenging) signal model of LLP pair production
from the decay of a Higgs-like scalar in \sref{htoXX}. We now
generalize this method to outline a possible comprehensive search
program for LLPs.

There are a large number of theories that yield LLPs with
detector-scale lifetimes.  We begin by surveying several of the best
motivated classes of such theories, and then discuss how they can be
mapped onto a simpler signature space for displaced decays.  This
signature space then naturally suggests a set of signal-like and
control-like regions defined by observables $Y_i$, which can be used
to implement flexible and model-independent searches for displaced
decays in the MS.  We conclude this section by commenting on how this
approach could be extended to searches for displaced decays in other
detector systems.

\subsection{Theories yielding long-lived signatures}

A wide variety of well-motivated theories of BSM physics predict LLPs.
Perhaps the most familiar framework yielding LLPs is the MSSM, which
can easily yield displaced superpartner
decays 
through a variety of mechanisms:
\begin{itemize}

\item In split SUSY, the decay rate of (sub-)TeV gaugino and higgsino
  superpartners is suppressed by heavy sfermions with masses in excess
  of 1000 TeV, leading to displaced decays \cite{Arvanitaki:2012ps,
    ArkaniHamed:2012gw,Rolbiecki:2015gsa}.

\item In models of gauge-mediated SUSY-breaking (GMSB), NLSP decays
  can become displaced when the scale of supersymmetry breaking is
  sufficiently high \cite{Giudice:1998bp}.

\item Models with $R$-parity violation (RPV) \cite{Barbier:2004ez}
  often feature very small couplings that in some cases can be
  generated dynamically \cite{Csaki:2013jza}, leading to
  detector-scale LSP lifetimes.
  Such small baryon-number-violating RPV couplings are independently
  well-motivated by models of baryogenesis \cite{Bouquet:1986mq,
    Campbell:1990fa, Cui:2012jh, Barry:2013nva, Ipek:2016bpf}.

\item In anomaly-mediated SUSY-breaking (AMSB) \cite{Chen:1996ap,
    Giudice:1998xp, Randall:1998uk}, the neutral wino LSP
  is nearly degenerate with the charged wino, resulting in a
  macroscopic lifetime for the charged state.
 
\end{itemize} 
With this panoply of well-motivated mechanisms to produce displaced
decays, the MSSM is a very effective generator of displaced
signatures: most prompt SUSY signatures can be readily translated into
well-motivated displaced signatures through one of these mechanisms
simply by giving the last stage of the cascade decay a macroscopic
lifetime.

Another large class of models that generically features long-lived
particles are weak-scale hidden sectors (Hidden Valleys)
\cite{Strassler:2006im,Strassler:2006ri,Strassler:2006qa,Han:2007ae,Strassler:2008bv,Strassler:2008fv},
which include the theories of Neutral Naturalness discussed in the
last section.  Possible portals into hidden sectors are provided by
the $h$ and $Z$ bosons coupling or mixing with the hidden states, as
well as through BSM states that have tree-level couplings to both
sectors. Such BSM mediator particles can be either singly produced
(e.g., a $Z'$), or pair-produced (e.g., new vector-like fermions),
corresponding to resonant and non-resonant LLP pair production.  These
models frequently contain strong dynamics in the hidden sector, which
can lead to variable and potentially high LLP multiplicities through
hidden sector showers.  It is also common in these models for hidden
sector states to exhibit a hierarchy of proper lifetimes.

Combining these two classes of ideas to extend the MSSM with a
(softly-broken supersymmetric) weak-scale hidden sector is
well-motivated by dark matter model-building \cite{Baumgart:2009tn,
  Chan:2011aa} and as a way to reconcile natural SUSY with constraints
from LHC Run-1 (Stealth SUSY)~\cite{Fan:2011yu}.  These theories can
lead to even more varied LLP phenomenology.  Typically in these
theories, it is the LSP (or, more exactly, the Lightest Ordinary
Superpartner ``LOSP'') that mediates decays into the hidden sector
\cite{Strassler:2006qa}.  Here displacement can arise in the decay of
the LOSP into the hidden sector as well as in the decay of one or more
of the hidden sector states back to the SM, and the detector signature
of the displaced decay is largely controlled by the detailed content
of the hidden sector.

Cosmology provides additional motivation for displaced decays at
colliders.  As already mentioned, baryogenesis can motivate long-lived
particles that decay via tiny baryon-number-violating interactions
\cite{Cui:2012jh, Barry:2013nva, Cui:2014twa, Ipek:2016bpf}. A
detector-scale lifetime can also be directly related to the DM relic
abundance in some models of freeze-in DM \cite{Co:2015pka}, or as a
consequence of a small mass splitting between two dark states to
enable efficient coannihilation in the early
universe~\cite{Falkowski:2014sma}.  Models with
heavy ($m\gtrsim 10$ GeV) sterile neutrinos generically predict
macroscopic decay lengths for the heavy right-handed neutrinos.
Depending on the details of the model, sterile neutrinos can be
dominantly singly-produced through charged-current interactions~\cite{Helo:2013esa, Antusch:2016vyf} or pair-produced through a
mediator such as the SM Higgs or a BSM vector
boson~\cite{Graesser:2007yj, Graesser:2007pc, Maiezza:2015lza,
  Batell:2016zod}.

\subsection{A signature space for displaced searches}
\label{ss.programoutline}

From the point of view of designing a flexible search program, what
matters is not the theoretical motivation for the displaced decay but
the detector signature. To a very large degree, this is controlled by
two features of any given model of new physics: (i) production, and
(ii) decay.  As the variety of theories listed above suggests, the
lifetime of any given LLP can effectively be regarded as a free
parameter.  Focusing on displaced decays in the MS as the case of
interest, we  further note that here searches are insensitive to
the fine details of the LLP decay.  Thus for searches in the MS, the
detector signatures will largely be controlled by the production
mechanism, which determines overall expectations for signal rates well
as the number and type of AOs.  These AOs in turn will control the
useful choice(s) of $Y$, the variable that along with the distinction
between iso- and non-iso-events defines the signal and control
regions in \fref{SRCR}.

Common production modes for LLPs $X$ include:
\begin{itemize}

\item The pair-production of a parent particle $P$ that then decays
  to $X$ + SM particles.  This production mode includes the vast
  majority of SUSY models, as well as models that pair-produce BSM
  mediator states, including some Hidden Valley theories and the
  cosmologically motivated models of Refs.~\cite{Co:2015pka,
    Falkowski:2014sma}.  If $P$ is colored, or if it is produced in
  the decays of colored particles, then the typical AOs are jets; if $P$
  can only be produced through its electroweak interactions with the
  SM, then the most useful AOs are likely to be leptons or photons.

\item The production of a single parent particle $P$ that decays via
  $P\to XX$.  This production mode can dominate in many Hidden Valley
  theories, including theories of Neutral Naturalness.  If $P$ is a
  Higgs-like state (either the SM-like Higgs, or a state that mixes
  with it) as in \sref{htoXX}, there are no distinctive AOs.  If $P$
  is a vector $Z'$ that mixes with SM gauge bosons via either mass or
  kinetic mixing, Drell-Yan (DY)-like production dominates, again
  without distinctive AOs (see e.g., \cite{Curtin:2014cca,
    Clarke:2015ala, Arguelles:2016ney, Aad:2014yea}). However, in
  fermiophobic $Z'$ scenarios production could require an associated
  SM $W, Z$ or $\gamma$.
  
\item LLPs can also be singly produced, generally in combination with
  some AOs in the event. For example, if BSM states $(X_1, X_2)$ form
  an $SU(2)_L$ doublet where $X_1$ is the neutral LLP, they can be
  produced in a DY-like process $W^* \to X_1 X_2 \to X_1 X_1 + $ soft,
  as in AMSB.  Another way to singly produce an LLP $X$ at a
  non-negligible rate is the decay of a parent $P$ into a hidden
  sector, such as $h\to X_1 X_2$, with $X_1$ long-lived and $X_2$
  promptly decaying to the SM.  Since it is generic for different
  states in a confining hidden sector to have order(s) of magnitude
  differences in their lifetimes, this possibility should not be
  overlooked.  Single production is also common in
  models of weak-scale sterile neutrinos, where the dominant
  production mechanism can be $pp\to N \ell$.

\end{itemize}
This, together with our findings from \sref{htoXX} for LLPs produced
in the decay of a Higgs-like scalar, suggest the following simple
dictionary of choices for the variable $Y$:
\begin{itemize}
\item The number of leptons $N_\ell$.
  The simplest SR$_Y$ is defined by requiring $N_\ell > 0$.
\item The number of light or heavy flavor jets $N_j, N_b$.  It may also be useful to include kinematic properties
  such as VBF tagging or $p_T$ cuts.  The simplest SR$_Y$ is defined
  by requiring $N_j$ or $N_b > 0$.
\item The number of tagged/reconstructed SM $W$, $Z$, $\gamma$ bosons,
  with the simplest SR$_Y$ requiring at least one of these.
\item To target signal models that dominantly do not produce AOs, the
  kinematic variable $\deltaphi$ and the veto on unusual objects in
  the calorimeters or tracker, as discussed in \ssref{SRCRdiscussion}.
\end{itemize}
Our toy analysis in \sref{htoXX} estimated the sensitivity of an
analysis using $Y = \deltaphi$. In more straightforward cases, simple
scaling arguments can give an idea of the achievable
sensitivity. Consider the case of $Y$ = number of leptons or $W, Z$.
The \emph{inclusive} QCD cross section for events with one DV in the
MS is $\sim O(100 \,\mathrm{fb})$ at 13 TeV, see
\fref{spectrabeforeSRCR}. The inclusive production cross section for
$W, Z$ bosons, which are also the dominant source of leptons, is about
$10^{-5}$ times smaller than inclusive QCD jet
production~\cite{Aad:2011rr}. We can therefore expect a signal region
defined by requiring a lepton, $W$, or $Z$ to have fewer background
events than the $h \to X X$ analysis by a similar factor. With such a
$\sim 10^{-3} \ \mathrm{fb}$ background cross section, searches for a
DV in the MS + lepton or $Z, W$ are likely to be nearly
background-free even with more than 300 $\ifb$ of luminosity. On the
other hand, a search for LLPs produced in association with at least
one $b$-jet defines a signal region with $Y = N_b \geq 1$. In this
case, the background reduction is a more modest factor of $\sim
30$~\cite{Chatrchyan:2012dk}. This will give sensitivity to the
pair-production of colored parent particles $P$ with masses in excess
of 1 TeV with only 30 $\ifb$ of LHC13 luminosity.  The sensitivity for
LLPs dominantly produced in association with light jets is more
challenging to estimate, and will depend on the spectrum.
In the presence of large mass splittings in the decay chain that
produces the LLPs, variables sensitive to the acoplanarity of the jets
+ DV system are attractive candidates for defining a robust
signal-like/control-like region split.
On the other hand, in compressed regions of parameter space where
associated jets are soft, $\deltaphi$ can be the most useful
choice of $Y$.

To realize maximum sensitivity, a detailed analysis along the lines of
that sketched in \sref{htoXX} would be implemented for each
signal model under consideration.  However, it is worth emphasizing
that for each of the above choices of $Y$, with associated definitions
of the regions A,B,C,D in \fref{SRCR}, BSM contributions to regions A
and C can be made visible simply by examining the ratio
\begin{equation}
\label{e.magicratio}
\mathcal{R}_Y(H_T^\prime, \ldots) = \frac{r_\nonisotoiso^{\mathrm{CR}_Y}(H_T^\prime, \ldots)}{r_\nonisotoiso^{\mathrm{SR}_Y}(H_T^\prime,\ldots)} 
\end{equation}
of rescaling functions as computed in SR$_Y$ and CR$_Y$ without
accounting for any BSM contributions. The $\ldots$ indicates
that this ratio could be observed as a function of other variables as
well, if statistics are sufficient. An excess in A but not in C (or
vice versa) would show up as a positive (negative) deviation in the
$\mathcal{R}_Y$ distribution from unity.
This is a flexible and fully model-independent way of searching for
deviations from SM expectations, with the potential to direct future
targeted analyses if interesting deviations are observed.

\subsection{Generalization to other Detector Systems}

The discussion in this paper centers on the detection of LLPs decaying
in the MS, due to the availability of both signal and orthogonal
triggers at ATLAS, and the unique advantage conferred by such searches
in probing LLPs with very long lifetimes.  However, the general
strategy we outline in \sref{overview}, and a model-independent set of
searches as suggested in \ssref{programoutline}, could be adapted to
LLP searches using other detector subsystems as well.

For LLPs decaying in the calorimeters, the isolation criteria used to
distinguish between iso- and non-iso-events operate similarly to the
criteria for the MS by quantifying the amount of activity upstream of
the DV-candidate. A signal trigger already exists at ATLAS
\cite{MuonRoITrigger}, and an orthogonal trigger for non-iso-events
could in principle be implemented as well.

In general, this search strategy can also be implemented in any
detector subsystem where the triggering strategy does not rely on the
LLP decay directly, which would be the case, e.g., when using the
single lepton trigger to search for displaced decays in the tracker
(as suggested for exotic Higgs decays \cite{Curtin:2015fna,
  Csaki:2015fba}).  In that case, different offline reconstruction
criteria could separate iso from non-iso-events, in the cases where
the background rate is large enough to necessitate implementation of
the data-driven background estimation strategy presented here.  This
would likely be relevant, for example, when reconstructing macroscopic
decay lengths less than a mm. Such an analysis would be very
challenging, but is highly motivated in many models, see, e.g.,
\cite{Curtin:2015fna}.

\section{Conclusions}
\label{s.conclusions}

Searches for long-lived particles (LLPs) are motivated in a large
variety of BSM scenarios connected to naturalness, dark matter,
baryogenesis, and other fundamental mysteries of particle physics. In
this paper we suggest use of existing ATLAS triggers to conduct a
search for a single LLP decaying in the MS, which offers great
improvements in sensitivity over existing searches for long proper
lifetimes. Such a search has to contend with sizable SM background
from QCD jets faking DVs in the MS, and we propose a data-driven
approach to obtain predictions for those backgrounds that can be made
differential in important kinematic variables.

We explicitly implement our strategy for the specific case where two
LLPs are pair-produced in the exotic decay of the 125 GeV Higgs boson
(or a general Higgs-like scalar). This is a very well-motivated
scenario that arises, e.g., in theories of Neutral Naturalness or
general Hidden Valleys. It also represents the most \emph{challenging}
application of our strategy, since the inclusive production mode
prohibits a straightforward definition of signal and control regions
by, for example, selecting for accompanying prompt leptons or
jets. Even so, our study demonstrates large sensitivity gains at long
lifetimes compared to a background-free search for two DVs in the
MS. This leads to significantly expanded reach in the parameter space
of BSM models, as we have explicitly demonstrated in the case of
Neutral Naturalness for the Fraternal Twin Higgs model.

Our strategy lends itself to the formulation of a model-independent
search program for LLPs decaying in the MS, in analogy to the
simplified model framework for prompt searches.  In our approach,
different signal model classes are categorized by the production mode
of the LLP(s), and deviations from the SM expectation are
parameterized as deviations of a data-driven ratio $\mathcal{R}_Y$
from unity, see \eref{magicratio}. Furthermore, our approach can be
generalized to LLP searches in other detector systems such as the
calorimeters and the tracker. This has the potential to significantly
expand both the breadth and the sensitivity of the search program for
long-lived particles at the LHC.

\subsection*{Acknowledgements}
D.C. would like to thank Sarah Eno for useful conversation.
A.C. acknowledges the support of the Swiss National Science Foundation under the grant 200020\_156083.
D.C. is supported by National Science Foundation grant No. PHY-1315155 and the Maryland Center for Fundamental Physics.  
H.L. and H.R. acknowledge the support of National Science Foundation grant PHY-1509257.
The work of J.S. is supported in part by DOE grant DE-SC0015655.

\bibliography{MS_DV}

\begin{thebibliography}{93}%
\makeatletter
\providecommand \@ifxundefined [1]{%
 \@ifx{#1\undefined}
}%
\providecommand \@ifnum [1]{%
 \ifnum #1\expandafter \@firstoftwo
 \else \expandafter \@secondoftwo
 \fi
}%
\providecommand \@ifx [1]{%
 \ifx #1\expandafter \@firstoftwo
 \else \expandafter \@secondoftwo
 \fi
}%
\providecommand \natexlab [1]{#1}%
\providecommand \enquote  [1]{``#1''}%
\providecommand \bibnamefont  [1]{#1}%
\providecommand \bibfnamefont [1]{#1}%
\providecommand \citenamefont [1]{#1}%
\providecommand \href@noop [0]{\@secondoftwo}%
\providecommand \href [0]{\begingroup \@sanitize@url \@href}%
\providecommand \@href[1]{\@@startlink{#1}\@@href}%
\providecommand \@@href[1]{\endgroup#1\@@endlink}%
\providecommand \@sanitize@url [0]{\catcode `\\12\catcode `\$12\catcode
  `\&12\catcode `\#12\catcode `\^12\catcode `\_12\catcode `\%12\relax}%
\providecommand \@@startlink[1]{}%
\providecommand \@@endlink[0]{}%
\providecommand \url  [0]{\begingroup\@sanitize@url \@url }%
\providecommand \@url [1]{\endgroup\@href {#1}{\urlprefix }}%
\providecommand \urlprefix  [0]{URL }%
\providecommand \Eprint [0]{\href }%
\providecommand \doibase [0]{http://dx.doi.org/}%
\providecommand \selectlanguage [0]{\@gobble}%
\providecommand \bibinfo  [0]{\@secondoftwo}%
\providecommand \bibfield  [0]{\@secondoftwo}%
\providecommand \translation [1]{[#1]}%
\providecommand \BibitemOpen [0]{}%
\providecommand \bibitemStop [0]{}%
\providecommand \bibitemNoStop [0]{.\EOS\space}%
\providecommand \EOS [0]{\spacefactor3000\relax}%
\providecommand \BibitemShut  [1]{\csname bibitem#1\endcsname}%
\let\auto@bib@innerbib\@empty
\bibitem [{\citenamefont {Arvanitaki}\ \emph {et~al.}(2013)\citenamefont
  {Arvanitaki}, \citenamefont {Craig}, \citenamefont {Dimopoulos},\ and\
  \citenamefont {Villadoro}}]{Arvanitaki:2012ps}%
  \BibitemOpen
  \bibfield  {author} {\bibinfo {author} {\bibfnamefont {A.}~\bibnamefont
  {Arvanitaki}}, \bibinfo {author} {\bibfnamefont {N.}~\bibnamefont {Craig}},
  \bibinfo {author} {\bibfnamefont {S.}~\bibnamefont {Dimopoulos}}, \ and\
  \bibinfo {author} {\bibfnamefont {G.}~\bibnamefont {Villadoro}},\ }\href
  {\doibase 10.1007/JHEP02(2013)126} {\bibfield  {journal} {\bibinfo  {journal}
  {JHEP}\ }\textbf {\bibinfo {volume} {02}},\ \bibinfo {pages} {126} (\bibinfo
  {year} {2013})},\ \Eprint {http://arxiv.org/abs/1210.0555} {arXiv:1210.0555
  [hep-ph]} \BibitemShut {NoStop}%
\bibitem [{\citenamefont {Arkani-Hamed}\ \emph {et~al.}(2012)\citenamefont
  {Arkani-Hamed}, \citenamefont {Gupta}, \citenamefont {Kaplan}, \citenamefont
  {Weiner},\ and\ \citenamefont {Zorawski}}]{ArkaniHamed:2012gw}%
  \BibitemOpen
  \bibfield  {author} {\bibinfo {author} {\bibfnamefont {N.}~\bibnamefont
  {Arkani-Hamed}}, \bibinfo {author} {\bibfnamefont {A.}~\bibnamefont {Gupta}},
  \bibinfo {author} {\bibfnamefont {D.~E.}\ \bibnamefont {Kaplan}}, \bibinfo
  {author} {\bibfnamefont {N.}~\bibnamefont {Weiner}}, \ and\ \bibinfo {author}
  {\bibfnamefont {T.}~\bibnamefont {Zorawski}},\ }\href@noop {} {\  (\bibinfo
  {year} {2012})},\ \Eprint {http://arxiv.org/abs/1212.6971} {arXiv:1212.6971
  [hep-ph]} \BibitemShut {NoStop}%
\bibitem [{\citenamefont {Giudice}\ and\ \citenamefont
  {Rattazzi}(1999)}]{Giudice:1998bp}%
  \BibitemOpen
  \bibfield  {author} {\bibinfo {author} {\bibfnamefont {G.~F.}\ \bibnamefont
  {Giudice}}\ and\ \bibinfo {author} {\bibfnamefont {R.}~\bibnamefont
  {Rattazzi}},\ }\href {\doibase 10.1016/S0370-1573(99)00042-3} {\bibfield
  {journal} {\bibinfo  {journal} {Phys. Rept.}\ }\textbf {\bibinfo {volume}
  {322}},\ \bibinfo {pages} {419} (\bibinfo {year} {1999})},\ \Eprint
  {http://arxiv.org/abs/hep-ph/9801271} {arXiv:hep-ph/9801271 [hep-ph]}
  \BibitemShut {NoStop}%
\bibitem [{\citenamefont {Barbier}\ \emph {et~al.}(2005)\citenamefont {Barbier}
  \emph {et~al.}}]{Barbier:2004ez}%
  \BibitemOpen
  \bibfield  {author} {\bibinfo {author} {\bibfnamefont {R.}~\bibnamefont
  {Barbier}} \emph {et~al.},\ }\href {\doibase 10.1016/j.physrep.2005.08.006}
  {\bibfield  {journal} {\bibinfo  {journal} {Phys. Rept.}\ }\textbf {\bibinfo
  {volume} {420}},\ \bibinfo {pages} {1} (\bibinfo {year} {2005})},\ \Eprint
  {http://arxiv.org/abs/hep-ph/0406039} {arXiv:hep-ph/0406039 [hep-ph]}
  \BibitemShut {NoStop}%
\bibitem [{\citenamefont {Csaki}\ \emph {et~al.}(2014)\citenamefont {Csaki},
  \citenamefont {Kuflik},\ and\ \citenamefont {Volansky}}]{Csaki:2013jza}%
  \BibitemOpen
  \bibfield  {author} {\bibinfo {author} {\bibfnamefont {C.}~\bibnamefont
  {Csaki}}, \bibinfo {author} {\bibfnamefont {E.}~\bibnamefont {Kuflik}}, \
  and\ \bibinfo {author} {\bibfnamefont {T.}~\bibnamefont {Volansky}},\ }\href
  {\doibase 10.1103/PhysRevLett.112.131801} {\bibfield  {journal} {\bibinfo
  {journal} {Phys. Rev. Lett.}\ }\textbf {\bibinfo {volume} {112}},\ \bibinfo
  {pages} {131801} (\bibinfo {year} {2014})},\ \Eprint
  {http://arxiv.org/abs/1309.5957} {arXiv:1309.5957 [hep-ph]} \BibitemShut
  {NoStop}%
\bibitem [{\citenamefont {Fan}\ \emph {et~al.}(2011)\citenamefont {Fan},
  \citenamefont {Reece},\ and\ \citenamefont {Ruderman}}]{Fan:2011yu}%
  \BibitemOpen
  \bibfield  {author} {\bibinfo {author} {\bibfnamefont {J.}~\bibnamefont
  {Fan}}, \bibinfo {author} {\bibfnamefont {M.}~\bibnamefont {Reece}}, \ and\
  \bibinfo {author} {\bibfnamefont {J.~T.}\ \bibnamefont {Ruderman}},\ }\href
  {\doibase 10.1007/JHEP11(2011)012} {\bibfield  {journal} {\bibinfo  {journal}
  {JHEP}\ }\textbf {\bibinfo {volume} {1111}},\ \bibinfo {pages} {012}
  (\bibinfo {year} {2011})},\ \Eprint {http://arxiv.org/abs/1105.5135}
  {arXiv:1105.5135 [hep-ph]} \BibitemShut {NoStop}%
\bibitem [{\citenamefont {Bouquet}\ and\ \citenamefont
  {Salati}(1987)}]{Bouquet:1986mq}%
  \BibitemOpen
  \bibfield  {author} {\bibinfo {author} {\bibfnamefont {A.}~\bibnamefont
  {Bouquet}}\ and\ \bibinfo {author} {\bibfnamefont {P.}~\bibnamefont
  {Salati}},\ }\href {\doibase 10.1016/0550-3213(87)90050-2} {\bibfield
  {journal} {\bibinfo  {journal} {Nucl. Phys.}\ }\textbf {\bibinfo {volume}
  {B284}},\ \bibinfo {pages} {557} (\bibinfo {year} {1987})}\BibitemShut
  {NoStop}%
\bibitem [{\citenamefont {Campbell}\ \emph {et~al.}(1991)\citenamefont
  {Campbell}, \citenamefont {Davidson}, \citenamefont {Ellis},\ and\
  \citenamefont {Olive}}]{Campbell:1990fa}%
  \BibitemOpen
  \bibfield  {author} {\bibinfo {author} {\bibfnamefont {B.~A.}\ \bibnamefont
  {Campbell}}, \bibinfo {author} {\bibfnamefont {S.}~\bibnamefont {Davidson}},
  \bibinfo {author} {\bibfnamefont {J.~R.}\ \bibnamefont {Ellis}}, \ and\
  \bibinfo {author} {\bibfnamefont {K.~A.}\ \bibnamefont {Olive}},\ }\href
  {\doibase 10.1016/0370-2693(91)91795-W} {\bibfield  {journal} {\bibinfo
  {journal} {Phys. Lett.}\ }\textbf {\bibinfo {volume} {B256}},\ \bibinfo
  {pages} {484} (\bibinfo {year} {1991})}\BibitemShut {NoStop}%
\bibitem [{\citenamefont {Cui}\ and\ \citenamefont
  {Sundrum}(2013)}]{Cui:2012jh}%
  \BibitemOpen
  \bibfield  {author} {\bibinfo {author} {\bibfnamefont {Y.}~\bibnamefont
  {Cui}}\ and\ \bibinfo {author} {\bibfnamefont {R.}~\bibnamefont {Sundrum}},\
  }\href {\doibase 10.1103/PhysRevD.87.116013} {\bibfield  {journal} {\bibinfo
  {journal} {Phys. Rev.}\ }\textbf {\bibinfo {volume} {D87}},\ \bibinfo {pages}
  {116013} (\bibinfo {year} {2013})},\ \Eprint {http://arxiv.org/abs/1212.2973}
  {arXiv:1212.2973 [hep-ph]} \BibitemShut {NoStop}%
\bibitem [{\citenamefont {Barry}\ \emph {et~al.}(2014)\citenamefont {Barry},
  \citenamefont {Graham},\ and\ \citenamefont {Rajendran}}]{Barry:2013nva}%
  \BibitemOpen
  \bibfield  {author} {\bibinfo {author} {\bibfnamefont {K.}~\bibnamefont
  {Barry}}, \bibinfo {author} {\bibfnamefont {P.~W.}\ \bibnamefont {Graham}}, \
  and\ \bibinfo {author} {\bibfnamefont {S.}~\bibnamefont {Rajendran}},\ }\href
  {\doibase 10.1103/PhysRevD.89.054003} {\bibfield  {journal} {\bibinfo
  {journal} {Phys. Rev.}\ }\textbf {\bibinfo {volume} {D89}},\ \bibinfo {pages}
  {054003} (\bibinfo {year} {2014})},\ \Eprint {http://arxiv.org/abs/1310.3853}
  {arXiv:1310.3853 [hep-ph]} \BibitemShut {NoStop}%
\bibitem [{\citenamefont {Ipek}\ and\ \citenamefont
  {March-Russell}(2016)}]{Ipek:2016bpf}%
  \BibitemOpen
  \bibfield  {author} {\bibinfo {author} {\bibfnamefont {S.}~\bibnamefont
  {Ipek}}\ and\ \bibinfo {author} {\bibfnamefont {J.}~\bibnamefont
  {March-Russell}},\ }\href@noop {} {\  (\bibinfo {year} {2016})},\ \Eprint
  {http://arxiv.org/abs/1604.00009} {arXiv:1604.00009 [hep-ph]} \BibitemShut
  {NoStop}%
\bibitem [{\citenamefont {Strassler}\ and\ \citenamefont
  {Zurek}(2007)}]{Strassler:2006im}%
  \BibitemOpen
  \bibfield  {author} {\bibinfo {author} {\bibfnamefont {M.~J.}\ \bibnamefont
  {Strassler}}\ and\ \bibinfo {author} {\bibfnamefont {K.~M.}\ \bibnamefont
  {Zurek}},\ }\href {\doibase 10.1016/j.physletb.2007.06.055} {\bibfield
  {journal} {\bibinfo  {journal} {Phys.Lett.}\ }\textbf {\bibinfo {volume}
  {B651}},\ \bibinfo {pages} {374} (\bibinfo {year} {2007})},\ \Eprint
  {http://arxiv.org/abs/hep-ph/0604261} {arXiv:hep-ph/0604261 [hep-ph]}
  \BibitemShut {NoStop}%
\bibitem [{\citenamefont {Strassler}\ and\ \citenamefont
  {Zurek}(2008)}]{Strassler:2006ri}%
  \BibitemOpen
  \bibfield  {author} {\bibinfo {author} {\bibfnamefont {M.~J.}\ \bibnamefont
  {Strassler}}\ and\ \bibinfo {author} {\bibfnamefont {K.~M.}\ \bibnamefont
  {Zurek}},\ }\href {\doibase 10.1016/j.physletb.2008.02.008} {\bibfield
  {journal} {\bibinfo  {journal} {Phys.Lett.}\ }\textbf {\bibinfo {volume}
  {B661}},\ \bibinfo {pages} {263} (\bibinfo {year} {2008})},\ \Eprint
  {http://arxiv.org/abs/hep-ph/0605193} {arXiv:hep-ph/0605193 [hep-ph]}
  \BibitemShut {NoStop}%
\bibitem [{\citenamefont {Strassler}(2006)}]{Strassler:2006qa}%
  \BibitemOpen
  \bibfield  {author} {\bibinfo {author} {\bibfnamefont {M.~J.}\ \bibnamefont
  {Strassler}},\ }\href@noop {} {\  (\bibinfo {year} {2006})},\ \Eprint
  {http://arxiv.org/abs/hep-ph/0607160} {arXiv:hep-ph/0607160 [hep-ph]}
  \BibitemShut {NoStop}%
\bibitem [{\citenamefont {Han}\ \emph {et~al.}(2008)\citenamefont {Han},
  \citenamefont {Si}, \citenamefont {Zurek},\ and\ \citenamefont
  {Strassler}}]{Han:2007ae}%
  \BibitemOpen
  \bibfield  {author} {\bibinfo {author} {\bibfnamefont {T.}~\bibnamefont
  {Han}}, \bibinfo {author} {\bibfnamefont {Z.}~\bibnamefont {Si}}, \bibinfo
  {author} {\bibfnamefont {K.~M.}\ \bibnamefont {Zurek}}, \ and\ \bibinfo
  {author} {\bibfnamefont {M.~J.}\ \bibnamefont {Strassler}},\ }\href {\doibase
  10.1088/1126-6708/2008/07/008} {\bibfield  {journal} {\bibinfo  {journal}
  {JHEP}\ }\textbf {\bibinfo {volume} {0807}},\ \bibinfo {pages} {008}
  (\bibinfo {year} {2008})},\ \Eprint {http://arxiv.org/abs/0712.2041}
  {arXiv:0712.2041 [hep-ph]} \BibitemShut {NoStop}%
\bibitem [{\citenamefont {Strassler}(2008{\natexlab{a}})}]{Strassler:2008bv}%
  \BibitemOpen
  \bibfield  {author} {\bibinfo {author} {\bibfnamefont {M.~J.}\ \bibnamefont
  {Strassler}},\ }\href@noop {} {\  (\bibinfo {year} {2008}{\natexlab{a}})},\
  \Eprint {http://arxiv.org/abs/0801.0629} {arXiv:0801.0629 [hep-ph]}
  \BibitemShut {NoStop}%
\bibitem [{\citenamefont {Strassler}(2008{\natexlab{b}})}]{Strassler:2008fv}%
  \BibitemOpen
  \bibfield  {author} {\bibinfo {author} {\bibfnamefont {M.~J.}\ \bibnamefont
  {Strassler}},\ }\href@noop {} {\  (\bibinfo {year} {2008}{\natexlab{b}})},\
  \Eprint {http://arxiv.org/abs/0806.2385} {arXiv:0806.2385 [hep-ph]}
  \BibitemShut {NoStop}%
\bibitem [{\citenamefont {Curtin}\ \emph {et~al.}(2015)\citenamefont {Curtin},
  \citenamefont {Essig}, \citenamefont {Gori},\ and\ \citenamefont
  {Shelton}}]{Curtin:2014cca}%
  \BibitemOpen
  \bibfield  {author} {\bibinfo {author} {\bibfnamefont {D.}~\bibnamefont
  {Curtin}}, \bibinfo {author} {\bibfnamefont {R.}~\bibnamefont {Essig}},
  \bibinfo {author} {\bibfnamefont {S.}~\bibnamefont {Gori}}, \ and\ \bibinfo
  {author} {\bibfnamefont {J.}~\bibnamefont {Shelton}},\ }\href {\doibase
  10.1007/JHEP02(2015)157} {\bibfield  {journal} {\bibinfo  {journal} {JHEP}\
  }\textbf {\bibinfo {volume} {1502}},\ \bibinfo {pages} {157} (\bibinfo {year}
  {2015})},\ \Eprint {http://arxiv.org/abs/1412.0018} {arXiv:1412.0018
  [hep-ph]} \BibitemShut {NoStop}%
\bibitem [{\citenamefont {Clarke}(2015)}]{Clarke:2015ala}%
  \BibitemOpen
  \bibfield  {author} {\bibinfo {author} {\bibfnamefont {J.~D.}\ \bibnamefont
  {Clarke}},\ }\href {\doibase 10.1007/JHEP10(2015)061} {\bibfield  {journal}
  {\bibinfo  {journal} {JHEP}\ }\textbf {\bibinfo {volume} {10}},\ \bibinfo
  {pages} {061} (\bibinfo {year} {2015})},\ \Eprint
  {http://arxiv.org/abs/1505.00063} {arXiv:1505.00063 [hep-ph]} \BibitemShut
  {NoStop}%
\bibitem [{\citenamefont {Argüelles}\ \emph {et~al.}(2016)\citenamefont
  {Argüelles}, \citenamefont {He}, \citenamefont {Ovanesyan}, \citenamefont
  {Peng},\ and\ \citenamefont {Ramsey-Musolf}}]{Arguelles:2016ney}%
  \BibitemOpen
  \bibfield  {author} {\bibinfo {author} {\bibfnamefont {C.~A.}\ \bibnamefont
  {Argüelles}}, \bibinfo {author} {\bibfnamefont {X.-G.}\ \bibnamefont {He}},
  \bibinfo {author} {\bibfnamefont {G.}~\bibnamefont {Ovanesyan}}, \bibinfo
  {author} {\bibfnamefont {T.}~\bibnamefont {Peng}}, \ and\ \bibinfo {author}
  {\bibfnamefont {M.~J.}\ \bibnamefont {Ramsey-Musolf}},\ }\href@noop {} {\
  (\bibinfo {year} {2016})},\ \Eprint {http://arxiv.org/abs/1604.00044}
  {arXiv:1604.00044 [hep-ph]} \BibitemShut {NoStop}%
\bibitem [{\citenamefont {Burdman}\ \emph {et~al.}(2007)\citenamefont
  {Burdman}, \citenamefont {Chacko}, \citenamefont {Goh},\ and\ \citenamefont
  {Harnik}}]{Burdman:2006tz}%
  \BibitemOpen
  \bibfield  {author} {\bibinfo {author} {\bibfnamefont {G.}~\bibnamefont
  {Burdman}}, \bibinfo {author} {\bibfnamefont {Z.}~\bibnamefont {Chacko}},
  \bibinfo {author} {\bibfnamefont {H.-S.}\ \bibnamefont {Goh}}, \ and\
  \bibinfo {author} {\bibfnamefont {R.}~\bibnamefont {Harnik}},\ }\href
  {\doibase 10.1088/1126-6708/2007/02/009} {\bibfield  {journal} {\bibinfo
  {journal} {JHEP}\ }\textbf {\bibinfo {volume} {0702}},\ \bibinfo {pages}
  {009} (\bibinfo {year} {2007})},\ \Eprint
  {http://arxiv.org/abs/hep-ph/0609152} {arXiv:hep-ph/0609152 [hep-ph]}
  \BibitemShut {NoStop}%
\bibitem [{\citenamefont {Cai}\ \emph {et~al.}(2009)\citenamefont {Cai},
  \citenamefont {Cheng},\ and\ \citenamefont {Terning}}]{Cai:2008au}%
  \BibitemOpen
  \bibfield  {author} {\bibinfo {author} {\bibfnamefont {H.}~\bibnamefont
  {Cai}}, \bibinfo {author} {\bibfnamefont {H.-C.}\ \bibnamefont {Cheng}}, \
  and\ \bibinfo {author} {\bibfnamefont {J.}~\bibnamefont {Terning}},\ }\href
  {\doibase 10.1088/1126-6708/2009/05/045} {\bibfield  {journal} {\bibinfo
  {journal} {JHEP}\ }\textbf {\bibinfo {volume} {0905}},\ \bibinfo {pages}
  {045} (\bibinfo {year} {2009})},\ \Eprint {http://arxiv.org/abs/0812.0843}
  {arXiv:0812.0843 [hep-ph]} \BibitemShut {NoStop}%
\bibitem [{\citenamefont {Chacko}\ \emph {et~al.}(2006)\citenamefont {Chacko},
  \citenamefont {Goh},\ and\ \citenamefont {Harnik}}]{Chacko:2005pe}%
  \BibitemOpen
  \bibfield  {author} {\bibinfo {author} {\bibfnamefont {Z.}~\bibnamefont
  {Chacko}}, \bibinfo {author} {\bibfnamefont {H.-S.}\ \bibnamefont {Goh}}, \
  and\ \bibinfo {author} {\bibfnamefont {R.}~\bibnamefont {Harnik}},\ }\href
  {\doibase 10.1103/PhysRevLett.96.231802} {\bibfield  {journal} {\bibinfo
  {journal} {Phys.Rev.Lett.}\ }\textbf {\bibinfo {volume} {96}},\ \bibinfo
  {pages} {231802} (\bibinfo {year} {2006})},\ \Eprint
  {http://arxiv.org/abs/hep-ph/0506256} {arXiv:hep-ph/0506256 [hep-ph]}
  \BibitemShut {NoStop}%
\bibitem [{\citenamefont {Cui}\ and\ \citenamefont
  {Shuve}(2015)}]{Cui:2014twa}%
  \BibitemOpen
  \bibfield  {author} {\bibinfo {author} {\bibfnamefont {Y.}~\bibnamefont
  {Cui}}\ and\ \bibinfo {author} {\bibfnamefont {B.}~\bibnamefont {Shuve}},\
  }\href {\doibase 10.1007/JHEP02(2015)049} {\bibfield  {journal} {\bibinfo
  {journal} {JHEP}\ }\textbf {\bibinfo {volume} {02}},\ \bibinfo {pages} {049}
  (\bibinfo {year} {2015})},\ \Eprint {http://arxiv.org/abs/1409.6729}
  {arXiv:1409.6729 [hep-ph]} \BibitemShut {NoStop}%
\bibitem [{\citenamefont {Craig}\ \emph {et~al.}(2015)\citenamefont {Craig},
  \citenamefont {Katz}, \citenamefont {Strassler},\ and\ \citenamefont
  {Sundrum}}]{Craig:2015pha}%
  \BibitemOpen
  \bibfield  {author} {\bibinfo {author} {\bibfnamefont {N.}~\bibnamefont
  {Craig}}, \bibinfo {author} {\bibfnamefont {A.}~\bibnamefont {Katz}},
  \bibinfo {author} {\bibfnamefont {M.}~\bibnamefont {Strassler}}, \ and\
  \bibinfo {author} {\bibfnamefont {R.}~\bibnamefont {Sundrum}},\ }\href@noop
  {} {\  (\bibinfo {year} {2015})},\ \Eprint {http://arxiv.org/abs/1501.05310}
  {arXiv:1501.05310 [hep-ph]} \BibitemShut {NoStop}%
\bibitem [{\citenamefont {Liu}\ and\ \citenamefont
  {Tweedie}(2015)}]{Liu:2015bma}%
  \BibitemOpen
  \bibfield  {author} {\bibinfo {author} {\bibfnamefont {Z.}~\bibnamefont
  {Liu}}\ and\ \bibinfo {author} {\bibfnamefont {B.}~\bibnamefont {Tweedie}},\
  }\href {\doibase 10.1007/JHEP06(2015)042} {\bibfield  {journal} {\bibinfo
  {journal} {JHEP}\ }\textbf {\bibinfo {volume} {06}},\ \bibinfo {pages} {042}
  (\bibinfo {year} {2015})},\ \Eprint {http://arxiv.org/abs/1503.05923}
  {arXiv:1503.05923 [hep-ph]} \BibitemShut {NoStop}%
\bibitem [{\citenamefont {Csaki}\ \emph
  {et~al.}(2015{\natexlab{a}})\citenamefont {Csaki}, \citenamefont {Kuflik},
  \citenamefont {Lombardo}, \citenamefont {Slone},\ and\ \citenamefont
  {Volansky}}]{Csaki:2015uza}%
  \BibitemOpen
  \bibfield  {author} {\bibinfo {author} {\bibfnamefont {C.}~\bibnamefont
  {Csaki}}, \bibinfo {author} {\bibfnamefont {E.}~\bibnamefont {Kuflik}},
  \bibinfo {author} {\bibfnamefont {S.}~\bibnamefont {Lombardo}}, \bibinfo
  {author} {\bibfnamefont {O.}~\bibnamefont {Slone}}, \ and\ \bibinfo {author}
  {\bibfnamefont {T.}~\bibnamefont {Volansky}},\ }\href {\doibase
  10.1007/JHEP08(2015)016} {\bibfield  {journal} {\bibinfo  {journal} {JHEP}\
  }\textbf {\bibinfo {volume} {08}},\ \bibinfo {pages} {016} (\bibinfo {year}
  {2015}{\natexlab{a}})},\ \Eprint {http://arxiv.org/abs/1505.00784}
  {arXiv:1505.00784 [hep-ph]} \BibitemShut {NoStop}%
\bibitem [{\citenamefont {Curtin}\ and\ \citenamefont
  {Verhaaren}(2015)}]{Curtin:2015fna}%
  \BibitemOpen
  \bibfield  {author} {\bibinfo {author} {\bibfnamefont {D.}~\bibnamefont
  {Curtin}}\ and\ \bibinfo {author} {\bibfnamefont {C.~B.}\ \bibnamefont
  {Verhaaren}},\ }\href@noop {} {\  (\bibinfo {year} {2015})},\ \Eprint
  {http://arxiv.org/abs/1506.06141} {arXiv:1506.06141 [hep-ph]} \BibitemShut
  {NoStop}%
\bibitem [{\citenamefont {Evans}\ and\ \citenamefont
  {Shelton}(2016)}]{Evans:2016zau}%
  \BibitemOpen
  \bibfield  {author} {\bibinfo {author} {\bibfnamefont {J.~A.}\ \bibnamefont
  {Evans}}\ and\ \bibinfo {author} {\bibfnamefont {J.}~\bibnamefont
  {Shelton}},\ }\href {\doibase 10.1007/JHEP04(2016)056} {\bibfield  {journal}
  {\bibinfo  {journal} {JHEP}\ }\textbf {\bibinfo {volume} {04}},\ \bibinfo
  {pages} {056} (\bibinfo {year} {2016})},\ \Eprint
  {http://arxiv.org/abs/1601.01326} {arXiv:1601.01326 [hep-ph]} \BibitemShut
  {NoStop}%
\bibitem [{\citenamefont {{ATLAS
  Collaboration}}(2012{\natexlab{a}})}]{ATLAS:2012av}%
  \BibitemOpen
  \bibfield  {author} {\bibinfo {author} {\bibnamefont {{ATLAS
  Collaboration}}},\ }\href {\doibase 10.1103/PhysRevLett.108.251801}
  {\bibfield  {journal} {\bibinfo  {journal} {Phys. Rev. Lett.}\ }\textbf
  {\bibinfo {volume} {108}},\ \bibinfo {pages} {251801} (\bibinfo {year}
  {2012}{\natexlab{a}})},\ \Eprint {http://arxiv.org/abs/1203.1303}
  {arXiv:1203.1303 [hep-ex]} \BibitemShut {NoStop}%
\bibitem [{\citenamefont {{ATLAS
  Collaboration}}(2013{\natexlab{a}})}]{Aad:2012kw}%
  \BibitemOpen
  \bibfield  {author} {\bibinfo {author} {\bibnamefont {{ATLAS
  Collaboration}}},\ }\href {\doibase 10.1016/j.physletb.2013.02.058}
  {\bibfield  {journal} {\bibinfo  {journal} {Phys. Lett.}\ }\textbf {\bibinfo
  {volume} {B721}},\ \bibinfo {pages} {32} (\bibinfo {year}
  {2013}{\natexlab{a}})},\ \Eprint {http://arxiv.org/abs/1210.0435}
  {arXiv:1210.0435 [hep-ex]} \BibitemShut {NoStop}%
\bibitem [{\citenamefont {{ATLAS
  Collaboration}}(2013{\natexlab{b}})}]{Aad:2012zx}%
  \BibitemOpen
  \bibfield  {author} {\bibinfo {author} {\bibnamefont {{ATLAS
  Collaboration}}},\ }\href {\doibase 10.1016/j.physletb.2013.01.042}
  {\bibfield  {journal} {\bibinfo  {journal} {Phys. Lett.}\ }\textbf {\bibinfo
  {volume} {B719}},\ \bibinfo {pages} {280} (\bibinfo {year}
  {2013}{\natexlab{b}})},\ \Eprint {http://arxiv.org/abs/1210.7451}
  {arXiv:1210.7451 [hep-ex]} \BibitemShut {NoStop}%
\bibitem [{\citenamefont {{ATLAS
  Collaboration}}(2015{\natexlab{a}})}]{Aad:2015uaa}%
  \BibitemOpen
  \bibfield  {author} {\bibinfo {author} {\bibnamefont {{ATLAS
  Collaboration}}},\ }\href {\doibase 10.1103/PhysRevD.92.012010} {\bibfield
  {journal} {\bibinfo  {journal} {Phys. Rev.}\ }\textbf {\bibinfo {volume}
  {D92}},\ \bibinfo {pages} {15} (\bibinfo {year} {2015}{\natexlab{a}})},\
  \Eprint {http://arxiv.org/abs/1504.03634v2} {arXiv:1504.03634v2} \BibitemShut
  {NoStop}%
\bibitem [{\citenamefont {{ATLAS
  Collaboration}}(2015{\natexlab{b}})}]{Aad:2015asa}%
  \BibitemOpen
  \bibfield  {author} {\bibinfo {author} {\bibnamefont {{ATLAS
  Collaboration}}},\ }\href {\doibase 10.1016/j.physletb.2015.02.015}
  {\bibfield  {journal} {\bibinfo  {journal} {Phys.Lett.}\ }\textbf {\bibinfo
  {volume} {B743}},\ \bibinfo {pages} {15} (\bibinfo {year}
  {2015}{\natexlab{b}})},\ \Eprint {http://arxiv.org/abs/1501.04020}
  {arXiv:1501.04020 [hep-ex]} \BibitemShut {NoStop}%
\bibitem [{\citenamefont {{ATLAS
  Collaboration}}(2014{\natexlab{a}})}]{Aad:2014yea}%
  \BibitemOpen
  \bibfield  {author} {\bibinfo {author} {\bibnamefont {{ATLAS
  Collaboration}}},\ }\href {\doibase 10.1007/JHEP11(2014)088} {\bibfield
  {journal} {\bibinfo  {journal} {JHEP}\ }\textbf {\bibinfo {volume} {11}},\
  \bibinfo {pages} {088} (\bibinfo {year} {2014}{\natexlab{a}})},\ \Eprint
  {http://arxiv.org/abs/1409.0746} {arXiv:1409.0746 [hep-ex]} \BibitemShut
  {NoStop}%
\bibitem [{\citenamefont {{ATLAS
  Collaboration}}(2015{\natexlab{c}})}]{Aad:2015rba}%
  \BibitemOpen
  \bibfield  {author} {\bibinfo {author} {\bibnamefont {{ATLAS
  Collaboration}}},\ }\href {\doibase 10.1103/PhysRevD.92.072004} {\bibfield
  {journal} {\bibinfo  {journal} {Phys. Rev.}\ }\textbf {\bibinfo {volume}
  {D92}},\ \bibinfo {pages} {072004} (\bibinfo {year} {2015}{\natexlab{c}})},\
  \Eprint {http://arxiv.org/abs/1504.05162} {arXiv:1504.05162 [hep-ex]}
  \BibitemShut {NoStop}%
\bibitem [{\citenamefont {{CMS
  Collaboration}}(2012{\natexlab{a}})}]{Chatrchyan:2012sp}%
  \BibitemOpen
  \bibfield  {author} {\bibinfo {author} {\bibnamefont {{CMS Collaboration}}},\
  }\href {\doibase 10.1016/j.physletb.2012.06.023} {\bibfield  {journal}
  {\bibinfo  {journal} {Phys. Lett.}\ }\textbf {\bibinfo {volume} {B713}},\
  \bibinfo {pages} {408} (\bibinfo {year} {2012}{\natexlab{a}})},\ \Eprint
  {http://arxiv.org/abs/1205.0272} {arXiv:1205.0272 [hep-ex]} \BibitemShut
  {NoStop}%
\bibitem [{\citenamefont {{CMS Collaboration}}(2013)}]{Chatrchyan:2012jna}%
  \BibitemOpen
  \bibfield  {author} {\bibinfo {author} {\bibnamefont {{CMS Collaboration}}},\
  }\href {\doibase 10.1007/JHEP02(2013)085} {\bibfield  {journal} {\bibinfo
  {journal} {JHEP}\ }\textbf {\bibinfo {volume} {02}},\ \bibinfo {pages} {085}
  (\bibinfo {year} {2013})},\ \Eprint {http://arxiv.org/abs/1211.2472}
  {arXiv:1211.2472 [hep-ex]} \BibitemShut {NoStop}%
\bibitem [{\citenamefont {{CMS
  Collaboration}}(2015{\natexlab{a}})}]{CMS:2014wda}%
  \BibitemOpen
  \bibfield  {author} {\bibinfo {author} {\bibnamefont {{CMS Collaboration}}},\
  }\href {\doibase 10.1103/PhysRevD.91.012007} {\bibfield  {journal} {\bibinfo
  {journal} {Phys.Rev.}\ }\textbf {\bibinfo {volume} {D91}},\ \bibinfo {pages}
  {012007} (\bibinfo {year} {2015}{\natexlab{a}})},\ \Eprint
  {http://arxiv.org/abs/1411.6530} {arXiv:1411.6530 [hep-ex]} \BibitemShut
  {NoStop}%
\bibitem [{\citenamefont {{CMS
  Collaboration}}(2015{\natexlab{b}})}]{CMS:2014hka}%
  \BibitemOpen
  \bibfield  {author} {\bibinfo {author} {\bibnamefont {{CMS Collaboration}}},\
  }\href {\doibase 10.1103/PhysRevD.91.052012} {\bibfield  {journal} {\bibinfo
  {journal} {Phys. Rev.}\ }\textbf {\bibinfo {volume} {D91}},\ \bibinfo {pages}
  {052012} (\bibinfo {year} {2015}{\natexlab{b}})},\ \Eprint
  {http://arxiv.org/abs/1411.6977} {arXiv:1411.6977 [hep-ex]} \BibitemShut
  {NoStop}%
\bibitem [{\citenamefont {Aaij}\ \emph {et~al.}(2015)\citenamefont {Aaij} \emph
  {et~al.}}]{Aaij:2014nma}%
  \BibitemOpen
  \bibfield  {author} {\bibinfo {author} {\bibfnamefont {R.}~\bibnamefont
  {Aaij}} \emph {et~al.} (\bibinfo {collaboration} {LHCb Collaboration}),\
  }\href {\doibase 10.1140/epjc/s10052-015-3344-6} {\bibfield  {journal}
  {\bibinfo  {journal} {Eur. Phys. J.}\ }\textbf {\bibinfo {volume} {C75}},\
  \bibinfo {pages} {152} (\bibinfo {year} {2015})},\ \Eprint
  {http://arxiv.org/abs/1412.3021} {arXiv:1412.3021 [hep-ex]} \BibitemShut
  {NoStop}%
\bibitem [{\citenamefont {{CMS
  Collaboration}}(2015{\natexlab{c}})}]{Khachatryan:2014mea}%
  \BibitemOpen
  \bibfield  {author} {\bibinfo {author} {\bibnamefont {{CMS Collaboration}}},\
  }\href {\doibase 10.1103/PhysRevLett.114.061801} {\bibfield  {journal}
  {\bibinfo  {journal} {Phys. Rev. Lett.}\ }\textbf {\bibinfo {volume} {114}},\
  \bibinfo {pages} {061801} (\bibinfo {year} {2015}{\natexlab{c}})},\ \Eprint
  {http://arxiv.org/abs/1409.4789} {arXiv:1409.4789 [hep-ex]} \BibitemShut
  {NoStop}%
\bibitem [{\citenamefont {{ATLAS
  Collaboration}}(2013{\natexlab{c}})}]{MuonRoITrigger}%
  \BibitemOpen
  \bibfield  {author} {\bibinfo {author} {\bibnamefont {{ATLAS
  Collaboration}}},\ }\href {\doibase 10.1088/1748-0221/8/07/P07015} {\bibfield
   {journal} {\bibinfo  {journal} {JINST}\ }\textbf {\bibinfo {volume} {8}},\
  \bibinfo {pages} {P07015} (\bibinfo {year} {2013}{\natexlab{c}})},\ \Eprint
  {http://arxiv.org/abs/1305.2284} {arXiv:1305.2284} \BibitemShut {NoStop}%
\bibitem [{\citenamefont {{ATLAS
  Collaboration}}(2014{\natexlab{b}})}]{MSVXRECO}%
  \BibitemOpen
  \bibfield  {author} {\bibinfo {author} {\bibnamefont {{ATLAS
  Collaboration}}},\ }\href {\doibase 10.1088/1748-0221/9/02/P02001} {\bibfield
   {journal} {\bibinfo  {journal} {JINST}\ }\textbf {\bibinfo {volume} {9}},\
  \bibinfo {pages} {P02001} (\bibinfo {year} {2014}{\natexlab{b}})},\ \Eprint
  {http://arxiv.org/abs/1311.7070} {arXiv:1311.7070} \BibitemShut {NoStop}%
\bibitem [{Note1()}]{Note1}%
  \BibitemOpen
  \bibinfo {note} {In case of bandwidth limitations, the non-iso trigger rate
  can be adjusted to populate the control regions with the necessary
  statistics.}\BibitemShut {Stop}%
\bibitem [{Note2()}]{Note2}%
  \BibitemOpen
  \bibinfo {note} {For example, it may be that $\protect \mathrm {MET}^\prime $
  is the more useful variable, since it more directly isolates the properties
  of the individual jet producing the DV, but this detail does not affect our
  discussion.}\BibitemShut {Stop}%
\bibitem [{Note3()}]{Note3}%
  \BibitemOpen
  \bibinfo {note} {If a bin is so small that its expected occupation is much
  less than 1, then the experimentally derived upper limit on its expected
  occupation will significantly overestimate $r_\protect \mathrm {non\mhyphen
  iso\to iso }$. The minimum useful bin width thus depends on the background
  rate in a given search.}\BibitemShut {Stop}%
\bibitem [{\citenamefont {Aad}\ \emph {et~al.}(2016)\citenamefont {Aad} \emph
  {et~al.}}]{Aad:2015sms}%
  \BibitemOpen
  \bibfield  {author} {\bibinfo {author} {\bibfnamefont {G.}~\bibnamefont
  {Aad}} \emph {et~al.} (\bibinfo {collaboration} {ATLAS}),\ }\href {\doibase
  10.1007/JHEP02(2016)062} {\bibfield  {journal} {\bibinfo  {journal} {JHEP}\
  }\textbf {\bibinfo {volume} {02}},\ \bibinfo {pages} {062} (\bibinfo {year}
  {2016})},\ \Eprint {http://arxiv.org/abs/1511.05542} {arXiv:1511.05542
  [hep-ex]} \BibitemShut {NoStop}%
\bibitem [{\citenamefont {Curtin}\ \emph {et~al.}(2014)\citenamefont {Curtin},
  \citenamefont {Essig}, \citenamefont {Gori}, \citenamefont {Jaiswal},
  \citenamefont {Katz} \emph {et~al.}}]{Curtin:2013fra}%
  \BibitemOpen
  \bibfield  {author} {\bibinfo {author} {\bibfnamefont {D.}~\bibnamefont
  {Curtin}}, \bibinfo {author} {\bibfnamefont {R.}~\bibnamefont {Essig}},
  \bibinfo {author} {\bibfnamefont {S.}~\bibnamefont {Gori}}, \bibinfo {author}
  {\bibfnamefont {P.}~\bibnamefont {Jaiswal}}, \bibinfo {author} {\bibfnamefont
  {A.}~\bibnamefont {Katz}},  \emph {et~al.},\ }\href {\doibase
  10.1103/PhysRevD.90.075004} {\bibfield  {journal} {\bibinfo  {journal}
  {Phys.Rev.}\ }\textbf {\bibinfo {volume} {D90}},\ \bibinfo {pages} {075004}
  (\bibinfo {year} {2014})},\ \Eprint {http://arxiv.org/abs/1312.4992}
  {arXiv:1312.4992 [hep-ph]} \BibitemShut {NoStop}%
\bibitem [{\citenamefont {{ATLAS
  Collaboration}}(2015{\natexlab{d}})}]{ATL-PHYS-PUB-2015-023}%
  \BibitemOpen
  \bibfield  {author} {\bibinfo {author} {\bibnamefont {{ATLAS
  Collaboration}}},\ }\href {http://cds.cern.ch/record/2037700} {\emph
  {\bibinfo {title} {{Expected performance of missing transverse momentum
  reconstruction for the ATLAS detector at $\sqrt{s}= 13~\text{TeV}$}}}},\
  \bibinfo {type} {Tech. Rep.}\ \bibinfo {number} {ATL-PHYS-PUB-2015-023}\
  (\bibinfo  {institution} {CERN},\ \bibinfo {address} {Geneva},\ \bibinfo
  {year} {2015})\BibitemShut {NoStop}%
\bibitem [{\citenamefont {Andersen}\ \emph {et~al.}(2013)\citenamefont
  {Andersen} \emph {et~al.}}]{Heinemeyer:2013tqa}%
  \BibitemOpen
  \bibfield  {author} {\bibinfo {author} {\bibfnamefont {J.~R.}\ \bibnamefont
  {Andersen}} \emph {et~al.} (\bibinfo {collaboration} {LHC Higgs Cross Section
  Working Group}),\ }\href {\doibase 10.5170/CERN-2013-004} {\  (\bibinfo
  {year} {2013}),\ 10.5170/CERN-2013-004},\ \Eprint
  {http://arxiv.org/abs/1307.1347} {arXiv:1307.1347 [hep-ph]} \BibitemShut
  {NoStop}%
\bibitem [{\citenamefont {Alwall}\ \emph {et~al.}(2014)\citenamefont {Alwall},
  \citenamefont {Frederix}, \citenamefont {Frixione}, \citenamefont {Hirschi},
  \citenamefont {Maltoni}, \citenamefont {Mattelaer}, \citenamefont {Shao},
  \citenamefont {Stelzer}, \citenamefont {Torrielli},\ and\ \citenamefont
  {Zaro}}]{Alwall:2014hca}%
  \BibitemOpen
  \bibfield  {author} {\bibinfo {author} {\bibfnamefont {J.}~\bibnamefont
  {Alwall}}, \bibinfo {author} {\bibfnamefont {R.}~\bibnamefont {Frederix}},
  \bibinfo {author} {\bibfnamefont {S.}~\bibnamefont {Frixione}}, \bibinfo
  {author} {\bibfnamefont {V.}~\bibnamefont {Hirschi}}, \bibinfo {author}
  {\bibfnamefont {F.}~\bibnamefont {Maltoni}}, \bibinfo {author} {\bibfnamefont
  {O.}~\bibnamefont {Mattelaer}}, \bibinfo {author} {\bibfnamefont {H.~S.}\
  \bibnamefont {Shao}}, \bibinfo {author} {\bibfnamefont {T.}~\bibnamefont
  {Stelzer}}, \bibinfo {author} {\bibfnamefont {P.}~\bibnamefont {Torrielli}},
  \ and\ \bibinfo {author} {\bibfnamefont {M.}~\bibnamefont {Zaro}},\ }\href
  {\doibase 10.1007/JHEP07(2014)079} {\bibfield  {journal} {\bibinfo  {journal}
  {JHEP}\ }\textbf {\bibinfo {volume} {07}},\ \bibinfo {pages} {079} (\bibinfo
  {year} {2014})},\ \Eprint {http://arxiv.org/abs/1405.0301} {arXiv:1405.0301
  [hep-ph]} \BibitemShut {NoStop}%
\bibitem [{\citenamefont {Sjostrand}\ \emph {et~al.}(2008)\citenamefont
  {Sjostrand}, \citenamefont {Mrenna},\ and\ \citenamefont
  {Skands}}]{Sjostrand:2007gs}%
  \BibitemOpen
  \bibfield  {author} {\bibinfo {author} {\bibfnamefont {T.}~\bibnamefont
  {Sjostrand}}, \bibinfo {author} {\bibfnamefont {S.}~\bibnamefont {Mrenna}}, \
  and\ \bibinfo {author} {\bibfnamefont {P.~Z.}\ \bibnamefont {Skands}},\
  }\href {\doibase 10.1016/j.cpc.2008.01.036} {\bibfield  {journal} {\bibinfo
  {journal} {Comput.Phys.Commun.}\ }\textbf {\bibinfo {volume} {178}},\
  \bibinfo {pages} {852} (\bibinfo {year} {2008})},\ \Eprint
  {http://arxiv.org/abs/0710.3820} {arXiv:0710.3820 [hep-ph]} \BibitemShut
  {NoStop}%
\bibitem [{Note4()}]{Note4}%
  \BibitemOpen
  \bibinfo {note} {In \cite {Curtin:2014cca}, it was shown that the matched
  production of $h$ with one extra jet within the EFT framework of the Madgraph
  model gives a surprisingly accurate representation of the Higgs $p_T$
  spectrum.}\BibitemShut {Stop}%
\bibitem [{\citenamefont {{ATLAS
  Collaboration}}(2012{\natexlab{b}})}]{8tevpileup}%
  \BibitemOpen
  \bibfield  {author} {\bibinfo {author} {\bibnamefont {{ATLAS
  Collaboration}}},\ }\href
  {https://twiki.cern.ch/twiki/bin/view/AtlasPublic/LuminosityPublicResults}
  {\emph {\bibinfo {title} {{ATLAS run 1 Luminosity Public Results}}}},\
  \bibinfo {type} {Tech. Rep.}\ (\bibinfo  {institution} {CERN},\ \bibinfo
  {address} {Geneva},\ \bibinfo {year} {2012})\BibitemShut {NoStop}%
\bibitem [{\citenamefont {{ATLAS
  Collaboration}}(2015{\natexlab{e}})}]{ATL-PHYS-PUB-2015-008}%
  \BibitemOpen
  \bibfield  {author} {\bibinfo {author} {\bibnamefont {{ATLAS
  Collaboration}}},\ }\href {http://cds.cern.ch/record/2008700} {\emph
  {\bibinfo {title} {{An imaging algorithm for vertex reconstruction for ATLAS
  Run-2}}}},\ \bibinfo {type} {Tech. Rep.}\ \bibinfo {number}
  {ATL-PHYS-PUB-2015-008}\ (\bibinfo  {institution} {CERN},\ \bibinfo {address}
  {Geneva},\ \bibinfo {year} {2015})\BibitemShut {NoStop}%
\bibitem [{\citenamefont {{ATLAS
  Collaboration}}(2015{\natexlab{f}})}]{13tevpileup}%
  \BibitemOpen
  \bibfield  {author} {\bibinfo {author} {\bibnamefont {{ATLAS
  Collaboration}}},\ }\href
  {https://twiki.cern.ch/twiki/bin/view/AtlasPublic/LuminosityPublicResultsRun2}
  {\emph {\bibinfo {title} {{ATLAS run 2 Luminosity Public Results}}}},\
  \bibinfo {type} {Tech. Rep.}\ (\bibinfo  {institution} {CERN},\ \bibinfo
  {address} {Geneva},\ \bibinfo {year} {2015})\BibitemShut {NoStop}%
\bibitem [{\citenamefont {{ATLAS
  Collaboration}}(2014{\natexlab{c}})}]{PLOT-UPGRADE-2014-003}%
  \BibitemOpen
  \bibfield  {author} {\bibinfo {author} {\bibnamefont {{ATLAS
  Collaboration}}},\ }\href
  {http://atlas.web.cern.ch/Atlas/GROUPS/PHYSICS/UPGRADE/PLOT-UPGRADE-2014-003/}
  {\emph {\bibinfo {title} {{Primary Vertex and b-Tagging Performance under
  HL-LHC conditions}}}},\ \bibinfo {type} {Tech. Rep.}\ \bibinfo {number}
  {ATL-PLOT-UPGRADE-2014-003}\ (\bibinfo  {institution} {CERN},\ \bibinfo
  {address} {Geneva},\ \bibinfo {year} {2014})\BibitemShut {NoStop}%
\bibitem [{Note5()}]{Note5}%
  \BibitemOpen
  \bibinfo {note} {Systematic uncertainties from weak correlations will be
  subdominant, as discussed in Section~\ref
  {ss.importantconsiderations}.}\BibitemShut {Stop}%
\bibitem [{Note6()}]{Note6}%
  \BibitemOpen
  \bibinfo {note} {The actual experimental sensitivity in the short lifetime
  regime will also depend on any differences between the contributions to the
  MET from QCD jets versus $X$ particles that decay before the MS, as well as a
  more detailed treatment of the DV reconstruction efficiency for $X$ particles
  decaying in the outer regions of the HCAL.}\BibitemShut {Stop}%
\bibitem [{\citenamefont {Morningstar}\ and\ \citenamefont
  {Peardon}(1999)}]{Morningstar:1999rf}%
  \BibitemOpen
  \bibfield  {author} {\bibinfo {author} {\bibfnamefont {C.~J.}\ \bibnamefont
  {Morningstar}}\ and\ \bibinfo {author} {\bibfnamefont {M.~J.}\ \bibnamefont
  {Peardon}},\ }\href {\doibase 10.1103/PhysRevD.60.034509} {\bibfield
  {journal} {\bibinfo  {journal} {Phys.Rev.}\ }\textbf {\bibinfo {volume}
  {D60}},\ \bibinfo {pages} {034509} (\bibinfo {year} {1999})},\ \Eprint
  {http://arxiv.org/abs/hep-lat/9901004} {arXiv:hep-lat/9901004 [hep-lat]}
  \BibitemShut {NoStop}%
\bibitem [{\citenamefont {Juknevich}\ \emph {et~al.}(2009)\citenamefont
  {Juknevich}, \citenamefont {Melnikov},\ and\ \citenamefont
  {Strassler}}]{Juknevich:2009ji}%
  \BibitemOpen
  \bibfield  {author} {\bibinfo {author} {\bibfnamefont {J.~E.}\ \bibnamefont
  {Juknevich}}, \bibinfo {author} {\bibfnamefont {D.}~\bibnamefont {Melnikov}},
  \ and\ \bibinfo {author} {\bibfnamefont {M.~J.}\ \bibnamefont {Strassler}},\
  }\href {\doibase 10.1088/1126-6708/2009/07/055} {\bibfield  {journal}
  {\bibinfo  {journal} {JHEP}\ }\textbf {\bibinfo {volume} {0907}},\ \bibinfo
  {pages} {055} (\bibinfo {year} {2009})},\ \Eprint
  {http://arxiv.org/abs/0903.0883} {arXiv:0903.0883 [hep-ph]} \BibitemShut
  {NoStop}%
\bibitem [{\citenamefont {Juknevich}(2010)}]{Juknevich:2009gg}%
  \BibitemOpen
  \bibfield  {author} {\bibinfo {author} {\bibfnamefont {J.~E.}\ \bibnamefont
  {Juknevich}},\ }\href {\doibase 10.1007/JHEP08(2010)121} {\bibfield
  {journal} {\bibinfo  {journal} {JHEP}\ }\textbf {\bibinfo {volume} {1008}},\
  \bibinfo {pages} {121} (\bibinfo {year} {2010})},\ \Eprint
  {http://arxiv.org/abs/0911.5616} {arXiv:0911.5616 [hep-ph]} \BibitemShut
  {NoStop}%
\bibitem [{\citenamefont {Curtin}\ and\ \citenamefont {Tsai}(2016)}]{fthpheno}%
  \BibitemOpen
  \bibfield  {author} {\bibinfo {author} {\bibfnamefont {D.}~\bibnamefont
  {Curtin}}\ and\ \bibinfo {author} {\bibfnamefont {Y.}~\bibnamefont {Tsai}},\
  }\href@noop {} {\  (\bibinfo {year} {2016})},\ \Eprint
  {http://arxiv.org/abs/TO APPEAR} {arXiv:TO APPEAR [hep-ph]} \BibitemShut
  {NoStop}%
\bibitem [{\citenamefont {Kang}\ and\ \citenamefont
  {Luty}(2009)}]{Kang:2008ea}%
  \BibitemOpen
  \bibfield  {author} {\bibinfo {author} {\bibfnamefont {J.}~\bibnamefont
  {Kang}}\ and\ \bibinfo {author} {\bibfnamefont {M.~A.}\ \bibnamefont
  {Luty}},\ }\href {\doibase 10.1088/1126-6708/2009/11/065} {\bibfield
  {journal} {\bibinfo  {journal} {JHEP}\ }\textbf {\bibinfo {volume} {0911}},\
  \bibinfo {pages} {065} (\bibinfo {year} {2009})},\ \Eprint
  {http://arxiv.org/abs/0805.4642} {arXiv:0805.4642 [hep-ph]} \BibitemShut
  {NoStop}%
\bibitem [{\citenamefont {Burdman}\ \emph {et~al.}(2008)\citenamefont
  {Burdman}, \citenamefont {Chacko}, \citenamefont {Goh}, \citenamefont
  {Harnik},\ and\ \citenamefont {Krenke}}]{Burdman:2008ek}%
  \BibitemOpen
  \bibfield  {author} {\bibinfo {author} {\bibfnamefont {G.}~\bibnamefont
  {Burdman}}, \bibinfo {author} {\bibfnamefont {Z.}~\bibnamefont {Chacko}},
  \bibinfo {author} {\bibfnamefont {H.-S.}\ \bibnamefont {Goh}}, \bibinfo
  {author} {\bibfnamefont {R.}~\bibnamefont {Harnik}}, \ and\ \bibinfo {author}
  {\bibfnamefont {C.~A.}\ \bibnamefont {Krenke}},\ }\href {\doibase
  10.1103/PhysRevD.78.075028} {\bibfield  {journal} {\bibinfo  {journal}
  {Phys.Rev.}\ }\textbf {\bibinfo {volume} {D78}},\ \bibinfo {pages} {075028}
  (\bibinfo {year} {2008})},\ \Eprint {http://arxiv.org/abs/0805.4667}
  {arXiv:0805.4667 [hep-ph]} \BibitemShut {NoStop}%
\bibitem [{\citenamefont {Harnik}\ and\ \citenamefont
  {Wizansky}(2009)}]{Harnik:2008ax}%
  \BibitemOpen
  \bibfield  {author} {\bibinfo {author} {\bibfnamefont {R.}~\bibnamefont
  {Harnik}}\ and\ \bibinfo {author} {\bibfnamefont {T.}~\bibnamefont
  {Wizansky}},\ }\href {\doibase 10.1103/PhysRevD.80.075015} {\bibfield
  {journal} {\bibinfo  {journal} {Phys.Rev.}\ }\textbf {\bibinfo {volume}
  {D80}},\ \bibinfo {pages} {075015} (\bibinfo {year} {2009})},\ \Eprint
  {http://arxiv.org/abs/0810.3948} {arXiv:0810.3948 [hep-ph]} \BibitemShut
  {NoStop}%
\bibitem [{\citenamefont {Harnik}\ \emph {et~al.}(2011)\citenamefont {Harnik},
  \citenamefont {Kribs},\ and\ \citenamefont {Martin}}]{Harnik:2011mv}%
  \BibitemOpen
  \bibfield  {author} {\bibinfo {author} {\bibfnamefont {R.}~\bibnamefont
  {Harnik}}, \bibinfo {author} {\bibfnamefont {G.~D.}\ \bibnamefont {Kribs}}, \
  and\ \bibinfo {author} {\bibfnamefont {A.}~\bibnamefont {Martin}},\ }\href
  {\doibase 10.1103/PhysRevD.84.035029} {\bibfield  {journal} {\bibinfo
  {journal} {Phys.Rev.}\ }\textbf {\bibinfo {volume} {D84}},\ \bibinfo {pages}
  {035029} (\bibinfo {year} {2011})},\ \Eprint {http://arxiv.org/abs/1106.2569}
  {arXiv:1106.2569 [hep-ph]} \BibitemShut {NoStop}%
\bibitem [{\citenamefont {Burdman}\ \emph {et~al.}(2015)\citenamefont
  {Burdman}, \citenamefont {Chacko}, \citenamefont {Harnik}, \citenamefont
  {de~Lima},\ and\ \citenamefont {Verhaaren}}]{Burdman:2014zta}%
  \BibitemOpen
  \bibfield  {author} {\bibinfo {author} {\bibfnamefont {G.}~\bibnamefont
  {Burdman}}, \bibinfo {author} {\bibfnamefont {Z.}~\bibnamefont {Chacko}},
  \bibinfo {author} {\bibfnamefont {R.}~\bibnamefont {Harnik}}, \bibinfo
  {author} {\bibfnamefont {L.}~\bibnamefont {de~Lima}}, \ and\ \bibinfo
  {author} {\bibfnamefont {C.~B.}\ \bibnamefont {Verhaaren}},\ }\href {\doibase
  10.1103/PhysRevD.91.055007} {\bibfield  {journal} {\bibinfo  {journal}
  {Phys.Rev.}\ }\textbf {\bibinfo {volume} {D91}},\ \bibinfo {pages} {055007}
  (\bibinfo {year} {2015})},\ \Eprint {http://arxiv.org/abs/1411.3310}
  {arXiv:1411.3310 [hep-ph]} \BibitemShut {NoStop}%
\bibitem [{\citenamefont {Chacko}\ \emph {et~al.}(2015)\citenamefont {Chacko},
  \citenamefont {Curtin},\ and\ \citenamefont {Verhaaren}}]{Chacko:2015fbc}%
  \BibitemOpen
  \bibfield  {author} {\bibinfo {author} {\bibfnamefont {Z.}~\bibnamefont
  {Chacko}}, \bibinfo {author} {\bibfnamefont {D.}~\bibnamefont {Curtin}}, \
  and\ \bibinfo {author} {\bibfnamefont {C.~B.}\ \bibnamefont {Verhaaren}},\
  }\href@noop {} {\  (\bibinfo {year} {2015})},\ \Eprint
  {http://arxiv.org/abs/1512.05782} {arXiv:1512.05782 [hep-ph]} \BibitemShut
  {NoStop}%
\bibitem [{Note7()}]{Note7}%
  \BibitemOpen
  \bibinfo {note} {Thermodynamic arguments \cite {JuknevichPhD} indicate that
  $\protect \mathcal {O}(50\%)$ of produced glueballs end up in the $0^{++}$
  state, but those predictions are highly uncertain due to our ignorance of
  pure-gauge hadronization. More sophisticated methods of parameterizing this
  ignorance have been proposed in \cite {Chacko:2015fbc} but they are not
  suitable for exotic Higgs decays where glueball masses cannot be
  neglected.}\BibitemShut {Stop}%
\bibitem [{\citenamefont {Cowan}\ \emph {et~al.}(2011)\citenamefont {Cowan},
  \citenamefont {Cranmer}, \citenamefont {Gross},\ and\ \citenamefont
  {Vitells}}]{Cowan:2010js}%
  \BibitemOpen
  \bibfield  {author} {\bibinfo {author} {\bibfnamefont {G.}~\bibnamefont
  {Cowan}}, \bibinfo {author} {\bibfnamefont {K.}~\bibnamefont {Cranmer}},
  \bibinfo {author} {\bibfnamefont {E.}~\bibnamefont {Gross}}, \ and\ \bibinfo
  {author} {\bibfnamefont {O.}~\bibnamefont {Vitells}},\ }\href {\doibase
  10.1140/epjc/s10052-011-1554-0, 10.1140/epjc/s10052-013-2501-z} {\bibfield
  {journal} {\bibinfo  {journal} {Eur. Phys. J.}\ }\textbf {\bibinfo {volume}
  {C71}},\ \bibinfo {pages} {1554} (\bibinfo {year} {2011})},\ \bibinfo {note}
  {[Erratum: Eur. Phys. J.C73,2501(2013)]},\ \Eprint
  {http://arxiv.org/abs/1007.1727} {arXiv:1007.1727 [physics.data-an]}
  \BibitemShut {NoStop}%
\bibitem [{\citenamefont {Vidal}(2014)}]{Marono:2014jta}%
  \BibitemOpen
  \bibfield  {author} {\bibinfo {author} {\bibfnamefont {M.}~\bibnamefont
  {Vidal}} (\bibinfo {collaboration} {CMS Collaboration}),\ }in\ \href
  {http://inspirehep.net/record/1315323/files/arXiv:1409.1711.pdf} {\emph
  {\bibinfo {booktitle} {{Proceedings, 2nd Conference on Large Hadron Collider
  Physics Conference (LHCP 2014)}}}}\ (\bibinfo {year} {2014})\ \Eprint
  {http://arxiv.org/abs/1409.1711} {arXiv:1409.1711 [hep-ex]} \BibitemShut
  {NoStop}%
\bibitem [{\citenamefont {Flechl}(2015)}]{Flechl:2015foa}%
  \BibitemOpen
  \bibfield  {author} {\bibinfo {author} {\bibfnamefont {M.}~\bibnamefont
  {Flechl}} (\bibinfo {collaboration} {ATLAS Collaboration, CMS
  Collaboration}),\ }\bibfield  {booktitle} {\emph {\bibinfo {booktitle}
  {{Proceedings, 4th Symposium on Prospects in the Physics of Discrete
  Symmetries (DISCRETE 2014)}}},\ }\href {\doibase
  10.1088/1742-6596/631/1/012028} {\bibfield  {journal} {\bibinfo  {journal}
  {J. Phys. Conf. Ser.}\ }\textbf {\bibinfo {volume} {631}},\ \bibinfo {pages}
  {012028} (\bibinfo {year} {2015})},\ \Eprint
  {http://arxiv.org/abs/1503.00632} {arXiv:1503.00632 [hep-ex]} \BibitemShut
  {NoStop}%
\bibitem [{Note8()}]{Note8}%
  \BibitemOpen
  \bibinfo {note} {We use the most optimistic projections for the precision of
  Higgs coupling measurements (CMS) and assume they apply for both ATLAS and
  CMS, in order to demonstrate the large gain in sensitivity achieved by the
  direct searches for displaced vertices.}\BibitemShut {Stop}%
\bibitem [{\citenamefont {Rolbiecki}\ and\ \citenamefont
  {Sakurai}(2015)}]{Rolbiecki:2015gsa}%
  \BibitemOpen
  \bibfield  {author} {\bibinfo {author} {\bibfnamefont {K.}~\bibnamefont
  {Rolbiecki}}\ and\ \bibinfo {author} {\bibfnamefont {K.}~\bibnamefont
  {Sakurai}},\ }\href {\doibase 10.1007/JHEP11(2015)091} {\bibfield  {journal}
  {\bibinfo  {journal} {JHEP}\ }\textbf {\bibinfo {volume} {11}},\ \bibinfo
  {pages} {091} (\bibinfo {year} {2015})},\ \Eprint
  {http://arxiv.org/abs/1506.08799} {arXiv:1506.08799 [hep-ph]} \BibitemShut
  {NoStop}%
\bibitem [{\citenamefont {Chen}\ \emph {et~al.}(1997)\citenamefont {Chen},
  \citenamefont {Drees},\ and\ \citenamefont {Gunion}}]{Chen:1996ap}%
  \BibitemOpen
  \bibfield  {author} {\bibinfo {author} {\bibfnamefont {C.~H.}\ \bibnamefont
  {Chen}}, \bibinfo {author} {\bibfnamefont {M.}~\bibnamefont {Drees}}, \ and\
  \bibinfo {author} {\bibfnamefont {J.~F.}\ \bibnamefont {Gunion}},\ }\href
  {\doibase 10.1103/PhysRevD.60.039901, 10.1103/PhysRevD.55.330} {\bibfield
  {journal} {\bibinfo  {journal} {Phys. Rev.}\ }\textbf {\bibinfo {volume}
  {D55}},\ \bibinfo {pages} {330} (\bibinfo {year} {1997})},\ \bibinfo {note}
  {[Erratum: Phys. Rev.D60,039901(1999)]},\ \Eprint
  {http://arxiv.org/abs/hep-ph/9607421} {arXiv:hep-ph/9607421 [hep-ph]}
  \BibitemShut {NoStop}%
\bibitem [{\citenamefont {Giudice}\ \emph {et~al.}(1998)\citenamefont
  {Giudice}, \citenamefont {Luty}, \citenamefont {Murayama},\ and\
  \citenamefont {Rattazzi}}]{Giudice:1998xp}%
  \BibitemOpen
  \bibfield  {author} {\bibinfo {author} {\bibfnamefont {G.~F.}\ \bibnamefont
  {Giudice}}, \bibinfo {author} {\bibfnamefont {M.~A.}\ \bibnamefont {Luty}},
  \bibinfo {author} {\bibfnamefont {H.}~\bibnamefont {Murayama}}, \ and\
  \bibinfo {author} {\bibfnamefont {R.}~\bibnamefont {Rattazzi}},\ }\href
  {\doibase 10.1088/1126-6708/1998/12/027} {\bibfield  {journal} {\bibinfo
  {journal} {JHEP}\ }\textbf {\bibinfo {volume} {12}},\ \bibinfo {pages} {027}
  (\bibinfo {year} {1998})},\ \Eprint {http://arxiv.org/abs/hep-ph/9810442}
  {arXiv:hep-ph/9810442 [hep-ph]} \BibitemShut {NoStop}%
\bibitem [{\citenamefont {Randall}\ and\ \citenamefont
  {Sundrum}(1999)}]{Randall:1998uk}%
  \BibitemOpen
  \bibfield  {author} {\bibinfo {author} {\bibfnamefont {L.}~\bibnamefont
  {Randall}}\ and\ \bibinfo {author} {\bibfnamefont {R.}~\bibnamefont
  {Sundrum}},\ }\href {\doibase 10.1016/S0550-3213(99)00359-4} {\bibfield
  {journal} {\bibinfo  {journal} {Nucl. Phys.}\ }\textbf {\bibinfo {volume}
  {B557}},\ \bibinfo {pages} {79} (\bibinfo {year} {1999})},\ \Eprint
  {http://arxiv.org/abs/hep-th/9810155} {arXiv:hep-th/9810155 [hep-th]}
  \BibitemShut {NoStop}%
\bibitem [{\citenamefont {Baumgart}\ \emph {et~al.}(2009)\citenamefont
  {Baumgart}, \citenamefont {Cheung}, \citenamefont {Ruderman}, \citenamefont
  {Wang},\ and\ \citenamefont {Yavin}}]{Baumgart:2009tn}%
  \BibitemOpen
  \bibfield  {author} {\bibinfo {author} {\bibfnamefont {M.}~\bibnamefont
  {Baumgart}}, \bibinfo {author} {\bibfnamefont {C.}~\bibnamefont {Cheung}},
  \bibinfo {author} {\bibfnamefont {J.~T.}\ \bibnamefont {Ruderman}}, \bibinfo
  {author} {\bibfnamefont {L.-T.}\ \bibnamefont {Wang}}, \ and\ \bibinfo
  {author} {\bibfnamefont {I.}~\bibnamefont {Yavin}},\ }\href {\doibase
  10.1088/1126-6708/2009/04/014} {\bibfield  {journal} {\bibinfo  {journal}
  {JHEP}\ }\textbf {\bibinfo {volume} {04}},\ \bibinfo {pages} {014} (\bibinfo
  {year} {2009})},\ \Eprint {http://arxiv.org/abs/0901.0283} {arXiv:0901.0283
  [hep-ph]} \BibitemShut {NoStop}%
\bibitem [{\citenamefont {Chan}\ \emph {et~al.}(2012)\citenamefont {Chan},
  \citenamefont {Low}, \citenamefont {Morrissey},\ and\ \citenamefont
  {Spray}}]{Chan:2011aa}%
  \BibitemOpen
  \bibfield  {author} {\bibinfo {author} {\bibfnamefont {Y.~F.}\ \bibnamefont
  {Chan}}, \bibinfo {author} {\bibfnamefont {M.}~\bibnamefont {Low}}, \bibinfo
  {author} {\bibfnamefont {D.~E.}\ \bibnamefont {Morrissey}}, \ and\ \bibinfo
  {author} {\bibfnamefont {A.~P.}\ \bibnamefont {Spray}},\ }\href {\doibase
  10.1007/JHEP05(2012)155} {\bibfield  {journal} {\bibinfo  {journal} {JHEP}\
  }\textbf {\bibinfo {volume} {05}},\ \bibinfo {pages} {155} (\bibinfo {year}
  {2012})},\ \Eprint {http://arxiv.org/abs/1112.2705} {arXiv:1112.2705
  [hep-ph]} \BibitemShut {NoStop}%
\bibitem [{\citenamefont {Co}\ \emph {et~al.}(2015)\citenamefont {Co},
  \citenamefont {D'Eramo}, \citenamefont {Hall},\ and\ \citenamefont
  {Pappadopulo}}]{Co:2015pka}%
  \BibitemOpen
  \bibfield  {author} {\bibinfo {author} {\bibfnamefont {R.~T.}\ \bibnamefont
  {Co}}, \bibinfo {author} {\bibfnamefont {F.}~\bibnamefont {D'Eramo}},
  \bibinfo {author} {\bibfnamefont {L.~J.}\ \bibnamefont {Hall}}, \ and\
  \bibinfo {author} {\bibfnamefont {D.}~\bibnamefont {Pappadopulo}},\ }\href
  {\doibase 10.1088/1475-7516/2015/12/024} {\bibfield  {journal} {\bibinfo
  {journal} {JCAP}\ }\textbf {\bibinfo {volume} {1512}},\ \bibinfo {pages}
  {024} (\bibinfo {year} {2015})},\ \Eprint {http://arxiv.org/abs/1506.07532}
  {arXiv:1506.07532 [hep-ph]} \BibitemShut {NoStop}%
\bibitem [{\citenamefont {Falkowski}\ \emph {et~al.}(2014)\citenamefont
  {Falkowski}, \citenamefont {Hochberg},\ and\ \citenamefont
  {Ruderman}}]{Falkowski:2014sma}%
  \BibitemOpen
  \bibfield  {author} {\bibinfo {author} {\bibfnamefont {A.}~\bibnamefont
  {Falkowski}}, \bibinfo {author} {\bibfnamefont {Y.}~\bibnamefont {Hochberg}},
  \ and\ \bibinfo {author} {\bibfnamefont {J.~T.}\ \bibnamefont {Ruderman}},\
  }\href {\doibase 10.1007/JHEP11(2014)140} {\bibfield  {journal} {\bibinfo
  {journal} {JHEP}\ }\textbf {\bibinfo {volume} {11}},\ \bibinfo {pages} {140}
  (\bibinfo {year} {2014})},\ \Eprint {http://arxiv.org/abs/1409.2872}
  {arXiv:1409.2872 [hep-ph]} \BibitemShut {NoStop}%
\bibitem [{\citenamefont {Helo}\ \emph {et~al.}(2014)\citenamefont {Helo},
  \citenamefont {Hirsch},\ and\ \citenamefont {Kovalenko}}]{Helo:2013esa}%
  \BibitemOpen
  \bibfield  {author} {\bibinfo {author} {\bibfnamefont {J.~C.}\ \bibnamefont
  {Helo}}, \bibinfo {author} {\bibfnamefont {M.}~\bibnamefont {Hirsch}}, \ and\
  \bibinfo {author} {\bibfnamefont {S.}~\bibnamefont {Kovalenko}},\ }\href
  {\doibase 10.1103/PhysRevD.89.073005, 10.1103/PhysRevD.93.099902} {\bibfield
  {journal} {\bibinfo  {journal} {Phys. Rev.}\ }\textbf {\bibinfo {volume}
  {D89}},\ \bibinfo {pages} {073005} (\bibinfo {year} {2014})},\ \bibinfo
  {note} {[Erratum: Phys. Rev.D93,no.9,099902(2016)]},\ \Eprint
  {http://arxiv.org/abs/1312.2900} {arXiv:1312.2900 [hep-ph]} \BibitemShut
  {NoStop}%
\bibitem [{\citenamefont {Antusch}\ \emph {et~al.}(2016)\citenamefont
  {Antusch}, \citenamefont {Cazzato},\ and\ \citenamefont
  {Fischer}}]{Antusch:2016vyf}%
  \BibitemOpen
  \bibfield  {author} {\bibinfo {author} {\bibfnamefont {S.}~\bibnamefont
  {Antusch}}, \bibinfo {author} {\bibfnamefont {E.}~\bibnamefont {Cazzato}}, \
  and\ \bibinfo {author} {\bibfnamefont {O.}~\bibnamefont {Fischer}},\
  }\href@noop {} {\  (\bibinfo {year} {2016})},\ \Eprint
  {http://arxiv.org/abs/1604.02420} {arXiv:1604.02420 [hep-ph]} \BibitemShut
  {NoStop}%
\bibitem [{\citenamefont {Graesser}(2007{\natexlab{a}})}]{Graesser:2007yj}%
  \BibitemOpen
  \bibfield  {author} {\bibinfo {author} {\bibfnamefont {M.~L.}\ \bibnamefont
  {Graesser}},\ }\href {\doibase 10.1103/PhysRevD.76.075006} {\bibfield
  {journal} {\bibinfo  {journal} {Phys. Rev.}\ }\textbf {\bibinfo {volume}
  {D76}},\ \bibinfo {pages} {075006} (\bibinfo {year} {2007}{\natexlab{a}})},\
  \Eprint {http://arxiv.org/abs/0704.0438} {arXiv:0704.0438 [hep-ph]}
  \BibitemShut {NoStop}%
\bibitem [{\citenamefont {Graesser}(2007{\natexlab{b}})}]{Graesser:2007pc}%
  \BibitemOpen
  \bibfield  {author} {\bibinfo {author} {\bibfnamefont {M.~L.}\ \bibnamefont
  {Graesser}},\ }\href@noop {} {\  (\bibinfo {year} {2007}{\natexlab{b}})},\
  \Eprint {http://arxiv.org/abs/0705.2190} {arXiv:0705.2190 [hep-ph]}
  \BibitemShut {NoStop}%
\bibitem [{\citenamefont {Maiezza}\ \emph {et~al.}(2015)\citenamefont
  {Maiezza}, \citenamefont {Nemevšek},\ and\ \citenamefont
  {Nesti}}]{Maiezza:2015lza}%
  \BibitemOpen
  \bibfield  {author} {\bibinfo {author} {\bibfnamefont {A.}~\bibnamefont
  {Maiezza}}, \bibinfo {author} {\bibfnamefont {M.}~\bibnamefont {Nemevšek}},
  \ and\ \bibinfo {author} {\bibfnamefont {F.}~\bibnamefont {Nesti}},\ }\href
  {\doibase 10.1103/PhysRevLett.115.081802} {\bibfield  {journal} {\bibinfo
  {journal} {Phys. Rev. Lett.}\ }\textbf {\bibinfo {volume} {115}},\ \bibinfo
  {pages} {081802} (\bibinfo {year} {2015})},\ \Eprint
  {http://arxiv.org/abs/1503.06834} {arXiv:1503.06834 [hep-ph]} \BibitemShut
  {NoStop}%
\bibitem [{\citenamefont {Batell}\ \emph {et~al.}(2016)\citenamefont {Batell},
  \citenamefont {Pospelov},\ and\ \citenamefont {Shuve}}]{Batell:2016zod}%
  \BibitemOpen
  \bibfield  {author} {\bibinfo {author} {\bibfnamefont {B.}~\bibnamefont
  {Batell}}, \bibinfo {author} {\bibfnamefont {M.}~\bibnamefont {Pospelov}}, \
  and\ \bibinfo {author} {\bibfnamefont {B.}~\bibnamefont {Shuve}},\
  }\href@noop {} {\  (\bibinfo {year} {2016})},\ \Eprint
  {http://arxiv.org/abs/1604.06099} {arXiv:1604.06099 [hep-ph]} \BibitemShut
  {NoStop}%
\bibitem [{\citenamefont {{ATLAS
  Collaboration}}(2012{\natexlab{c}})}]{Aad:2011rr}%
  \BibitemOpen
  \bibfield  {author} {\bibinfo {author} {\bibnamefont {{ATLAS
  Collaboration}}},\ }\href {\doibase 10.1016/j.physletb.2011.12.054}
  {\bibfield  {journal} {\bibinfo  {journal} {Phys. Lett.}\ }\textbf {\bibinfo
  {volume} {B707}},\ \bibinfo {pages} {438} (\bibinfo {year}
  {2012}{\natexlab{c}})},\ \Eprint {http://arxiv.org/abs/1109.0525}
  {arXiv:1109.0525 [hep-ex]} \BibitemShut {NoStop}%
\bibitem [{\citenamefont {{CMS
  Collaboration}}(2012{\natexlab{b}})}]{Chatrchyan:2012dk}%
  \BibitemOpen
  \bibfield  {author} {\bibinfo {author} {\bibnamefont {{CMS Collaboration}}},\
  }\href {\doibase 10.1007/JHEP04(2012)084} {\bibfield  {journal} {\bibinfo
  {journal} {JHEP}\ }\textbf {\bibinfo {volume} {04}},\ \bibinfo {pages} {084}
  (\bibinfo {year} {2012}{\natexlab{b}})},\ \Eprint
  {http://arxiv.org/abs/1202.4617} {arXiv:1202.4617 [hep-ex]} \BibitemShut
  {NoStop}%
\bibitem [{\citenamefont {Csaki}\ \emph
  {et~al.}(2015{\natexlab{b}})\citenamefont {Csaki}, \citenamefont {Kuflik},
  \citenamefont {Lombardo},\ and\ \citenamefont {Slone}}]{Csaki:2015fba}%
  \BibitemOpen
  \bibfield  {author} {\bibinfo {author} {\bibfnamefont {C.}~\bibnamefont
  {Csaki}}, \bibinfo {author} {\bibfnamefont {E.}~\bibnamefont {Kuflik}},
  \bibinfo {author} {\bibfnamefont {S.}~\bibnamefont {Lombardo}}, \ and\
  \bibinfo {author} {\bibfnamefont {O.}~\bibnamefont {Slone}},\ }\href
  {\doibase 10.1103/PhysRevD.92.073008} {\bibfield  {journal} {\bibinfo
  {journal} {Phys. Rev.}\ }\textbf {\bibinfo {volume} {D92}},\ \bibinfo {pages}
  {073008} (\bibinfo {year} {2015}{\natexlab{b}})},\ \Eprint
  {http://arxiv.org/abs/1508.01522} {arXiv:1508.01522 [hep-ph]} \BibitemShut
  {NoStop}%
\bibitem [{\citenamefont {Juknevich}()}]{JuknevichPhD}%
  \BibitemOpen
  \bibfield  {author} {\bibinfo {author} {\bibfnamefont {J.}~\bibnamefont
  {Juknevich}},\ }\emph {\bibinfo {title} {{Phenomenology of pure-gauge hidden
  valleys at Hadron colliders}}},\ \href {\doibase 10.7282/T34F1QHM} {Ph.D.
  thesis},\ \bibinfo  {school} {Rutgers U., Piscataway}\BibitemShut {NoStop}%
\end{thebibliography}%

\end{document}